\documentclass[12pt]{article}
\usepackage[a4paper,margin=1in]{geometry}
\usepackage[T1]{fontenc}
\usepackage[utf8]{inputenc}
\IfFileExists{lmodern.sty}{\usepackage{lmodern}\usepackage{microtype}}{}
\usepackage{setspace}
\usepackage{amsmath,amssymb,amsthm}
\usepackage{booktabs}
\usepackage{threeparttable}
\usepackage{multirow}
\usepackage{array}
\usepackage{dcolumn}
\newcolumntype{d}[1]{D{.}{.}{#1}}

\usepackage{graphicx}
\graphicspath{{figures/}}
\usepackage{caption}
\usepackage{float}
\usepackage{flafter}
\usepackage{rotating}

\usepackage[authoryear,round]{natbib}
\usepackage{xr-hyper}
\makeatletter
\newcommand*{\addFileDependency}[1]{%
  \typeout{(#1)}%
  \@addtofilelist{#1}%
  \IfFileExists{#1}{}{\typeout{No file #1.}}%
}
\makeatother

\usepackage[colorlinks=true,linkcolor=blue!60!black,citecolor=blue!60!black,urlcolor=blue!60!black]{hyperref}

\usepackage{xcolor}
\usepackage[colorinlistoftodos,textsize=footnotesize]{todonotes}
\usepackage{enumitem}
\setlist{nosep}
\usepackage[most]{tcolorbox}
\usepackage{listings}
\newtcolorbox{importantbox}{colback=red!4!white,colframe=red!55!black,boxrule=0.7pt,
  left=6pt,right=6pt,top=5pt,bottom=5pt,arc=1pt}
\newif\ifshowpending
\showpendingfalse   

\newcommand{\HD}{HD}
\newcommand{\HDlong}{Hidden Dissent}
\newcommand{\Mzero}{M0}
\newcommand{\Mone}{M1}
\newcommand{\Mtwo}{M2}
\newcommand{\Mthree}{M3}

\newcommand{\Qtau}{Q_{\tau}}

\newcommand{\pinball}{\rho_{\tau}}
\newcommand{\gain}{\mathrm{Gain}}

\theoremstyle{definition}



\begin{document}
\title{\Large\textbf{Reading the Room with AI:\\
FOMC Deliberation and Growth-at-Risk}
\thanks{We thank Kwok Ping Tsang and Zichao Yang for sharing the speaker-level scores of their Hidden Dissent index, and Stephen Hansen (University College London) and Domenico Giannone (Johns Hopkins University) for helpful discussions. The views expressed are those of the authors and do not necessarily reflect those of BBVA. All errors are ours.}}

\author{%
\normalsize \'Alvaro Ortiz\thanks{Corresponding author. BBVA Research, Madrid. E-mail: \href{mailto:alvaro.ortiz@bbva.com}{alvaro.ortiz@bbva.com}. ORCID: \href{https://orcid.org/0000-0003-4373-0815}{0000-0003-4373-0815}.}\\
{\footnotesize BBVA Research}
\and
\normalsize Tomasa Rodrigo\thanks{BBVA Research, Madrid. E-mail: \href{mailto:tomasa.rodrigo@bbva.com}{tomasa.rodrigo@bbva.com}. ORCID: \href{https://orcid.org/0009-0003-4573-0736}{0009-0003-4573-0736}.}\\
{\footnotesize BBVA Research}}

\date{\vspace{-0.6em}November 2026}
\maketitle
\vspace{-1.2em}

\begin{abstract}
\singlespacing\small\noindent
Macro-at-Risk models forecast the growth distribution from public data; policy is set by a committee arguing over their meaning. From FOMC Minutes, public three weeks after each meeting, we add several dimensions of deliberation to a growth quantile model on FRED-MD's 123 series. Only disagreement present in the discussion but absent from the vote lowers out-of-sample loss by 4--6 percent, chiefly in the center, beyond professional, staff and Committee forecasts. Much of deliberation's informational value lies in the recorded discussion itself, and in the disagreement it reveals, rather than in the risk assessments the Committee chooses to state.

\end{abstract}
\thispagestyle{empty}
\newpage
\setcounter{page}{1}

\section{Introduction}\label{sec:intro}

Monetary policy in most advanced economies is set by a committee whose members receive the same staff briefing, observe the same public data and deliberate before they vote. Whether a committee's deliberation produces information that its members did not bring to the room, and whether what the committee says adds to what its staff already knows, are long-standing questions \citep{BlinderMorgan2005,RomerRomer2008,HansenMcMahonPrat2018}. This paper gives it an out-of-sample answer for the Federal Open Market Committee. We measure several dimensions of its deliberation from the Minutes, which are public three weeks after each meeting, and ask whether each improves the forecast of the distribution of future output growth beyond a data-rich Macro-at-Risk benchmark, an approach central banks increasingly use to assess the risks around their outlook \citep{ABG2019,ECBOP396,AGLW2026risks}. One dimension does: the disagreement that is present in the recorded discussion but absent from the vote. The Committee's published projections, its stated balance of risks and the other dimensions of the discussion do not. A large part of the informational value of central bank deliberation lies in the recorded discussion itself, and in the disagreement it reveals, rather than in the assessments of risk the Committee chooses to state.

The question sits in a long debate about what monetary policy committees add. One view is that a committee adds information: by deliberating, its members pool interpretations that no one of them has alone. Groups make better policy decisions than individuals \citep{BlinderMorgan2005}, and the FOMC's deliberation, in which all nineteen participants take part before its twelve voting members decide, is best understood as members interpreting a state they all observe rather than revealing private data \citep{HansenMcMahonPrat2018}. If this view is right, the information is in the discussion, and the natural place to look for it is the disagreement the discussion generates. Much of that disagreement never reaches the vote: dissent in voice is far more common than dissent in the record \citep{Meade2005}, because members who care about their reputation present a united front \citep{VisserSwank2007,Levy2007,SwankSwankVisser2008}, and where dissent is recorded it anticipates policy and moves markets \citep{GerlachKristen2004,MadeiraMadeira2019}. Policymakers' perceptions of risk, read from the transcripts, shape the policy stance beyond the staff forecast \citep{CHMX2023}, and the Minutes carry much of the transcripts' information to the public three weeks after each meeting \citep{ApelBlixGrimaldiHull2022}.

The other view is that the committee's own voice adds little to what its staff already knows. The Federal Reserve's informational advantage lies with its staff, the Board economists who prepare the forecast before each meeting and do not vote: that forecast beats the market's \citep{RomerRomer2000}. The projections that the FOMC participants, the governors and Reserve Bank presidents who deliberate and publish their own outlook, add to it carry little information beyond the staff forecast \citep{RomerRomer2008}; one reading is that the projections a policymaker publishes are not a full statement of what he or she thinks \citep{Tillmann2011}.

The two views can both be right. The first is about the discussion; the second is about what the Committee publishes when the discussion is over. Nothing contradictory follows if the information is in the discussion and does not reach the published projections. We test this: we take the public data, ask whether the Committee's published outlook improves the forecast built from those data, and then ask whether the disagreement in the discussion does, not for the point forecast of growth but for its whole distribution, the object the Growth-at-Risk literature uses to assess macroeconomic risk.

Two obstacles stand in the way of a credible answer. The first is the standard of comparison. Most of the Growth-at-Risk literature asks whether a signal improves a quantile regression of future growth on current growth and a financial-conditions index, the specification of \citet{ABG2019}. That standard is informative about the signal's usefulness but not about its content: the regression summarizes the economy in two variables, so a signal that improves on it may be adding information that is public but outside those two, rather than information that is not in the public data at all. We compare instead with the machine-learning version of Growth-at-Risk of \citet{CFGGW2025}, a regularized quantile regression on the 123 macroeconomic and financial series of FRED-MD \citep{McCrackenNg2016}, which forecasts as well as deep quantile networks on the same data. A signal that improves on it carries predictive content beyond what the benchmark extracts from the public data. We estimate this benchmark with and without the financial-conditions index; the two-variable regressions are special cases of it, and we report every comparison against all four. The second obstacle is the freedom to choose after seeing the results. With one country, quarterly data and seventy test targets, an author can try benchmarks, penalties, quantiles and samples until the signal looks good, and the reader cannot tell. We closed that door: the measures, the benchmarks, the penalties, the evaluation sample and the tests were fixed in dated records before any forecast was generated; every pre-specified test is reported, including one that fails; and every exercise added afterwards is labelled as such.

We measure the deliberative cycle from the Minutes, so that every measure is available three weeks after the meeting, and date each one conservatively, using at each forecast origin only the Minutes of meetings held in the previous quarter. The measures follow the stages of a meeting: the Committee's stated balance of risks, the insurance motive of its decisions, the salience of growth in the discussion, three readings of how much the room disagrees---the Hidden Dissent index of \citet{TsangYang2025}, trained on the language of formal dissents, a transparent measure of the views the Minutes attribute to groups of participants, and the gap between the balance of risks as stated and as weighted by its quantifiers---and, as a placebo, the length of the Minutes. Five public measures of uncertainty and risk, the professional consensus forecast and the Committee's published growth projections, from 1983, provide the external comparison. Forecasts are made out of sample for average growth over the next four quarters, at 19 quantiles, for the 70 targets from 2008Q4 to 2026Q1.

The evidence supports two findings about the forecast, of different strength, and a third about where the content lies.

\emph{First, deliberation carries information about the distribution of growth, mainly about its body (the central part of the distribution rather than the tails), and the information is in the disagreement that does not reach the vote.} Against the public panel, Hidden Dissent lowers the out-of-sample quantile loss by 4.1 percent over the whole distribution and by 5.6 percent in its center; once financial conditions are also in the benchmark, by 5.9 and 8.3 percent. The whole-distribution comparison is one of the two tests fixed in advance, and it survives adjustment for multiple testing. The contribution does not depend on how the benchmark is built: the central gain is the same, to within a tenth of a point, in the canonical Growth-at-Risk regressions; it rises when the benchmark is re-tuned so that the public panel works in the center of the distribution; it is unchanged when the public panel is rebuilt from the data vintages available at each date; and it survives adding to the public panel the professional consensus forecast, against which it is 12.6 percent, the staff's own Greenbook forecast (13.6 percent, over the 50 targets the Greenbook covers) and the regional dispersion of the Beige Book (6.7 percent). No other dimension of the deliberation reproduces it. The stated balance of risks, the insurance motive and the salience of growth add nothing, the placebo makes the forecast worse, and the quantifier-based measure of disagreement does not replicate the result; the transparent measure of attributed views improves the forecast under the pre-specified design, but most of its gain disappears when the benchmark is re-tuned.

\emph{Second, deliberation adds to the lower region of the distribution, but only conditionally.} Once financial conditions and the full public panel are both in the benchmark, the index improves the lower block of quantiles by 4.5 percent---the second test fixed in advance---and the gain survives real-time data and a shorter public history. But it is absent against current growth, financial conditions or the public panel separately; it sits at the twentieth and twenty-fifth percentiles rather than at the fifth; it does not survive treating the index as one more series of the regularized panel; and conditional on the professional consensus it is not distinguishable from zero. Downside risk remains, in the first instance, a matter of financial conditions; deliberation adds to it only once they are accounted for. Nor do the public proxies for uncertainty reproduce it: under real-time information, against the benchmark with financial conditions, Hidden Dissent is the only one of twelve signals whose lower-block contribution is established. Nor is either region's content in what the Committee declares. Conditional on the central tendency of its members' growth projections, or on its width, Hidden Dissent improves the center by 7.6 and 7.0 percent and the lower block by 7.3 and 7.7 percent, over the same 70 targets; the central gains are spread over time, the lower ones rest more on a few episodes.

\emph{Third, as far as the evidence can tell, the content is in the disagreement the Committee records, not in the words spoken in the room.} With the speaker-level scores of \citet{TsangYang2025}, computed from the transcripts of every meeting from 1976 to 2018, disagreement measured from what was said does not improve the public-data forecast over the 41 targets the transcripts cover, whether among voters, non-voters or all participants, while the Minutes index does, and it predicts the center of the distribution better than the voters' own transcript scores ($p=0.01$). Who votes does not matter either: the same president disagrees as much when he votes as when he does not, and neither the rotation of Reserve Bank presidents nor the vacancies on the Board detectably moves the index, though these tests have limited power. The transparency reform of 1993 halved recorded disagreement without detectably changing what it says about growth, and a one-standard-deviation surprise in the index moves the two-year Treasury yield by less than a basis point.

We read these facts as saying that committee disagreement tells us in which direction the forecast based on public data will miss, not how uncertain it is: when disagreement absent from the vote is high, growth comes in above what the public data imply; when it is low, below. Such a signal should shift the center of the predicted distribution rather than widen it, earn its gain there rather than in the tails, and help in ordinary quarters as much as in crises, which is what we find: the gain is not concentrated in the episodes of stress singled out in advance. What the room disagrees about is how to read the data. The comparison with professional forecasters says the same. Disagreement is highest when forecasters are most pessimistic, yet the index's coefficient keeps its sign with and without the consensus, and its gain is largest once the consensus is in the model: given what forecasters expect, high disagreement indicates that growth will come in above the consensus, and low disagreement below it.

\paragraph{Literature.}
The paper relates to three strands of the literature. To the literature on monetary policy committees, it brings out-of-sample evidence that the information of deliberation lies in its disagreement rather than in its stated conclusions, reconciling the evidence that committees aggregate information with the evidence that their declared forecasts add little. For Growth-at-Risk and Macro-at-Risk \citep{ABG2019,AdrianGrinbergLiangMalik2022,PlagborgMoller2020,BrownleesSouza2021,CarrieroClarkMarcellino2024}, it supplies a data-rich public benchmark in which the canonical regressions are nested, a new class of signal, and a finding about location: what deliberation adds is mainly about the center of the distribution, with the tail carried by financial conditions. For text as data in central banking \citep{HansenMcMahon2016,SteklerSymington2016,Ericsson2016,HuangKuan2021,SharpeSinhaHollrah2023}, it moves from the tone of what central banks write to the disagreement their records reveal, in a document public three weeks after the meeting, and against a benchmark of 123 predictors rather than point forecasts; closest to us, \citet{SharpeSinhaHollrah2023} show that the tone of the staff narrative predicts growth and its downside risks, while we find that the Committee's stated view of risks, which a tone measure summarizes, adds nothing to the public data and its disagreement does. The paper also relates to the literature on central-bank information effects \citep{NakamuraSteinsson2018,CieslakSchrimpf2019}---the content we measure is not what announcements convey, since Treasury yields show no detectable response to it on the day of the decision or on the day the Minutes are released (Section~\ref{sec:interpretation})---and to risk management in monetary policy \citep{KilianManganelli2008,EvansEtAl2015,AGLW2026risks}. Two implications follow. For communication, the Minutes are an informational product distinct from the statement and the projections: the disagreement they record carries content about the growth distribution that the declared outlook does not. For committee design, the evidence is consistent with recorded disagreement reflecting deliberation beyond the formal vote. Voting rights shape how presidents communicate and what the Committee decides \citep{EhrmannTietzVisser2021,FosXu2023}, and decisions follow a consensus rather than a majority rule \citep{RiboniRugeMurcia2010}, so that the vote does not record the disagreement and the Minutes do. What carries the information is the disagreement, not the text: in a committee whose members vote expressively, as at the Bank of England \citep{GerlachKristen2004}, the vote split may carry what the Minutes carry at the FOMC, a prediction we do not test. In our data, voting status does not detectably alter participants' expressed disagreement, and voting-composition measures have little explanatory power for the Minutes index; the implications for its predictive content remain tentative.

The paper proceeds as follows. Section~\ref{sec:setting} describes deliberation in the FOMC, the data and the measures. Section~\ref{sec:framework} sets out the benchmark and the evaluation. Section~\ref{sec:results} reports the results and Section~\ref{sec:robustness} how robust they are. Section~\ref{sec:interpretation} discusses what they mean, and Section~\ref{sec:conclusion} concludes. The Supplementary Material contains the construction of the data, the full inference, and every exercise summarized in the text.
\section{Deliberation in the FOMC and the measures}\label{sec:setting}

\subsection{What a committee's deliberation can reveal}\label{sec:deliberation}

A monetary policy committee turns the views of its members into a single decision. Each member arrives with an assessment of the outlook, formed from the same public data, the staff's analysis and, in the case of the Reserve Bank presidents, the information gathered in their districts. The meeting confronts those assessments: members state how they read the economy and the risks around it, respond to each other, and then state the policy they prefer. The decision compresses all of this into one vote, and the public record reconstructs part of it afterwards. The literature on committees gives two reasons why the discussion may contain information that the decision and the public statements do not. The first is aggregation: a committee pools interpretations that no single member holds, and the way members read a commonly observed state is itself informative \citep{BlinderMorgan2005,HansenMcMahonPrat2018}. The second is compression: a committee that decides by consensus, or under a strong chair, reports a single view even when the room is divided, so that disagreement present in the discussion is systematically absent from the vote \citep{Meade2005,RiboniRugeMurcia2010}. On the first view the information of deliberation is in how the room reads the data; on the second, much of it is in the disagreement that the decision hides. Neither view implies that the Committee's stated conclusions carry the information: a statement of the balance of risks, or a justification of the decision, is the output of the same compression, negotiated before it is published.

This gives the measurement strategy of the paper. Rather than choosing one signal, we read each stage of the meeting from the public record and ask which of them, if any, carries information the data do not. At the start of the cycle, what participants bring into the room is partly visible in their projections, published in the Summary of Economic Projections. In the discussion, the record shows how divided the room is---which we read three ways, from the latent disagreement in the language of the discussion, from the views the Minutes explicitly attribute to different groups of participants, and from the quantifiers attached to the Committee's assessment of risks---what risks the Committee says it sees, and how much of the discussion is about growth. At the decision, the record shows why the Committee says it acts, in particular whether it justifies policy as insurance against unlikely adverse outcomes. And the length of the record, which reflects none of these, serves as a placebo. If deliberation informs the outlook because it aggregates interpretations and the vote compresses them, the information should appear in the measures of disagreement rather than in the stated assessments; if the Committee's stated conclusions carry the information, the balance of risks and the insurance motive should do as well as any measure of disagreement.

\subsection{Central bank deliberation and its measures}\label{sec:cycle}

The FOMC meets eight times a year. In the days before each meeting the staff circulates its forecast, alternative scenarios and the policy options. At four meetings a year participants also submit their own projections of output growth, unemployment and inflation---and, since 2012, of the policy rate---which the Committee publishes as the Summary of Economic Projections (SEP). The meeting itself proceeds in two rounds. In the first, participants state their views on the economic outlook and on the risks around it; in the second, their preferred policy. The Committee then votes.

\begin{figure}[!t]
\centering
\includegraphics[width=\textwidth]{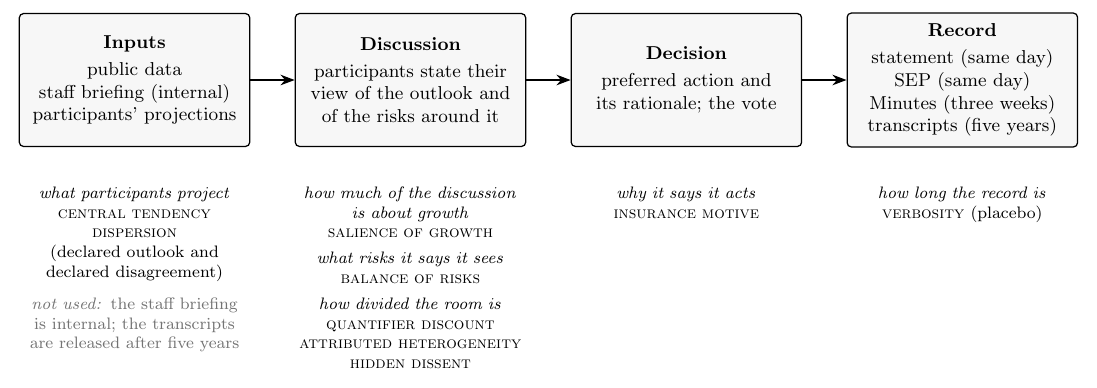}
\caption{One FOMC meeting, and what the paper reads from it. Under each stage, in italics, is the aspect of the meeting that the public record shows, and in small capitals the measure that reads it. The deliberative measures are extracted from the Minutes, published three weeks after the meeting; verbosity, the length of the record, is the placebo. The central tendency and the dispersion summarize the growth projections participants submit before the meeting. The staff briefing is internal and the transcripts are released after five years, so neither enters. A deliberative measure used at forecast origin $t$ comes from the Minutes of meetings held in quarter $t-1$.}
\label{fig:cycle}
\end{figure}

What the public learns, and when, is layered: the statement at the end of the meeting; the Minutes, which record the discussion in the Committee's institutional language---the range of views, how widely each was held, and the balance of risks---three weeks later; the projections, published twice a year in the Monetary Policy Report to the Congress until 2007 and in the SEP since (on the day of the meeting since April 2011, with the Minutes before then); and the verbatim transcripts, which record who said what, after five years. Figure~\ref{fig:cycle} lays one meeting out in four stages---inputs, discussion, decision and record---and places under each the aspect of the meeting that the public record shows and the measure that reads it. Every deliberative measure in the paper is read from the Minutes, and is therefore public three weeks after the meeting; the transcripts, which no forecaster could have read at the time, and the internal staff forecast are excluded, and the participants' projections enter in their own right, as the Committee's declared outlook.

A measure used at forecast origin $t$ is computed from the Minutes of the meetings held in quarter $t-1$. The rule is deliberately conservative: by the time the first estimate of growth in quarter $t$ is published, the Minutes of that quarter's meetings are public as well, and the rule leaves them unused. The question is therefore one of historical information content: whether the deliberation, dated this way, contained information about the future distribution of growth that the public data did not.

Table~\ref{tab:signals} lists the measures in the order of the meeting: what participants project before it, what the discussion is about and how divided it is, why the Committee says it acts, and how long the record is. The seven measures read from the Minutes form the deliberative family; the two read from the projections are the Committee's declared outlook, against which the family is compared (Supplementary Material~\ref*{app:data} and \ref*{app:properties} give the construction and the properties of every series).

\begin{table}[!t]
\centering
\caption{The measures}
\label{tab:signals}
\small
\resizebox{\textwidth}{!}{%
\begin{tabular}{lllc}
\toprule
Measure & What it reads & Construction & From \\
\midrule
\multicolumn{4}{l}{\emph{Inputs: what participants bring and project (Beige Book, Monetary Policy Reports and SEP)}} \\
Regional dispersion       & heterogeneity of the information brought to the meeting & dispersion across the twelve district reports of the tone of activity & 1983Q4 \\
Central tendency          & Committee's declared outlook    & midpoint of the central tendency, converted to the target & 1983Q1 \\
Dispersion                & Committee's declared disagreement & width of the central tendency, averaged over the target years & 1983Q1 \\
\multicolumn{4}{l}{\emph{Discussion: what it is about, and what risks the Committee says it sees (Minutes)}} \\
Salience of growth        & room the growth outlook occupies & share of paragraphs & 1994 \\
Balance of risks          & Committee's view of growth risks & Committee-level assessment, divided distinct from balanced & 1994 \\
\multicolumn{4}{l}{\emph{Discussion: how divided the room is (Minutes)}} \\
Quantifier discount       & disagreement about risks       & stated minus quantifier-weighted balance of risks & 1994 \\
Attributed heterogeneity  & explicitly attributed division & share of paragraphs attributing different views to different groups & 1994 \\
Hidden Dissent            & latent disagreement            & trained model of dissent language, \citet{TsangYang2025} & 1994 \\
\multicolumn{4}{l}{\emph{Decision: why the Committee says it acts (Minutes)}} \\
Insurance motive          & policy as insurance            & two-stage language-model classification & 1994 \\
\multicolumn{4}{l}{\emph{Record: how long it is (Minutes)}} \\
Verbosity (placebo)       & nothing                        & log word count & 1994 \\
\multicolumn{4}{l}{\emph{Outside the Committee}} \\
Current growth, FRED-MD   & public macro-financial state   & 123 series, \citet{McCrackenNg2016} & 1959 \\
NFCI                      & financial conditions           & Chicago Fed & 1971 \\
SPF consensus             & professional forecasters' outlook & median forecast of growth, averaged to the target & 1974Q4 \\
Five external signals     & uncertainty and risk           & SPF dispersion, geopolitical risk, policy uncertainty, VIX, oil price & various \\
\bottomrule
\end{tabular}}
\begin{minipage}{0.95\textwidth}\footnotesize\vspace{4pt}
Notes: The seven Minutes-based measures form the deliberative family; they are quarterly means over the meetings of the quarter, dated so that the value used at origin $t$ comes from meetings in quarter $t-1$, and cover the 70 test origins. The projection measures are dated by the publication of each Monetary Policy Report or SEP, the regional dispersion by the Beige Book's, published two weeks before each meeting. Minutes-based signals that start after the public history are standardized on their own available observations and set to zero before them; the projection measures enter in levels (Section~\ref{sec:protocols}). Supplementary Material~\ref*{app:data} gives the construction of every series.
\end{minipage}
\end{table}

The \emph{Committee's declared outlook} is taken from the projections the Committee publishes. In the SEP each participant projects the fourth-quarter-to-fourth-quarter growth of real GDP for the current year and the following two or three, and the Committee publishes the range of the projections and their central tendency, which excludes the three highest and the three lowest. The midpoint of the central tendency is the Committee's declared outlook, its own forecast of the variable we predict, formed by participants before the meeting and published under the Committee's name; the width of the central tendency is its declared disagreement. Neither is a measure of the deliberation, and together they let us ask whether what the Committee does not declare adds to what it does. Because the SEP begins only with the October 2007 meeting, both measures are extended back to 1983Q1 with the projections of the Monetary Policy Report to the Congress, published under the same rule; the calendar-year projections are converted to growth over the four quarters after each origin by a fixed rule, each round is dated by its publication on the same conservative clock as the Minutes, and both measures enter in levels, unpenalized (Supplementary Material~\ref*{app:sep}).

The discussion is read first for what it is about and what the Committee says it sees. The \emph{salience of growth} is the share of the discussion devoted to the growth outlook. The \emph{balance of risks} is the Committee's own assessment of the risks to growth---weighted to the downside, balanced, or weighted to the upside---read only from what the Minutes attribute to the Committee as a whole; a Committee whose members point in opposite directions is recorded as divided, not as balanced. Both, with the quantifier discount and the insurance motive below, follow the constructs with which \citet{CHMX2023} measure policymakers' perceptions of risk in the transcripts, rebuilt from the Minutes so that they are available three weeks after the meeting.

It is then read for how divided the room is, three ways, from the least to the most inferential. The \emph{quantifier discount}, independent of Hidden Dissent in construction, is the gap between the balance of risks as stated and as weighted by the quantifiers the Minutes attach to it (``a few'', ``several'', ``most'' participants). Where the transcripts let one observe the dispersion of individual views directly, the quantifiers are the only trace of that dispersion the Minutes leave.

\emph{Attributed heterogeneity} reads division from the way the Minutes distribute views across the room: when views differ, the drafters say so---``several participants'' saw the recovery as more fragile than ``most''; when they converge, the text speaks with one voice---``participants noted'', ``the Committee judged''. The measure is the share of the paragraphs of the discussion that attribute different views to different groups of participants; a language model extracts the attributions under fixed instructions, one paragraph at a time, and deterministic rules turn them into the measure, so that every observation can be traced to a sentence a reader can check. It was built after the results for the other measures were known and specified before any of its own results were computed.

\emph{Hidden Dissent} \citep{TsangYang2025}, the measure on which the paper's pre-specified tests are built, scores latent disagreement. A deep-learning model is trained to recognize the language of formal dissents and then applied to the Minutes of every meeting, where it scores the disagreement present in the discussion but absent from the vote. The name is the authors'; \emph{hidden} refers to the vote, not to the public: the disagreement does not appear as formal dissent, and it is read from Minutes that are published three weeks after the meeting. The authors show that the index tracks the dispersion of participants' projections in the SEP and predicts deviations of policy from standard rules. We use the series built from the Minutes, available from 1994, from the authors' public data site (\url{https://www.digitecon.org/hidden-dissent-index}).

The \emph{insurance motive}, read at the decision, is the intensity with which policy is justified as protection against unlikely adverse scenarios rather than as a response to the baseline. The record itself gives the placebo: \emph{verbosity}, the logarithm of the length of the Minutes, has no deliberative content and shows what a free regressor with nothing in it does to the forecast.

\subsection{Target, public information and external signals}\label{sec:public}

Following \citet{ABG2019}, the target is the average annualized growth rate of real GDP over the four quarters after the forecast origin,
\[
y_{t+4}=\frac{1}{4}\sum_{j=1}^{4} g_{t+j},
\]
where $g_q$ is the annualized growth rate of quarter $q$. Forecasts are made at the 70 origins from 2007Q4 to 2025Q1, for the targets ending in 2008Q4 to 2026Q1.

The public information set has three parts. Current growth, $g_t$, enters every model. The Chicago Fed National Financial Conditions Index (NFCI), a weighted average of more than a hundred indicators of risk, credit and leverage scaled so that positive values indicate tighter than average conditions, is the conditioning variable of the canonical Growth-at-Risk regression. The FRED-MD database of \citet{McCrackenNg2016} supplies the rest: 123 monthly macroeconomic and financial series, aggregated to quarterly frequency and transformed to stationarity with the database's own codes, of which 62 describe real activity, 23 prices, 34 money, credit and interest rates, and 4 market prices. The panel is assembled from 1973Q1, where the estimation history of the main specification begins; the comparisons with the consensus and with the Committee's projections use a common history from 1974Q4.

The primary results use the current vintage of every series. Section~\ref{sec:realtime} rebuilds the panel from the FRED-MD vintages available at each origin, and current growth from the real GDP vintages, to check that the results do not depend on data revisions.

Five public measures of uncertainty and risk from outside the Committee complete the comparison, the natural rivals of a disagreement index: the dispersion of professional forecasters' growth forecasts in the Survey of Professional Forecasters (SPF), the geopolitical risk index of \citet{CaldaraIacoviello2022}, the economic policy uncertainty index of \citet{BakerBloomDavis2016}, the implied volatility of the S\&P~500 (VIX), and the price of oil. They were specified as a set, before any was evaluated, and enter exactly as the deliberative measures do. Section~\ref{sec:robustness} also uses the SPF median forecast of growth over the four quarters after the origin, the professional consensus, as a benchmark in its own right.

\subsection{Samples}

The language-based measures are available from 1994 and the public history from 1973. The penalties of the benchmark were selected on 92 validation targets ending in 2007Q3; the test sample, the 70 quarterly targets from 2008Q4 to 2026Q1, was never used for a design choice. It contains the financial crisis at its start and the pandemic in its middle; the three pandemic targets, 2020Q1--2020Q3, are kept in estimation and evaluation, and every headline result is also reported without them.
\section{The benchmark and the evaluation}\label{sec:framework}

\subsection{A data-rich Growth-at-Risk benchmark}\label{sec:lineage}

The regression at the origin of the Growth-at-Risk literature, estimated quantile by quantile, is
\begin{equation}\label{eq:abg}
\Qtau\!\left(y_{t+4}\,\middle|\,\mathrm{GDP}_t,\mathrm{NFCI}_t\right)
 = \alpha_{\tau} + \beta_{\tau}\,\mathrm{GDP}_t + \gamma_{\tau}\,\mathrm{NFCI}_t .
\end{equation}
\citet{ABG2019} show that $\gamma_\tau$ is large and negative at the low quantiles and close to zero at the high ones: tighter financial conditions pull down the lower tail of future growth and leave the upper tail where it was. We call the growth-only version of \eqref{eq:abg} \Mzero{} and \eqref{eq:abg} itself \Mone{}.

The benchmark of this paper is the machine-learning version of Growth-at-Risk developed by \citet{CFGGW2025}. It keeps the structure of \eqref{eq:abg} and uses an entire macroeconomic database as regressors: the full public panel $X_t$, with a candidate signal $S_t$ added in the augmented models:
\begin{equation}\label{eq:qr}
\Qtau\!\left(y_{t+4}\,\middle|\,\Omega_t\right)
 = c_{\tau} + \beta_{\tau}\,\mathrm{GDP}_t + \gamma_{\tau}\,\mathrm{NFCI}_t + \Gamma_{\tau}' X_t + \delta_{\tau}\, S_t .
\end{equation}
With 123 regressors and a few hundred quarterly observations an unpenalized quantile regression fits the sample and forecasts nothing; the panel coefficients are therefore estimated by ridge quantile regression,
\begin{equation}\label{eq:ridge}
\min\;
\frac{1}{N_t}\sum_{s\le t-4}\pinball\!\left(y_{s+4}-c_{\tau}-\beta_{\tau}\mathrm{GDP}_s-\gamma_{\tau}\mathrm{NFCI}_s-\Gamma_{\tau}'X_s-\delta_{\tau}S_s\right)
\;+\;\lambda_{b(\tau)}\,\|\Gamma_{\tau}\|_2^2 ,
\end{equation}
where $\pinball(u)=u(\tau-\mathbf{1}\{u<0\})$ is the pinball loss. The penalty is ridge rather than lasso, and it applies to the panel only. Ridge is the choice of \citet{CFGGW2025}, whose regularized regression forecasts as well as deep quantile networks on the same panel, and the one the evidence on macroeconomic prediction supports: the information in a large panel is spread across many correlated series rather than concentrated in a few, so shrinking every coefficient forecasts better than selecting among them \citep{DeMolGiannoneReichlin2008,GiannoneLenzaPrimiceri2021}. The intercept, current growth, the NFCI and the signal are left unpenalized. The signal is the parameter under test, and a penalty chosen to keep 123 correlated series from overfitting would shrink a single regressor toward zero whatever its content; leaving the variable of interest free and regularizing only the high-dimensional controls is the treatment standard in inference with many controls \citep{BelloniChernozhukovHansen2014}. The benchmark is therefore as close as a transparent model gets to the best forecast the public data allow. As the penalty grows, \eqref{eq:ridge} collapses to \eqref{eq:abg}, so the canonical regression is nested in the benchmark. The four conditioning sets form a ladder:
\[
\Mzero:\{\mathrm{GDP}\}\ \subset\ \Mone:\{\mathrm{GDP},\mathrm{NFCI}\},\qquad \Mtwo:\{\mathrm{GDP},X\}\ \subset\ \Mthree:\{\mathrm{GDP},X,\mathrm{NFCI}\},
\]
with \Mzero{} nested in \Mtwo{} and \Mone{} in \Mthree{}. \Mtwo{} is the rich public benchmark and \Mthree{} adds financial conditions to it.

\subsection{Penalties}\label{sec:protocols}

The penalties were chosen once, on the public benchmark alone and before any test result existed, and every model that contains the panel inherits them; none was ever tuned for a signal. There is one penalty for each of three blocks of quantiles---lower (Q05--Q25), center (Q30--Q70) and upper (Q75--Q95)---selected on 92 validation targets ending in 2007Q3 by a one-standard-error rule that picks the strongest penalty statistically indistinguishable from the best. The selected values are $\lambda=0.56$ in the lower block, $\lambda=100$ in the center and $\lambda=100$ in the upper block. The second value matters for what the benchmark is. At $\lambda=100$ the panel barely moves the predicted center and upper quantiles, so in those blocks the rich benchmark is close to the canonical regression, and the panel does its work in the lower block. Section~\ref{sec:rich} therefore repeats the comparisons with the penalties that minimize validation loss, under which the panel works in every block. Minutes-based signals that start after the public history are standardized on their own available history and set to zero before it, and the projection measures enter in levels and are set to zero before 1983Q1, so that every model in a comparison is estimated on identical observations.

\subsection{Evaluation}

Forecasts are recursive and out of sample. The score is the pinball loss summed over the 70 targets and over the 19 quantiles $\tau=0.05,\dots,0.95$, or over the quantiles of one block; on the full grid it approximates the continuous ranked probability score, a proper score for the whole predictive distribution. The contribution of a signal $S$ to a benchmark $m$ is the percentage reduction in loss,
\begin{equation}\label{eq:gain}
\gain_S(m)=100\times\frac{L(m)-L(m{+}S)}{L(m)} .
\end{equation}
Inference is paired at the target level. For each target the loss of the benchmark is subtracted from that of the augmented model, averaging over the quantiles of the block, and the mean difference is tested with a HAC variance (Bartlett, bandwidth 3, since targets overlap at a four-quarter horizon) and the small-sample correction of \citet{HarveyLeybourneNewbold1997}; a moving-block bootstrap with blocks of four targets is reported as a check. With 70 overlapping targets the effective sample is about twenty independent observations, which is the arithmetic behind every interval in the paper.

Every signal is judged on the same two pre-specified tests: the whole distribution against the public panel, and the lower block against the panel with financial conditions; where a family of signals faces the same test, the $p$-values are adjusted by the Holm method within the family. The design, the signals, the penalties and the tests were fixed in dated records before the forecasts were generated, together with the two robustness exercises of Section~\ref{sec:prespec}. Every other exercise in Section~\ref{sec:robustness} was added after an audit of the results, to answer the objections a referee would raise; each was specified before its own forecasts were generated, is reported whatever it gave, and is labeled as such (Supplementary Material~\ref*{app:protocol}).
\section{Results}\label{sec:results}

\subsection{The family against the public panel}\label{sec:overview}

Table~\ref{tab:overview} puts every measure on the same footing. Each row adds one measure, unpenalized, to the rich public benchmark \Mtwo{} and reports the percentage reduction in out-of-sample loss over the whole distribution, over the whole distribution without the pandemic quarters, and in each block, together with the number of the 19 quantiles at which the forecast improves. Three rows show dashes where the measure's forecast sets were estimated at penalties or histories different from the benchmark's, so that no equal-footing comparison exists; their content is reported as descriptive in the Supplementary Material, and Hidden Dissent's gain conditional on each of them is in Table~\ref{tab:robust} and Figure~\ref{fig:sep}.

\begin{table}[!t]
\centering
\caption{The family at a glance: gain from adding each measure to the public benchmark \Mtwo{}}
\label{tab:overview}
\small
\resizebox{\textwidth}{!}{%
\begin{tabular}{l d{2.3} d{2.3} d{2.3} d{2.3} d{2.3} c}
\toprule
 & \multicolumn{1}{c}{All 19} & \multicolumn{1}{c}{All 19,} & \multicolumn{1}{c}{Lower} & \multicolumn{1}{c}{Center} & \multicolumn{1}{c}{Upper} & Quantiles \\
Measure added to \Mtwo{} & \multicolumn{1}{c}{quantiles} & \multicolumn{1}{c}{ex-pandemic} & \multicolumn{1}{c}{Q05--Q25} & \multicolumn{1}{c}{Q30--Q70} & \multicolumn{1}{c}{Q75--Q95} & improved \\
\midrule
\multicolumn{7}{l}{\emph{Panel A. What participants bring and project (Beige Book and published projections)}} \\
Regional dispersion$^{*}$ & \multicolumn{1}{c}{--} & \multicolumn{1}{c}{--} & \multicolumn{1}{c}{--} & \multicolumn{1}{c}{--} & \multicolumn{1}{c}{--} & -- \\
Central tendency$^{*}$ & \multicolumn{1}{c}{--} & \multicolumn{1}{c}{--} & \multicolumn{1}{c}{--} & -1.855 & \multicolumn{1}{c}{--} & -- \\
Dispersion$^{*}$       & \multicolumn{1}{c}{--} & \multicolumn{1}{c}{--} & -16.253 & -1.876 & \multicolumn{1}{c}{--} & -- \\
\midrule
\multicolumn{7}{l}{\emph{Panel B. The deliberative family (FOMC Minutes)}} \\
Hidden Dissent        &  4.066 &  3.958 & -0.044 &  5.555 &  4.303 & 15 of 19 \\
Attributed heterogeneity &  3.535 &  3.971 & -5.863 &  4.858 &  9.136 & 11 of 19 \\
Quantifier discount   & -1.131 & -1.424 &  1.284 & -2.121 & -0.988 &  5 of 19 \\
Balance of risks      &  1.492 &  1.445 &  0.377 &  2.133 &  0.978 & 13 of 19 \\
Salience of growth    & -1.872 & -1.638 & -1.793 & -1.897 & -1.886 &  2 of 19 \\
Insurance motive      & -2.420 & -1.981 & -5.386 & -1.736 & -1.303 &  3 of 19 \\
Verbosity (placebo)   & -1.039 & -2.044 & -3.748 & -0.266 & -0.377 &  7 of 19 \\
\midrule
\multicolumn{7}{l}{\emph{Reference: financial conditions}} \\
NFCI                  &  0.685 &  1.809 & 16.112 & -3.855 & -2.749 &  9 of 19 \\
Hidden Dissent, added to \Mtwo{}${}+{}$NFCI &  5.927 &  6.128 &  4.495 &  8.335 &  1.103 & 16 of 19 \\
\midrule
\multicolumn{7}{l}{\emph{Panel C. Public signals from outside the Committee}} \\
Forecaster dispersion (SPF) & -0.070 & -0.170 & -1.805 & -0.066 &  1.549 &  7 of 19 \\
Geopolitical risk     &  0.479 &  0.540 &  0.420 &  0.181 &  1.260 & 14 of 19 \\
Economic policy uncertainty & -3.215 & -3.740 & -2.187 & -2.973 & -4.767 &  4 of 19 \\
Implied volatility (VIX) &  0.382 &  0.680 &  4.036 & -1.394 &  1.275 & 13 of 19 \\
Oil price             & -0.847 & -1.810 & -18.291 &  0.102 & 13.220 &  8 of 19 \\
\bottomrule
\end{tabular}}
\begin{minipage}{0.97\textwidth}\footnotesize\vspace{4pt}
Notes: Entries are $100\times[L(\Mtwo{})-L(\Mtwo{}{+}s)]/L(\Mtwo{})$ on the block pinball loss over the 70 targets (2008Q4--2026Q1); the ex-pandemic column drops 2020Q1--2020Q3 from evaluation only. Every measure enters unpenalized, with the pre-specified penalties on the public panel. Paired inference for Hidden Dissent is in Table~\ref{tab:anatomy} and for the other rows in the Supplementary Material; after Holm adjustment within each block, the Hidden Dissent entries for the whole distribution, center and upper block survive, and among the external signals only the oil price's upper-block entry does. Attributed heterogeneity was added after the other rows were known. $^{*}$Measures that precede the discussion: the dispersion across the twelve district sections of the Beige Book of the coded tone of activity (from 1983Q4), and the midpoint and width of the central tendency of the Committee's growth projections (Monetary Policy Reports 1983Q1--2007Q3, SEP from 2007Q4). Each is measured against a benchmark estimated like it on the history from 1974Q4; dashes mark entries for which no benchmark estimated at the same penalties exists, reported as descriptive in Supplementary Material~\ref*{app:sep} and \ref*{app:robust}. No entry of these rows is distinguishable from zero. Hidden Dissent's gain conditional on each measure is in Table~\ref{tab:robust} and Figure~\ref{fig:sep}.
\end{minipage}
\end{table}

The Committee's declared outlook, listed first because the projections are the first stage of the meeting, does not improve the benchmark: the central tendency and the dispersion worsen the center by 1.9 percent each, the dispersion the lower block by 16 percent, and no entry is distinguishable from zero. Among the measures of the deliberation, four things stand out. First, Hidden Dissent improves the forecast at 15 of the 19 quantiles, by 4.1 percent over the whole distribution and 5.6 percent in the center, and the gain is essentially unchanged without the pandemic quarters. Second, the other measures of what the Committee says do not reproduce it: the stated balance of risks shows a small and imprecise gain in the center (two-sided $p=0.32$); the insurance motive, the salience of growth and the placebo worsen the forecast; and the quantifier discount, a second instrument for disagreement, does not replicate the result. Third, financial conditions do what the Growth-at-Risk literature says they should: the NFCI improves the lower block by 16 percent and worsens the center and upper blocks. Fourth, with financial conditions in the benchmark, Hidden Dissent's gain rises to 5.9 percent over the whole distribution and a lower-block contribution appears. Attributed heterogeneity improves the whole distribution by 3.5 percent under the pre-specified design, mostly in the upper block, a gain that does not survive re-tuning the benchmark (Section~\ref{sec:robustness}); among the public signals from outside the Committee, none improves the whole distribution by more than half a percentage point.

\subsection{Where in the distribution deliberation adds}\label{sec:anatomy}

Table~\ref{tab:anatomy} gives the inference for the two comparisons around which the paper is built: the whole distribution against the public panel, and the lower block against the panel with financial conditions.

\begin{table}[!t]
\centering
\caption{Hidden Dissent against the public panel and against the panel with financial conditions}
\label{tab:anatomy}
\small
\resizebox{\textwidth}{!}{%
\begin{tabular}{l d{2.3} d{2.4} d{1.4} d{2.3} d{1.4} l c}
\toprule
Block & \multicolumn{1}{c}{Gain (\%)} & \multicolumn{1}{c}{$\bar d$} & \multicolumn{1}{c}{HAC SE} & \multicolumn{1}{c}{HLN $t$} & \multicolumn{1}{c}{$p$, one-sided} & \multicolumn{1}{c}{95\% HAC interval} & Wins \\
\midrule
\multicolumn{8}{l}{\emph{Panel A. Added to the public panel \Mtwo{}}} \\
All 19 quantiles   &  4.066 & -0.0253 & 0.0098 & -2.467 & 0.0068 & $[-0.0444,\ -0.0062]$ & 53/70 \\
Lower (Q05--Q25)   & -0.044 &  0.0002 & 0.0116 &  0.018 & 0.5072 & $[-0.0226,\ +0.0230]$ & 33/70 \\
Center (Q30--Q70)  &  5.555 & -0.0407 & 0.0139 & -2.777 & 0.0027 & $[-0.0679,\ -0.0134]$ & 48/70 \\
Upper (Q75--Q95)   &  4.303 & -0.0233 & 0.0078 & -2.833 & 0.0023 & $[-0.0386,\ -0.0080]$ & 59/70 \\
\midrule
\multicolumn{8}{l}{\emph{Panel B. Added to the panel with financial conditions \Mthree{}}} \\
All 19 quantiles   &  5.927 & -0.0367 & 0.0069 & -5.037 & 0.0000 & $[-0.0502,\ -0.0231]$ & 53/70 \\
Lower (Q05--Q25)   &  4.495 & -0.0192 & 0.0072 & -2.533 & 0.0057 & $[-0.0333,\ -0.0051]$ & 39/70 \\
Center (Q30--Q70)  &  8.335 & -0.0634 & 0.0135 & -4.448 & 0.0000 & $[-0.0899,\ -0.0368]$ & 53/70 \\
Upper (Q75--Q95)   &  1.103 & -0.0061 & 0.0077 & -0.757 & 0.2244 & $[-0.0212,\ +0.0089]$ & 55/70 \\
\bottomrule
\end{tabular}}
\begin{minipage}{0.97\textwidth}\footnotesize\vspace{4pt}
Notes: 70 targets, 19 quantiles. $d_t$ is the loss of the augmented model minus that of the benchmark for target $t$, averaged over the quantiles of the block; negative values favor Hidden Dissent. Bartlett HAC of bandwidth 3 with the Harvey--Leybourne--Newbold correction; one-sided $p$ for a loss reduction. Wins: targets at which the augmented model has the lower block loss. The two pre-specified tests are Panel A, all 19 quantiles, and Panel B, lower block; after Holm adjustment across the family of Table~\ref{tab:overview} they stand at 0.048 and 0.045, and the center entries of both panels and the upper entry of Panel A also survive. Moving-block bootstrap intervals and per-quantile detail are in the Supplementary Material.
\end{minipage}
\end{table}

Against the public panel alone, the improvement lies in the body of the distribution: 5.6 percent in the center and 4.3 percent in the upper block, both established, and nothing in the lower block, where the point estimate is zero and the interval is centered on it. With financial conditions in the benchmark, the central gain rises to 8.3 percent and a lower-block contribution of 4.5 percent appears, also established; the upper-block gain falls to 1.1 percent and is no longer distinguishable from zero. The two facts fit together: financial conditions carry the lower tail and deliberation adds on top of them there, while in the upper block part of what the signal added against the public panel is information that the financial-conditions index already contains.

Within the lower block the gain is not spread evenly: $-1.3$ percent at the fifth percentile, 1.4 and 0.6 percent at the tenth and fifteenth, none distinguishable from zero, and 6.1 and 10.0 percent at the twentieth and twenty-fifth (one-sided $p$ of 0.015 and 0.014). The result is therefore a statement about the lower region of the distribution and not about its extreme, and we state it at that level throughout. Section~\ref{sec:penalty} shows that the whole-distribution and central gains are the same in the canonical regressions, where no coefficient is penalized, while the lower-block gain appears only against the richest benchmark.

\subsection{Is the comparison fair? Real-time public information}\label{sec:realtime}

The deliberative measures are read from documents as they were published, while the benchmark of the main specification uses the current vintage of the public data, whose history contains revisions no forecaster at the origin could have seen. We remove the asymmetry: the public panel is rebuilt from the FRED-MD vintages available at each origin under a release-lag rule fixed in advance, without backfilling or imputation, and the real-time and final-vintage benchmarks are restricted, origin by origin, to the same predictors, so that only the vintage of the data differs between them. The NFCI is placed on the same real-time clock and, in a last step, current growth is taken from the real GDP vintage available at each origin; the realized target stays at its final value. Everything else is that of the main specification (Supplementary Material~\ref*{app:realtime}).

\begin{table}[!t]
\centering
\caption{Hidden Dissent under final-vintage and real-time public information}
\label{tab:realtime}
\small
\resizebox{\textwidth}{!}{%
\begin{tabular}{l d{2.3} d{2.3} l l d{2.6}}
\toprule
 & \multicolumn{2}{c}{Gain (\%)} & \multicolumn{2}{c}{Real-time inference} & \multicolumn{1}{c}{Change in $\bar d$} \\
\cmidrule(lr){2-3}\cmidrule(lr){4-5}
Block & \multicolumn{1}{c}{Final} & \multicolumn{1}{c}{Real time} & \multicolumn{1}{c}{$p$, one-sided} & \multicolumn{1}{c}{95\% HAC interval} & \multicolumn{1}{c}{RT $-$ final} \\
\midrule
\multicolumn{6}{l}{\emph{Panel A. Against the public panel \Mtwo{}}} \\
All 19 quantiles  &  4.099 &  3.789 & 0.011 & $[-0.0423,\ -0.0050]$ &  0.001981 \\
Lower (Q05--Q25)  &  0.166 & -1.338 & 0.703 & $[-0.0169,\ +0.0305]$ &  0.007660 \\
Center (Q30--Q70) &  5.549 &  5.559 & 0.004 & $[-0.0680,\ -0.0134]$ & -0.000076 \\
Upper (Q75--Q95)  &  4.300 &  4.298 & 0.003 & $[-0.0386,\ -0.0080]$ &  0.000005 \\
\addlinespace
\multicolumn{6}{l}{\emph{Panel B. Against the public panel with financial conditions \Mthree{}}} \\
All 19 quantiles  &  6.036 &  5.855 & $<0.001$ & $[-0.0493,\ -0.0233]$ &  0.001079 \\
Lower (Q05--Q25)  &  5.099 &  4.123 & 0.008 & $[-0.0316,\ -0.0039]$ &  0.004075 \\
\quad with GDP vintages & \multicolumn{1}{c}{--} &  4.901 & 0.002 & $[-0.0346,\ -0.0079]$ & \multicolumn{1}{r}{$-0.003449^{a}$} \\
Center (Q30--Q70) &  8.334 &  8.337 & $<0.001$ & $[-0.0899,\ -0.0369]$ & -0.000032 \\
Upper (Q75--Q95)  &  1.109 &  1.094 & 0.226 & $[-0.0211,\ +0.0090]$ &  0.000083 \\
\bottomrule
\end{tabular}}
\begin{minipage}{0.97\textwidth}\footnotesize\vspace{4pt}
Notes: 70 targets, pre-specified penalties in both arms of every comparison, Hidden Dissent unpenalized; in Panel B the NFCI is unpenalized in both arms and on the same real-time clock. Both benchmarks of each panel are restricted, at each origin, to the predictors constructible from the historical vintage, so the final-vintage columns are matched benchmarks: in Panel A they reproduce the unrestricted losses of Table~\ref{tab:anatomy} to within a thousandth, in Panel B they place the lower-block gain at 5.099 rather than 4.495, a difference due to the common-support restriction alone (Supplementary Material~\ref*{app:realtime}). Inference as in Table~\ref{tab:anatomy}. Last column: paired change in the mean loss differential, positive when the signal's advantage weakens; a stability diagnostic, not an equivalence test. ``With GDP vintages'' takes current growth, in both arms, from the vintage available at each origin; $^{a}$paired change against the row above it (two-sided $p=0.389$).
\end{minipage}
\end{table}

Table~\ref{tab:realtime} reports the result. The contribution survives the change of information set, and the way it survives is informative. Against the public panel (Panel A), the whole-distribution gain falls from 4.1 to 3.8 percent and remains established ($p=0.011$). The blocks in which the signal's content sits do not move: the central gain is 5.55 percent with final data and 5.56 percent with real-time data, the upper-block gain 4.30 percent with both, and in each case the change in the underlying loss differential is less than a fifth of one percent of the differential itself. The whole of the modest attenuation is in the lower block, where the signal had no gain to begin with.

Panel B carries more weight, because it covers the one region of the distribution where the paper makes a claim about downside outcomes. With financial conditions in the benchmark and both the panel and the index on the real-time clock, the lower-block gain is 4.1 percent ($p=0.008$; moving-block intervals below zero at block lengths of four, six and eight targets); taking current growth from its own vintages as well leaves it at 4.9 percent ($p=0.002$), and the central gain is unchanged at 8.3 percent. The objection that the lower-region result depends on revised data is therefore not supported.

Two qualifications keep this statement at the level of the evidence: the lower block is more sensitive than the center to the composition of the panel (the common-support restriction alone moves the final-vintage lower-block gain from 4.5 to 5.1 percent, the central gain not at all), and between the matched final-vintage and real-time benchmarks the lower-block gain falls by about one percentage point, a change that is not detectable (two-sided $p=0.18$) but that, with seventy overlapping targets, we do not read as evidence that nothing changed. What the exercise establishes is the positive statement: the contribution is present, and established, when the public panel and current growth are held to the vintages available at the time. Supplementary Material~\ref*{app:realtime} reports the full inference and the same exercise for every other measure; attributed heterogeneity is in Supplementary Material~\ref*{app:ah}.

\subsection{Does any competing signal reproduce it?}\label{sec:competing}

Table~\ref{tab:rt-lower} asks whether any competing signal reproduces the lower-region result where it is made. It adds each signal to the real-time benchmark with financial conditions and compares it directly with Hidden Dissent on the same benchmark.

\begin{table}[!t]
\centering
\caption{Competing signals added to the real-time panel with financial conditions, lower block}
\label{tab:rt-lower}
\small
\resizebox{\textwidth}{!}{%
\begin{tabular}{l d{2.3} d{2.6} d{1.4} l d{2.6} d{1.4}}
\toprule
 & \multicolumn{4}{c}{Signal added to \Mthree{}, real time} & \multicolumn{2}{c}{Paired against Hidden Dissent} \\
\cmidrule(lr){2-5}\cmidrule(lr){6-7}
Signal & \multicolumn{1}{c}{Gain (\%)} & \multicolumn{1}{c}{$\bar d$} & \multicolumn{1}{c}{$p$, one-sided} & \multicolumn{1}{c}{95\% HAC interval} & \multicolumn{1}{c}{$\bar d^{\HD}-\bar d^{s}$} & \multicolumn{1}{c}{$p$, two-sided} \\
\midrule
\multicolumn{7}{l}{\emph{What participants bring and project (Beige Book and published projections)}} \\
Regional dispersion$^{*}$ &  -1.080 &  0.004804 & 0.9246 & $[-0.0015,\ +0.0111]$ & -0.035895 & 0.0479 \\
Central tendency$^{*}$ &   0.822 & -0.003660 & 0.4418 & $[-0.0500,\ +0.0427]$ & -0.028457 & 0.2516 \\
Dispersion$^{*}$       & -15.192 &  0.067597 & 0.9025 & $[-0.0286,\ +0.1638]$ & -0.107068 & 0.1408 \\
\addlinespace
\multicolumn{7}{l}{\emph{The deliberation (FOMC Minutes)}} \\
Hidden Dissent        &  4.123 & -0.017769 & 0.0084 & $[-0.0316,\ -0.0039]$ & \multicolumn{1}{c}{--} & \multicolumn{1}{c}{--} \\
Attributed heterogeneity & -11.088 &  0.047786 & 0.8730 & $[-0.0296,\ +0.1251]$ & -0.065555 & 0.1485 \\
Balance of risks      &  0.127 & -0.000546 & 0.4625 & $[-0.0114,\ +0.0103]$ & -0.017223 & 0.0671 \\
Quantifier discount   &  1.365 & -0.005884 & 0.3364 & $[-0.0318,\ +0.0201]$ & -0.011885 & 0.3724 \\
Salience of growth    &  1.391 & -0.005995 & 0.2600 & $[-0.0233,\ +0.0114]$ & -0.011774 & 0.2593 \\
Insurance motive      &  1.721 & -0.007415 & 0.2850 & $[-0.0317,\ +0.0169]$ & -0.010354 & 0.3816 \\
Verbosity             & -2.194 &  0.009456 & 0.9726 & $[+0.0003,\ +0.0186]$ & -0.027225 & 0.0016 \\
\addlinespace
\multicolumn{7}{l}{\emph{Public signals from outside the Committee}} \\
Forecaster dispersion (SPF) &  4.845 & -0.020879 & 0.1782 & $[-0.0630,\ +0.0213]$ &  0.003110 & 0.8755 \\
Geopolitical risk     &  1.729 & -0.007452 & 0.1766 & $[-0.0224,\ +0.0075]$ & -0.010317 & 0.2156 \\
Economic policy uncertainty & -0.266 &  0.001147 & 0.5902 & $[-0.0082,\ +0.0105]$ & -0.018915 & 0.0245 \\
Implied volatility (VIX) &  0.504 & -0.002174 & 0.4322 & $[-0.0259,\ +0.0215]$ & -0.015595 & 0.1891 \\
Oil price             &  2.068 & -0.008914 & 0.3456 & $[-0.0507,\ +0.0329]$ & -0.008855 & 0.6500 \\
\bottomrule
\end{tabular}}
\begin{minipage}{0.97\textwidth}\footnotesize\vspace{4pt}
Notes: Lower block (Q05--Q25), 70 targets, pre-specified penalties. The benchmark is the matched real-time public panel plus the NFCI on the same real-time clock, unpenalized; each signal enters unpenalized as the only difference. Inference as in Table~\ref{tab:anatomy}; moving-block intervals lie below zero for Hidden Dissent and contain zero for every other signal. The paired columns test $d^{\HD}_t-d^{s}_t$ on the same benchmark, negative values favoring Hidden Dissent; after Holm adjustment only the comparison with verbosity survives among the Minutes measures ($p_{\text{Holm}}=0.0096$), and none among the five external signals, pre-specified as a set. Attributed heterogeneity was added after the other rows were known. $^{*}$Added to the benchmark with financial conditions on final-vintage data and the history from 1974Q4 (Table~\ref{tab:overview}); Hidden Dissent has the lower loss in 44, 35 and 46 of 70 targets against them.
\end{minipage}
\end{table}

The table supports two statements of different strength. On incremental content, Hidden Dissent is the only one of the twelve signals whose lower-block improvement is established. On direct comparison the evidence is weaker: Hidden Dissent has detectably lower loss than the placebo after multiplicity adjustment, and than policy uncertainty at the unadjusted level, but against the other signals the differences favor it by amounts the sample cannot distinguish from zero; forecaster dispersion, with a point estimate as large as Hidden Dissent's but three times less precise, cannot be told apart from it. The table establishes that no competing signal reproduces the result, not that each is inferior to it. Nor are the Minutes-based measures three noisy readings of one disagreement: Hidden Dissent, the quantifier discount and attributed heterogeneity are weakly correlated with one another (0.31, $-0.13$ and 0.01 in levels; Supplementary Material~\ref*{app:properties}), and only one of them carries the lower-region content.

\subsection{When does deliberation help?}\label{sec:when}

If the deliberation carries information because the Committee reads the state of the economy differently from what the data alone imply, one would expect that information to matter most at turning points, when the data are hardest to read. The design fixed, before any result was seen, seven episodes of stress or transition---the global financial crisis, the euro-area sovereign crisis, the China and oil slowdown, the inflation and tightening episode of 2021, the energy and inflation shock of 2022, the banking stress of 2023 and the pandemic---covering 32 of the 70 targets, and stated the prediction: the gain should be larger inside the windows than outside them.

\begin{table}[!t]
\centering
\caption{Hidden Dissent added to \Mtwo{} inside and outside the pre-specified event windows}
\label{tab:events}
\small
\begin{tabular}{l d{2.3} d{2.3} d{2.3} d{2.3}}
\toprule
 & \multicolumn{1}{c}{All 19} & \multicolumn{1}{c}{Lower} & \multicolumn{1}{c}{Center} & \multicolumn{1}{c}{Upper} \\
 & \multicolumn{1}{c}{quantiles} & \multicolumn{1}{c}{Q05--Q25} & \multicolumn{1}{c}{Q30--Q70} & \multicolumn{1}{c}{Q75--Q95} \\
\midrule
\multicolumn{5}{l}{\emph{Event windows against their complement}} \\
Event ($N=32$)      &  3.657 & -0.515 &  5.146 &  4.065 \\
Non-event ($N=38$)  &  5.257 &  1.513 &  6.789 &  4.875 \\
Difference          & -1.600 & -2.028 & -1.643 & -0.810 \\
\midrule
\multicolumn{5}{l}{\emph{By window}} \\
Global financial crisis ($N=5$)      &  0.07 &  -1.48 &  0.87 &  0.44 \\
Euro-area sovereign crisis ($N=11$)  &  5.50 & -10.07 &  8.96 &  6.30 \\
China and oil slowdown ($N=4$)       &  9.03 &  -0.99 & 12.18 &  5.69 \\
Inflation and tightening ($N=3$)     &  4.89 &   5.90 &  5.63 &  3.29 \\
Energy and inflation shock ($N=4$)   &  6.39 &  -2.97 &  4.80 & 15.84 \\
Banking stress 2023 ($N=2$)          &  1.78 & -13.00 &  8.14 &  5.31 \\
Pandemic ($N=3$)                     &  4.84 &   1.03 &  7.06 &  4.70 \\
\bottomrule
\end{tabular}
\begin{minipage}{0.95\textwidth}\footnotesize\vspace{4pt}
Notes: Percentage reduction in block pinball loss from adding Hidden Dissent to \Mtwo{}, computed within each subset of targets on the forecasts of Table~\ref{tab:anatomy}, without re-estimation. The seven windows, which partition the 32 event targets, and their complement were fixed before any performance was examined and not altered afterwards. The augmented model wins 23 of 32 targets inside the windows and 30 of 38 outside. The event--non-event difference over the whole distribution is $-1.6$ percentage points (HAC/HLN two-sided $p=0.52$; HAC interval $[-6.5,\ 3.3]$, moving-block interval $[-7.2,\ 3.3]$); the third row is descriptive.
\end{minipage}
\end{table}

Table~\ref{tab:events} does not support the prediction. The gain is smaller inside the windows than outside in all four blocks, 3.7 against 5.3 percent over the whole distribution and 5.1 against 6.8 in the center; the difference is not distinguishable from zero (two-sided $p=0.52$), so the reverse statement is not established either, and no alternative dating was examined to recover the concentration the design looked for. Within the windows, which contain between two and eleven targets each, the central gain is positive in all seven and the lower-block gain negative in five: the location pattern of the full sample, if anything sharper inside the episodes. The result is not the product of a handful of dramatic quarters; it is earned in the ordinary quarters that make up more than half of the sample.

\subsection{With what public information does it combine?}\label{sec:with}

A signal that corrects the extrapolation from the public state should be worth more when that state is in the benchmark, and its value should depend on which part of the state it is read against. We ask which part by removing one group of public series at a time from the benchmark with financial conditions---real activity (62 series), credit and interest rates (34), prices (23) and market prices (4)---and measuring the change in the signal's lower-block contribution with the pre-specified penalty held fixed. Since any removal can lower the contribution simply by making the benchmark poorer, each is compared with 40 random removals of the same number of series.

\begin{figure}[!t]
\centering
\includegraphics[width=0.86\textwidth]{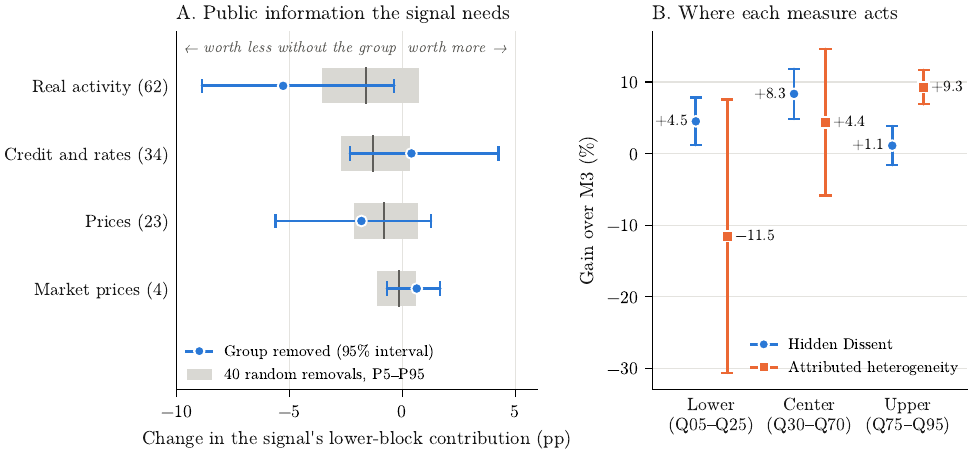}
\caption{Hidden Dissent: which public information it needs, and where in the distribution it acts. Panel A: change in the signal's lower-block contribution against the benchmark with financial conditions when each group of public series is removed from it, in percentage points of the benchmark's loss, with the pre-specified penalty held fixed; negative values mean that the signal is worth less without the group. Whiskers: 95 percent moving-block intervals (blocks of four targets); gray bars: 5th to 95th percentiles of the same change across 40 random removals of the same number of series, median marked. Panel B: gain from adding Hidden Dissent and attributed heterogeneity to the benchmark with financial conditions, by block, with 95 percent HAC intervals (Table~\ref{tab:anatomy}; Supplementary Material Table~\ref*{tab:ah-anatomy}); under the re-tuned penalties the upper-block gain of attributed heterogeneity disappears (Section~\ref{sec:onereading}). 70 targets, 2008Q4--2026Q1.}
\label{fig:groups}
\end{figure}

Panel A of Figure~\ref{fig:groups} reports the result. Removing the market or the credit series leaves the signal's contribution where it was, near the upper edge of what random removals of the same size produce: whatever the signal carries, it is not a repackaging of what markets have already priced. Removing prices lowers it by 1.8 points, within the range of random removals; removing real activity lowers it by 5.3 points, more than any of the 40 random removals of 62 series (randomization $p$-value of 0.024). The pattern survives doubling the penalty but not halving it, and the four contrasts do not survive an adjustment for multiple testing, so we read it as directional: the signal's content is read against the state of real activity. Panel B places the block profile of Table~\ref{tab:anatomy} beside that of attributed heterogeneity: the index acts in the center and the lower block, the transparent measure in the upper block, where its gain does not survive re-tuning the benchmark (Section~\ref{sec:onereading}). Supplementary Material~\ref*{sec:mechanism} gives the design and the full inference.
\section{Robustness}\label{sec:robustness}

A result of this kind invites a predictable set of objections: that it depends on the design, that the benchmark is too weak, that the signal repeats what forecasters or the Committee already publish, that a few extraordinary quarters drive it, or that it belongs to one reading of the Minutes. We take them in turn; Table~\ref{tab:robust} collects the evidence on the three comparisons that carry the paper's claims. Two exercises were specified with the main design, before any result existed; the others were added after an audit of the results, each specified before its own forecasts were generated and reported whatever it gave. The Supplementary Material gives the detail.

The exercises also fix the limits of what is claimed. The paper does not claim that the contribution is invariant to treating the signal as one more series of the regularized panel, that the signal improves the lower region against every benchmark or beyond the consensus, or that it measures tail risk: the extreme lower quantiles are carried by financial conditions, and the lower-block contribution rests, more than the central one, on a few episodes. It does not claim a second established dimension of the deliberation, private information, or causation: the evidence is about information content, in one country and seventy overlapping targets.

\begin{table}[!t]
\centering\small
\caption{Robustness of the three main comparisons}\label{tab:robust}
\begin{tabular}{lccc}
\toprule
 & All 19 & Center & Lower \\
Exercise & (vs.\ \Mtwo{}) & (vs.\ \Mtwo{}) & (vs.\ \Mthree{}) \\
\midrule
Main specification                     & 4.07 (0.007) & 5.56 (0.003) & 4.50 (0.006) \\
\addlinespace
\multicolumn{4}{l}{\emph{Pre-specified robustness}} \\
R-1: public history from 1986          & 5.75 (0.008) & -- & 6.07 (0.006) \\
R-2: signal penalized as the panel     & 0.03 (0.24) & -- & 0.11 (0.25) \\
\addlinespace
\multicolumn{4}{l}{\emph{Added after the audit}} \\
No penalty: canonical regressions      & 4.25 (0.003) & 5.46 (0.003) & $-0.32$ (0.60) \\
Penalties re-tuned (validation minimum) & 5.45 (0.010)$^{\dagger}$ & 6.51 (0.0001)$^{\dagger}$ & as main \\
Real-time public panel                 & 3.79 (0.011) & 5.56 (0.004) & 4.12 (0.008) \\
\quad + real-time current growth       & -- & -- & 4.90 (0.002) \\
Benchmark: SPF consensus               & 10.98 (0.019) & 12.54 (0.021) & 6.60 (0.20) \\
Benchmark: panel + SPF consensus       & 11.51 (0.008)$^{\dagger}$ & 12.64 (0.003)$^{\dagger}$ & 5.18 (0.15) \\
\addlinespace
\multicolumn{4}{l}{\emph{The Committee's published projections in the benchmark}} \\
\quad + central tendency               & -- & 7.60 (0.0002) & 7.28 (0.045)$^{\S}$ \\
\quad + dispersion                     & 4.69 (0.0001)$^{\S}$ & 7.03 (0.0002)$^{\S}$ & 7.70 (0.035) \\
\addlinespace
\multicolumn{4}{l}{\emph{The staff's own forecast in the benchmark}} \\
\quad + Greenbook forecast$^{\P}$       & 12.71 (0.013) & 13.59 (0.016) & 5.44 (0.077) \\
\addlinespace
\multicolumn{4}{l}{\emph{The information brought to the meeting in the benchmark}} \\
\quad + Beige Book regional dispersion & 5.21 (0.0002) & 6.69 (0.0003) & 4.89 (0.043) \\
\bottomrule
\end{tabular}
\begin{minipage}{0.97\textwidth}\footnotesize\vspace{4pt}
Notes: Percentage reduction in pinball loss from adding Hidden Dissent to the benchmark in each row. The lower-block column uses the benchmark with financial conditions (\Mone{} in the no-penalty row). One-sided HLN $p$-values in parentheses, except $^{\dagger}$ two-sided. A dash marks a comparison not computed in that exercise. ``As main'': the lower-block penalty is the same under both rules. The consensus and projection rows use a common history from 1974Q4; in the panel + SPF row the whole-distribution and center columns use the re-tuned penalties and the lower column adds financial conditions. The Greenbook and Beige Book rows compare two forecast sets estimated alike, on the history from 1974Q4 with $\lambda_U=1$ (Supplementary Material~\ref*{app:robust}). $^{\S}$Added after the two pre-specified comparisons with the projections were known. $^{\P}$50 targets, 2008Q4--2021Q1, the span of the published Greenbook/Tealbook (Section~\ref{sec:consensus}); Holm-adjusted $p$ of 0.025 and 0.077 for the all-19 and lower cells. Full inference in Supplementary Material~\ref*{app:realtime}, \ref*{app:robust} and \ref*{app:regularization}.
\end{minipage}
\end{table}

\subsection{The two pre-specified exercises}\label{sec:prespec}

The first exercise, R-1, starts the public history of every model in 1986 instead of 1973Q1, leaving out a more volatile regime. The result does not weaken: the whole-distribution gain against the public panel is 5.8 percent ($p=0.008$) and the lower-block gain against the panel with financial conditions 6.1 percent ($p=0.006$), with moving-block intervals below zero.

The second exercise, R-2, penalizes the signal as heavily as the public panel, at the block penalty chosen for the 123 series. It removes the result: the whole-distribution gain falls to 0.03 percent ($p=0.24$) and the lower-block gain to 0.11 percent ($p=0.25$). The reason is mechanical: a penalty chosen to keep 123 correlated series from overfitting imposes on a single regressor a strong prior at zero, and the penalized coefficient retains a median of 7 percent of its unpenalized value in the lower block and 0.07 percent in the center, never changing sign (Supplementary Material~\ref*{app:regularization}). R-2 therefore shows that the result depends on treating the signal as the variable under test, as Section~\ref{sec:lineage} argues it should be; it cannot distinguish an absence of content from the attenuation the penalty imposes.

\subsection{Does leaving the signal unpenalized create the gain?}\label{sec:penalty}

The benchmark penalizes the public panel and leaves the signal free; the asymmetry, not the signal, could produce the gain. The canonical Growth-at-Risk regressions settle the question, because they contain no panel and no penalty: the signal enters them exactly as current growth and financial conditions do. Supplementary Material Table~\ref*{tab:hd-ladder} adds the signal to all four rungs of the ladder.

Three facts follow. First, the central gain is the same with and without the regularized panel: 5.5 percent against the canonical regression on current growth and 5.6 percent against the public panel, and 8.3 percent in both once financial conditions are included; the whole-distribution gain against the rungs without financial conditions is 4.2 against 4.1 percent. The central result is therefore not a product of the regularization or of the asymmetry. Second, the upper block follows financial conditions: the signal improves it by 4.3 percent when the NFCI is absent from the benchmark and by about 1 percent, not established, when it is present. Third, the lower-block contribution is absent in both canonical regressions and against the public panel, and appears only against the panel with financial conditions, which is also the best lower-block benchmark of the four; this is why the whole-distribution gain is larger against the full benchmark than against the canonical regression with financial conditions, 5.9 against 5.0 percent. The lower-region result is conditional on the richest information set, and it is the one result that depends on the treatment of the signal.

Nor is a free coefficient an advantage in itself: every measure in Table~\ref{tab:overview} receives one, and the placebo and six of the ten other measures worsen the forecast with it.

\subsection{Is the benchmark rich enough?}\label{sec:rich}

Under the pre-specified penalties the public panel does its work in the lower block and barely moves the predicted center and upper quantiles, where the one-standard-error rule selects the strongest penalty that validation cannot distinguish from the best. A central gain measured against a benchmark so close to the canonical regression could be one that any cyclical indicator would deliver. We therefore re-estimate the benchmarks at the penalties that minimize the same validation criterion ($\lambda_C=5.6$ and $\lambda_U=0.32$; the lower-block penalty is the same under both rules). At these values the public panel moves the fitted central quantiles about twelve times more, and the benchmark with financial conditions improves its own central loss by 2.4 percent ($p=0.028$): under these penalties the public panel contributes to the center of the distribution, which is the benchmark the next comparison requires.

Against these benchmarks the signal's central gain does not fall; it rises, to 6.5 percent against the public panel and 8.4 percent against the panel with financial conditions (two-sided $p<0.001$), and its whole-distribution gain rises to 5.5 and 8.2 percent. The upper block is where the choice of penalty matters: the point gains rise, to 8.0 and 10.3 percent, but neither is established ($p=0.13$ and $0.09$). We therefore state the central result without qualification as to the penalty, and the upper-block result as established under the pre-specified penalties only. The same change in the benchmark removes most of the contribution of attributed heterogeneity (Section~\ref{sec:onereading}), a first indication that the index and the transparent measure do not carry the same information.

\subsection{Is it what forecasters, or the Committee, already expect?}\label{sec:consensus}

The most natural alternative source of the signal's content is the professional consensus: if the Committee's disagreement only echoed what forecasters already expected, the consensus would absorb it. We take the median SPF forecast of real GDP growth over the four quarters after the origin, from the latest survey released before the origin. Against the consensus alone the signal improves the whole distribution by 11.0 percent and the center by 12.5 percent (one-sided $p$ of 0.019 and 0.021); since that benchmark is weaker than the public panel, we also add the consensus to the re-tuned public panel, unpenalized, with every model estimated from 1974Q4, the first survey with all four horizons (on that history the central gain without the consensus is 6.3 percent against 6.5, $p=0.001$). Adding the consensus produces the benchmark with the lowest central loss in the paper; against it the signal's central gain is 12.6 percent (two-sided $p=0.003$), and its whole-distribution gain 11.5 percent ($p=0.008$).

The gain is larger against this benchmark than against the panel alone, and the signal and the consensus are negatively correlated in levels ($-0.62$ over the test origins), which raises the possibility of a suppression effect: a signal that matters only once the consensus is partialled out. The coefficients rule it out: the signal's coefficient in the central quantile regressions is positive in all 630 fits, with the consensus and without it, and conditioning on the consensus raises its median from 0.55 to 0.83. The consensus sharpens a relation that is present without it.

The lower region tells a different and weaker story. On the common history from 1974Q4, the signal improves the lower block of the panel with financial conditions by 7.0 percent (one-sided $p=0.036$), while the consensus barely changes that benchmark's lower-block loss (1.1 percent, $p=0.88$). Conditional on the consensus the signal's point gain is 5.2 percent, but the mean differential is no longer distinguishable from zero (one-sided $p=0.15$) and is concentrated: ten targets account for nine-tenths of it, and the 2008 target for half. Its lower-region contribution beyond the consensus is therefore not established in this sample (Supplementary Material~\ref*{app:robust}).

The Committee's own projections put the same question to the Committee itself: the central tendency of participants' growth projections is its declared outlook and its width its declared disagreement, while Hidden Dissent reads the disagreement it does not declare. If the index only repeated what the Committee says, conditioning on the projections would absorb it. Because the Summary of Economic Projections begins only in 2007Q4, we splice it with the central tendency and range of the growth projections that the Monetary Policy Report to the Congress published twice a year from 1983 to 2007 under the same rule; the two series run from 1983Q1 and cover the 70 targets (Supplementary Material~\ref*{app:sep}). Two tests were fixed before any forecast with these series was generated, the center against the public panel containing the central tendency and the lower block against the panel with financial conditions containing the dispersion; the two remaining pairings were added once those were known.

\begin{figure}[!t]
\centering
\includegraphics[width=0.8\textwidth]{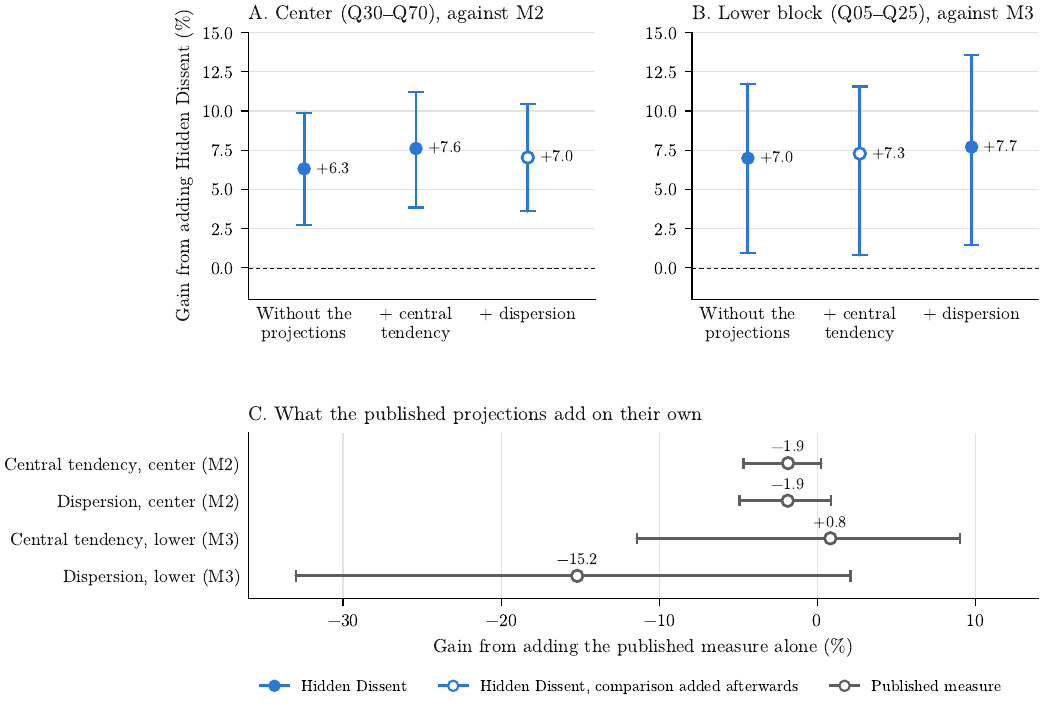}
\caption{Hidden Dissent and the Committee's published projections. Panels A and B: gain from adding Hidden Dissent to the benchmark without the projections and to the benchmark containing each projection measure. Panel C: gain from adding each projection measure to the benchmark without it. Bars are 95 percent moving-block bootstrap intervals ($L=4$), expressed as gains on the benchmark's loss. Every comparison holds the history (from 1974Q4) and the block penalty fixed across its two models (the center with the central tendency at the pre-specified penalty, the other center comparisons at the validation-minimizing one). The center with the central tendency and the lower block with the dispersion were fixed in advance (one-sided $p=0.0002$ and $0.035$, Holm 0.0003 and 0.035); hollow markers were added once those were known. 70 targets. Supplementary Material Table~\ref*{tab:sep-full} gives the full inference.}\label{fig:sep}
\end{figure}

Figure~\ref{fig:sep} gives the answer. With either measure in the benchmark, Hidden Dissent improves the center by 7.6 and 7.0 percent and the lower block by 7.3 and 7.7 percent, against 6.3 and 7.0 percent without them; both pre-specified tests pass after adjustment for multiple testing, and the moving-block intervals exclude zero in all six comparisons. Over the whole distribution, with the dispersion in the benchmark, the gain is 4.7 percent (one-sided $p=0.0001$). The projections themselves improve no block: the four gains in Panel C range from $-15.2$ to $+0.8$ percent, none distinguishable from zero. The long series matter here: with the SEP alone, from 2007Q4, the coefficient on the projections is estimated on very few years, and over the 45 targets from 2015Q1 the index's central gain is 0.6 percent with the SEP central outlook in the benchmark (two-sided $p=0.77$), against 6.1 without it. The evidence is spread over time in the center and concentrated in a few targets around the financial crisis and the pandemic in the lower block (Supplementary Material~\ref*{app:sep}). What the Committee declares neither improves the forecast nor contains what the index reads---in the center of the distribution clearly, and in its lower region with less precision.

Supplementary Material Table~\ref*{tab:declared} says why. The central tendency is largely a reading of the public state (correlations of 0.58 to 0.64 with current growth, the activity component of the public panel and the consensus), and the published dispersion a reading of public uncertainty (Section~\ref{sec:interpretation}). Hidden Dissent is correlated with none of these; it is negatively related to the consensus and, alone among the nine measures of Supplementary Material~\ref*{app:properties}, anticipates the error of the consensus.

A further candidate is what the Committee itself knows: the forecast of its staff (the Greenbook, the Tealbook since 2010), the standard measure of the Federal Reserve's informational advantage \citep{RomerRomer2000}, released after five years. We add it to both benchmarks as we add the consensus, with the tests fixed in advance. Over the 50 targets it covers it is a worse point forecast than the consensus and, added to the public panel, not distinguishable from zero (6.5 percent, $p=0.24$, descriptive). Hidden Dissent added to the panel containing it improves the whole distribution by 12.7 percent (one-sided $p=0.013$, Holm $0.025$) and the center by 13.6 percent; on the panel with financial conditions the lower-block gain is 5.4 percent ($p=0.077$). Conditioning on the staff forecast does not absorb the gain (Supplementary Material~\ref*{app:robust}): whatever the disagreement carries, it is not the staff's forecast.

The remaining candidate is the information the participants bring to the meeting. Two weeks before each meeting the twelve Reserve Banks report on their districts in the Beige Book, without coordination; we take the dispersion across districts of the tone of activity, coded on a five-point scale, as the heterogeneity of that information (Supplementary Material~\ref*{app:properties}). Added to the benchmark with financial conditions it contains nothing: $-1.1$ percent in the lower block, with an interval including zero (Table~\ref{tab:rt-lower}), and $-1.2$ percent against the public panel, a descriptive comparison (Supplementary Material~\ref*{app:robust}). Hidden Dissent added to the benchmark containing it improves the whole distribution by 5.2 percent (one-sided $p=0.0002$, Holm $0.0003$) and the center by 6.7 percent; in the lower block by 4.9 percent ($p=0.043$) against 7.0 without it, a paired difference of marginal significance: part of its lower-region content is shared with the regional information.

\paragraph{Not about prices.}
If the recorded disagreement is a reading of the state of activity, it should say little about inflation. We repeat the exercise for core and headline PCE inflation four quarters ahead, with every choice fixed in advance (Supplementary Material~\ref*{app:inflation}). The index adds nothing: against both benchmarks the central loss rises by 1 to 3 percent for core and by less than 1 percent for headline inflation, and the other blocks say the same. The recorded disagreement is information about activity, not about prices.

\subsection{Is it a few extraordinary quarters?}\label{sec:episodes}

A few quarters, the financial crisis and the pandemic among them, could drive everything. Three facts speak against it: without the three pandemic targets the whole-distribution gains are 4.0 and 6.1 percent and the lower-block gain 5.4; the gain is not concentrated in the seven episodes fixed in advance (Section~\ref{sec:when}), 3.7 percent inside them against 5.3 outside, with the central gain positive in each; and without the five targets of the financial crisis the real-time lower-block gain is 3.6 percent rather than 4.1. The qualification is the lower region beyond the consensus, concentrated around the crisis.

\subsection{Is it one reading of the Minutes?}\label{sec:onereading}

Does the result belong to the construct or to the instrument? The family answers in part: the quantifier discount, a second measure of disagreement built independently of the index, does not reproduce the result (Table~\ref{tab:overview}), so the gain is not generic to any reading of disagreement in the Minutes, and a measure of open conflict (groups attributed opposite directions on the same question) worsens the forecast of the whole distribution by 3.5 percent: what informs the distribution is the latent tension the index reads.

The transparent measure of attributed views is the most informative comparison. Under the pre-specified penalties it improves the whole distribution by 3.5 percent ($p=0.019$), mostly in the upper block (9.1 percent), and worsens the lower block; under the re-tuned penalties its whole-distribution contribution disappears (1.2 percent, one-sided $p=0.13$) and what remains is a central gain of 4.2 percent (two-sided $p=0.002$), below the index's 6.5. What it appeared to add in the upper region is public information the penalized benchmark was not using (Supplementary Material~\ref*{app:ah}). The stated balance of risks adds little in either construction (1.5 and 1.6 percent, neither established).
\section{What deliberation contributes}\label{sec:interpretation}

\paragraph{What the evidence rules out.}
The signal is not a downside-risk indicator in the Growth-at-Risk sense: financial conditions carry the extreme tail, and deliberation adds on top of them at the twentieth and twenty-fifth percentiles rather than at the fifth. Nor is it a product of the regularization, a crisis artifact, a repackaged measure of uncertainty or an artifact of revised data: the central gain is the same in the canonical regressions, smaller inside the episodes of stress than outside, not reproduced by any public proxy for uncertainty, and unchanged under real-time data.

\paragraph{Two kinds of disagreement.}
A committee can disagree in two ways that the data distinguish. Its members can put different numbers on the same outlook: the dispersion of the published projections, like that of the professional forecasters, rises with every public measure of uncertainty (Table~\ref{tab:declared}: 0.58 with the dispersion of the forecasters, 0.51 with policy uncertainty, 0.42 with the VIX) and improves no region of the forecast. Or the Committee can record that it read the state differently: the index is unrelated to those measures and correlated with the error the consensus went on to make. The two orderings do different work. Sorting the quarters by the index, the consensus turned out too high by 0.7 percentage points on average when disagreement was below its median and too low by 0.5 points when it was above; sorting them by declared dispersion separates the quarters of high public uncertainty but not the sign of the consensus error (Supplementary Material~\ref*{app:speakers}). We read the index, then, as saying in which direction the forecast implied by the public data will be revised---a reading, not an identified mechanism, and one that requires no private information: the Committee may read the same public news differently from the market \citep{BauerSwanson2023}. The lower-region contribution falls when the real-activity series are removed from the benchmark and not when market and credit series are (Figure~\ref{fig:groups}): a correction read against the state of activity, not a duplicate of what markets have priced, and a correction to the outlook for activity alone: the same index adds nothing to the forecast of inflation (Section~\ref{sec:consensus}). That the published projections carry none of it (Figure~\ref{fig:sep}) is what one expects if members present a united front in what they publish \citep{VisserSwank2007,Tillmann2011}.

\paragraph{The record, not the room.}
The index reads the Minutes, a summary the Committee approves, rather than the transcript of what was said. With the speaker-level scores of \citet{TsangYang2025}, available for every scheduled meeting from 1976 to 2018 (Supplementary Material~\ref*{app:speakers}), we can ask whether the disagreement as spoken carries the same content. It does not. Over the 41 targets of our window that the transcripts cover, the disagreement among voting members, among non-voting participants, among all participants and among the staff each raises the out-of-sample loss of the public benchmark rather than lowering it, while the Minutes index lowers it by 3.8 percent over the same targets and by 6.3 percent in the center; the Minutes beat the voters' own transcript scores in the center of the distribution ($p=0.01$). The Minutes index is, by construction, a prediction of the transcript index; that it forecasts better than what it predicts says that the record filters the discussion: what the Committee chooses to set down about its disagreement is what carries information about the economy, and the disagreement as spoken in the meeting does not. We note the limits: 41 targets, and transcript signals that are noisy in the lower tail. The staff's disagreement, read from the same transcripts, carries no content either, which is the direct form of the test that Section~\ref{sec:consensus} runs with the staff forecast.

\paragraph{Expressed and informative disagreement.}
The transparency change of October 1993, after which the Committee knew its transcripts would be published, offers a check on what the index measures (Supplementary Material~\ref*{app:properties}). Expressed disagreement fell by almost half, in the transcripts and more in the Minutes; what the remaining disagreement says about future growth did not: the index is unrelated to growth four quarters ahead before 1993 and positively related after, with no detectable break. Transparency changed how much the Committee disagrees on the record \citep{HansenMcMahonPrat2018}, not what that disagreement tells about the economy.

\paragraph{No market response, and not the path of policy.}
Two further tests locate it (Supplementary Material~\ref*{app:timing}). Daily yields do not respond to it: a one-standard-deviation surprise in the index moves the two-year Treasury yield by less than one basis point on the day the Minutes are released and on the day of the decision, no more than on placebo days. What the Minutes reveal is public three weeks after the meeting, and Treasury yields and breakeven inflation show no detectable response on the day of release. This does not show that prices never incorporate it, but is consistent with a forecast built on it gaining; it is not what the announcement-day revisions of \citet{NakamuraSteinsson2018} capture. Nor does it reach growth through policy. Higher disagreement precedes higher policy rates over the following year, though not beyond the term structure once adjusted for multiple testing, and precedes higher growth; a policy channel would require the opposite sign. Conditioning the association with growth on the realized change in the funds rate removes a third of it: policy responds to the growth it anticipates.

\paragraph{Not a matter of who is in the room.}
Two features of the FOMC change who votes without changing the economy: the annual rotation of four Reserve Bank presidents, fixed by statute, and the vacancies on the Board that depend on the timing of nominations. Neither detectably moves the index. With each president's disagreement measured by a fixed effect estimated on the transcripts, the mean fixed effect of the year's voters has no explanatory power for the Minutes index once the public state and the chair are controlled for (first-stage $F$ of 0.3, and below 1.6 without controls); Board vacancies lower recorded disagreement by about a tenth of a standard deviation per empty seat, with $F$ of 1.3 over 1994--2025 and 5.1 at most in any subperiod (Supplementary Material~\ref*{app:speakers}). The same president disagrees as much when he votes as when he does not (a difference of 0.001 on a 0--1 scale, with person and meeting fixed effects), and the gap between voters and non-voters since 1990 is composition. On this evidence, recorded disagreement reflects the Committee's deliberation beyond its formal vote: voting status does not detectably alter what a participant expresses, and the composition of the voting membership has little explanatory power for the index, with tests of limited power. Nor does the vote itself carry the content: the share of dissenting votes, added as the index is, does not improve the forecast, and with it in the model the index's whole-distribution gain is 4.4 and 7.5 percent, against 4.1 and 5.9 (added after the design; Supplementary Material~\ref*{sec:formaldissents}). Whether the index's predictive content varies with who votes is not tested here. The paper therefore has no exogenous shifter of the index, and we do not claim one.

\paragraph{Two readings of the evidence.}
Two readings fit every result: the value of deliberation may be the value of its form, reproducible by any committee that argues over the same public documents, or the value of what this Committee knows. The evidence narrows the second: the content is not what the Committee publishes, not what forecasters expect, not the forecast its staff puts before it, not the heterogeneity of the regional information its members bring to the meeting (Section~\ref{sec:consensus}), not the disagreement of its staff or the words of its members as spoken, and not its own future policy. Of the forms of private information that can be measured, none accounts for it; what remains is the disagreement the Committee records. The two new results lean toward the first reading: the content is tied to the procedure that produces the record, not to the identity of those who vote, though neither establishes it. Separating the two requires a committee deliberating on public information alone, such as the simulated FOMC of \citet{KazinnikSinclair2025}.
\section{Conclusion}\label{sec:conclusion}

What a monetary policy committee argues about contains information about the distribution of future output growth beyond what a broad public benchmark extracts. Against a benchmark built on 123 public macroeconomic and financial series, a measure of the disagreement that is absent from the vote, read from the FOMC's Minutes, public three weeks after each meeting, lowers the out-of-sample loss of the four-quarter-ahead growth distribution by about 4 to 6 percent, and by more in its center. That contribution is present in the canonical Growth-at-Risk regressions; it survives a benchmark re-tuned for the center, real-time vintages of the public data and the professional consensus; and it belongs to the disagreement rather than to what the Committee says about risks or declares in its projections. The contribution to the lower region is conditional: it appears once financial conditions and the full public panel are accounted for, and is sensitive to heavy shrinkage of the signal and to the consensus.

A large part of the informational value of central bank deliberation, as far as the Minutes allow it to be measured, therefore lies in the recorded discussion itself, and in the disagreement it reveals, rather than in the assessments of risk the Committee chooses to state. For forecasting, once a broad public benchmark is in place, the question for any new signal is no longer whether it helps but what part of its gain survives the benchmark and in which region of the distribution; for deliberation the answer is the center, while the lower tail remains the domain of financial conditions. For communication, a committee's output is not exhausted by its statement and its projections: the Minutes make that content available to any forecaster three weeks after each meeting. For committee design, the content is in the disagreement the Committee records rather than in the words of its members as spoken, and that disagreement does not detectably change with who votes; where members vote expressively, the vote itself may carry it.

What the paper cannot yet say is why deliberation helps, whether it is the form of the discussion or what the Committee knows. The evidence narrows the second reading without excluding it, since the contribution survives conditioning on the Committee's published outlook, the professional consensus, the forecast of its own staff, the regional information its members bring to the meeting and its own future policy; and the evidence on the record and on membership leans toward the first, from one committee and seventy overlapping targets, without establishing it. What remains, on this evidence, is the disagreement the Committee records.

\clearpage
{\footnotesize\singlespacing\setlength{\bibsep}{0pt}\bibliography{refs}}

\clearpage
\onehalfspacing
\begin{center}
{\Large\textbf{Supplementary Material}}\\[6pt]
{\large Reading the Room with AI: FOMC Deliberation and Growth-at-Risk}\\[4pt]
{\normalsize \'Alvaro Ortiz and Tomasa Rodrigo}
\end{center}
\vspace{1em}
\appendix
\numberwithin{table}{section}
\numberwithin{figure}{section}
\section{Estimation, verification and pre-specified objects}\label{app:protocol}

\paragraph{Estimation code.}
All results are produced by a single estimation and evaluation code. Quantile regressions are estimated by minimizing the pinball loss with an $\ell_2$ penalty on slopes (Section~\ref*{sec:framework}); the unpenalized path of the same code estimates the low-dimensional benchmarks \Mzero{} and \Mone{}. Density synthesis follows the official implementation of \citet{ABG2019}: the skew-$t$ of \citet{AzzaliniCapitanio2003} is fitted by matching the predicted quantiles, and the fit error is stored for every origin.

\paragraph{Verification against the official replication.}
The code's canonical path was verified against the official MATLAB implementation of \citet{ABG2019} on the authors' original sample: out-of-sample quantile forecasts coincide to machine precision at both horizons (maximum absolute differences of order $10^{-13}$), and in-sample skew-$t$ parameters coincide except on the flat valley of the likelihood at the degrees-of-freedom boundary, where parameters differ while the implied quantile and density curves agree to $10^{-4}$ and losses are identical (339 of 339 audited rows equivalent). The verification criterion is as follows. Out-of-sample forecasts are certified at a maximum absolute difference of $10^{-10}$; in-sample quantile and density curves at $10^{-3}$; skew-$t$ parameters at $10^{-4}$ for interior rows. At the degrees-of-freedom boundary ($\nu$ near 30) parameter equality is not required: rows are certified by equal objective values and fitted distribution curves. The audit classified all 339 rows as equivalent, with no row worse in either implementation and boundary differences in the curves below $4\times10^{-4}$.

\paragraph{Pre-specified objects.}
Public panel assembled from 1973Q1, which is the first training row of the main specification (136 observations at the first origin, series that start later entering under the available-only transformation); the common-history comparisons start at 1974Q4 (129 observations); language-based measures from 1994, under the available-only transformation for signals that start after the public history. Block penalties selected on the public benchmark alone, on the 92 validation targets ending in 2007Q3, by a one-standard-error rule with HAC standard errors (Bartlett, bandwidth 3): $\lambda_L=0.562$, $\lambda_C=100$, $\lambda_U=100$, inherited unchanged by every model containing the public panel in the main and real-time forecast sets and never re-selected for any signal; the Greenbook and Beige Book sets use $\lambda_U=1$ in both arms (Supplementary Material~\ref{app:robust}), and the common-history sets use the penalties stated where they appear; every comparison in the paper states whether its two arms were estimated at the same penalties, and those that were not are reported as descriptive. The minimizers of the same criterion, $\lambda_C=5.62$ and $\lambda_U=0.316$ ($\lambda_L$ identical), enter only the exercises added after the audit (Section~\ref*{sec:rich}). Test sample 2008Q4--2026Q1 (70 targets, $h=4$), never used for a design choice. Two tests applied to every signal: all 19 quantiles against the public panel, and the lower block (Q05--Q25) against the panel with financial conditions. Holm adjustment within each pre-specified family of signals. Moving-block bootstrap with block length $L=4$, and $L\in\{6,8\}$ as sensitivity. The seven event windows, fixed before any test-sample result was examined. The two robustness exercises R-1 (public history from 1986Q1) and R-2 (signal penalized at the block penalty), specified with the main design. The group contrast $C$ of Supplementary Material~\ref{sec:mechanism}, labeled exploratory. Forecasts were generated blind and time-stamped before any loss was computed, and every comparison is estimated on identical rows with identical realized targets.The validation period runs from target 1984Q4 to 2007Q3 (92 targets). Every design and every forecast set is recorded with a dated manifest and a SHA-256 hash in the replication package.

\subsection{Bootstrap details}\label{app:bootstrap}
For each comparison the per-target loss-differential series
\[
d_t = \frac{1}{|\mathcal{T}_b|}\sum_{\tau\in\mathcal{T}_b}\left[\pinball\!\left(y_{t+4}-\hat Q_{\tau,t}(m{+}S)\right) - \pinball\!\left(y_{t+4}-\hat Q_{\tau,t}(m)\right)\right],
\]
where $\mathcal{T}_b$ is the set of quantiles of block $b$ and negative values favor the signal, is resampled by the moving-block bootstrap of \citet{Kunsch1989} with non-circular blocks of $L$ consecutive targets and $B=10{,}000$ replicates; gains, family effects and the contrast $C$ are recomputed inside each replicate from the same resampled blocks, so that their joint dependence is preserved; percentile intervals are reported. The HAC test uses a Bartlett kernel of bandwidth $h-1=3$ on the same series. Inference is conditional on the pre-specified tuning and, for nested models, conservative \citep{ClarkMcCracken2001,ClarkWest2007}. For each endpoint-by-block-length cell the pseudorandom generator is initialized with seed 20260903; paired comparisons use the same resampled target indices; the main block length is 4, with 6 and 8 as sensitivity.
\section{Data details}\label{app:data}

\subsection{Information families}\label{app:families}
The 125 candidate FRED-MD series are partitioned into four families following the database's groups \citep{McCrackenNg2016}: \textsc{real} collects output and income, the labor market, housing, and orders and inventories (groups 1--4); \textsc{nominal} collects prices (group 7); \textsc{credit/rates} collects money and credit and interest rates and spreads (groups 5--6); \textsc{markets} collects stock-market and exchange-rate series (group 8 and the exchange-rate block of group 6). Table~\ref{tab:family-map} lists the series and their families. The 125 candidate series of the panel reduce to the 123 of the estimation panel after the exclusion, by the pre-specified data protocol, of UMCSENTx and TWEXAFEGSMTHx; the estimation counts are 62, 23, 34 and 4.

\begin{table}[!htbp]
\centering
\caption{The 125 candidate FRED-MD series and their families}
\label{tab:family-map}
\scriptsize
\begin{tabular}{llll}
\toprule
Series & Family & Series & Family \\
\midrule
RPI & Real & M2SL & Credit/rates \\
W875RX1 & Real & M2REAL & Credit/rates \\
DPCERA3M086SBEA & Real & BOGMBASE & Credit/rates \\
CMRMTSPLx & Real & TOTRESNS & Credit/rates \\
RETAILx & Real & NONBORRES & Credit/rates \\
INDPRO & Real & BUSLOANS & Credit/rates \\
IPFPNSS & Real & REALLN & Credit/rates \\
IPFINAL & Real & NONREVSL & Credit/rates \\
IPCONGD & Real & CONSPI & Credit/rates \\
IPDCONGD & Real & S\&P 500 & Markets \\
IPNCONGD & Real & S\&P div yield & Markets \\
IPBUSEQ & Real & S\&P PE ratio & Markets \\
IPMAT & Real & FEDFUNDS & Credit/rates \\
IPDMAT & Real & CP3Mx & Credit/rates \\
IPNMAT & Real & TB3MS & Credit/rates \\
IPMANSICS & Real & TB6MS & Credit/rates \\
IPB51222S & Real & GS1 & Credit/rates \\
IPFUELS & Real & GS5 & Credit/rates \\
CUMFNS & Real & GS10 & Credit/rates \\
HWI & Real & AAA & Credit/rates \\
HWIURATIO & Real & BAA & Credit/rates \\
CLF16OV & Real & COMPAPFFx & Credit/rates \\
CE16OV & Real & TB3SMFFM & Credit/rates \\
UNRATE & Real & TB6SMFFM & Credit/rates \\
UEMPMEAN & Real & T1YFFM & Credit/rates \\
UEMPLT5 & Real & T5YFFM & Credit/rates \\
UEMP5TO14 & Real & T10YFFM & Credit/rates \\
UEMP15OV & Real & AAAFFM & Credit/rates \\
UEMP15T26 & Real & BAAFFM & Credit/rates \\
UEMP27OV & Real & TWEXAFEGSMTHx & Credit/rates (excl.) \\
CLAIMSx & Real & EXSZUSx & Credit/rates \\
PAYEMS & Real & EXJPUSx & Credit/rates \\
USGOOD & Real & EXUSUKx & Credit/rates \\
CES1021000001 & Real & EXCAUSx & Credit/rates \\
USCONS & Real & WPSFD49207 & Nominal \\
MANEMP & Real & WPSFD49502 & Nominal \\
DMANEMP & Real & WPSID61 & Nominal \\
NDMANEMP & Real & WPSID62 & Nominal \\
SRVPRD & Real & OILPRICEx & Nominal \\
USTPU & Real & PPICMM & Nominal \\
USWTRADE & Real & CPIAUCSL & Nominal \\
USTRADE & Real & CPIAPPSL & Nominal \\
USFIRE & Real & CPITRNSL & Nominal \\
USGOVT & Real & CPIMEDSL & Nominal \\
CES0600000007 & Real & CUSR0000SAC & Nominal \\
AWOTMAN & Real & CUSR0000SAD & Nominal \\
AWHMAN & Real & CUSR0000SAS & Nominal \\
HOUST & Real & CPIULFSL & Nominal \\
HOUSTNE & Real & CUSR0000SA0L2 & Nominal \\
HOUSTMW & Real & CUSR0000SA0L5 & Nominal \\
HOUSTS & Real & PCEPI & Nominal \\
HOUSTW & Real & DDURRG3M086SBEA & Nominal \\
PERMIT & Real & DNDGRG3M086SBEA & Nominal \\
PERMITNE & Real & DSERRG3M086SBEA & Nominal \\
PERMITMW & Real & CES0600000008 & Nominal \\
PERMITS & Real & CES2000000008 & Nominal \\
PERMITW & Real & CES3000000008 & Nominal \\
AMDMNOx & Real & UMCSENTx & Real (excl.) \\
ANDENOx & Real & DTCOLNVHFNM & Credit/rates \\
AMDMUOx & Real & DTCTHFNM & Credit/rates \\
BUSINVx & Real & INVEST & Credit/rates \\
ISRATIOx & Real & VIXCLSx & Markets \\
M1SL & Credit/rates &  &  \\
\bottomrule
\end{tabular}
\begin{minipage}{0.95\textwidth}\scriptsize\vspace{3pt}
Notes: FRED-MD mnemonics in the order of the panel. Families: \textsc{real} 63, \textsc{nominal} 23, \textsc{credit/rates} 35, \textsc{markets} 4. The two series marked (excl.), UMCSENTx and TWEXAFEGSMTHx, are excluded by the pre-specified data protocol, leaving the 123 series of the estimation panel (62, 23, 34 and 4). Oil prices are in \textsc{nominal}; the VIX and the three S\&P series in \textsc{markets}; the four exchange-rate series in \textsc{credit/rates}. Current growth is a separate regressor and not part of the panel.
\end{minipage}
\end{table}

\subsection{The Committee's published projections: central outlook and dispersion}\label{app:sep}

\paragraph{Sources and coverage.}
The two measures are built from the central tendency of FOMC participants' projections of output growth, taken from two publications. From February 1983 to July 2007 the Monetary Policy Report to the Congress reported, twice a year, the range and the central tendency of the growth projections of Governors and Reserve Bank presidents; the February report projected the current year and the July report the current and the following year, and the variable was real GNP until 1991 and real GDP from February 1992. From November 2007 the Summary of Economic Projections (SEP) reports the same objects four times a year, for the current year and the following two or three: from the first SEP, prepared for the meeting of October 30--31, 2007 and published on November 20, 2007 as an addendum to the Minutes, to the SEP of December 2025, 73 releases and 251 release-year observations (no SEP was published in March 2020). From April 2011 the SEP central tendencies and ranges were released with the statement on the day of the meeting, and before then with the Minutes. In both publications each projection is the percentage change from the fourth quarter of the previous year to the fourth quarter of the projection year, and the central tendency excludes the three highest and the three lowest projections. The individual projections behind these summaries, released with a lag, are assembled by \citet{Romer2010}; we use only the published central tendencies. The quarterly series run from 1983Q1 to 2025Q1, 169 quarters: 99 from the Monetary Policy Reports and 70 from the SEP. The change from GNP to GDP is recorded, not harmonized: each round enters with the output concept it projected. The SEP also reports participants' assessments of uncertainty and of the balance of risks, and from September 2015 the median projection; none enters the paper.
The 50 Monetary Policy Report rounds are all on a fourth-quarter-over-fourth-quarter definition: 18 for real GNP (1983--1991) and 32 for real GDP (1992--2007). One round departs from the central-tendency rule: February 1994 excludes two observations at each end and rounds to the quarter point. The number of participants in each round is not recorded in the published tables and is not imputed.

\paragraph{Central outlook.}
For round $r$ and projection year $Y$, the declared outlook is the midpoint of the central tendency,
\[
m_{r,Y}=\tfrac12\left(\mathrm{CT}^{\mathrm{low}}_{r,Y}+\mathrm{CT}^{\mathrm{high}}_{r,Y}\right).
\]
It is the one summary of the projections available over the whole period. It is not the median projection, which the Committee did not publish before September 2015, and it is not the forecast of any individual participant.

\paragraph{Conversion to the target.}
The projections refer to calendar years, while the target is average growth over the four quarters after the origin. The conversion uses only information available on the day the round was published. Let $V_r$ be the latest vintage of real output dated on or before the publication date of round $r$, and let $q_r$ be the last quarter it contains, with level $X_{q_r}$. The projection for the current year $Y$ implies a fourth-quarter level $X_{Y:4}=X_{Y-1:4}\,(1+m_{r,Y}/100)$, where $X_{Y-1:4}$ is taken from $V_r$; the growth still to come between $q_r$ and the fourth quarter of $Y$ is spread evenly, with compounding, over the $k$ quarters in between, so that each has the gross growth factor $(X_{Y:4}/X_{q_r})^{1/k}$. Each later projection year is spread evenly over its four quarters, with gross quarterly factor $(1+m_{r,Y+1}/100)^{1/4}$, and so on. The implied quarterly path gives annualized growth rates $\hat g_{r,q}$, computed with the same convention as the target, and the measure used at origin $t$ is $\tfrac14\sum_{j=1}^{4}\hat g_{r(t),t+j}$, where $r(t)$ is the round assigned to origin $t$. When a February report gives no projection for the following year, the current-year projection is carried into it.
For the historical rounds the February carry-forward is used only when the projection for the following calendar year is absent and the four-quarter target requires it. GNP rounds use the real-time GNP vintage available at the origin and GDP rounds the corresponding GDP vintage; the published output concept is preserved, so GNP is not relabeled as GDP.

\begin{figure}[!htbp]
\centering
\includegraphics[width=0.95\textwidth]{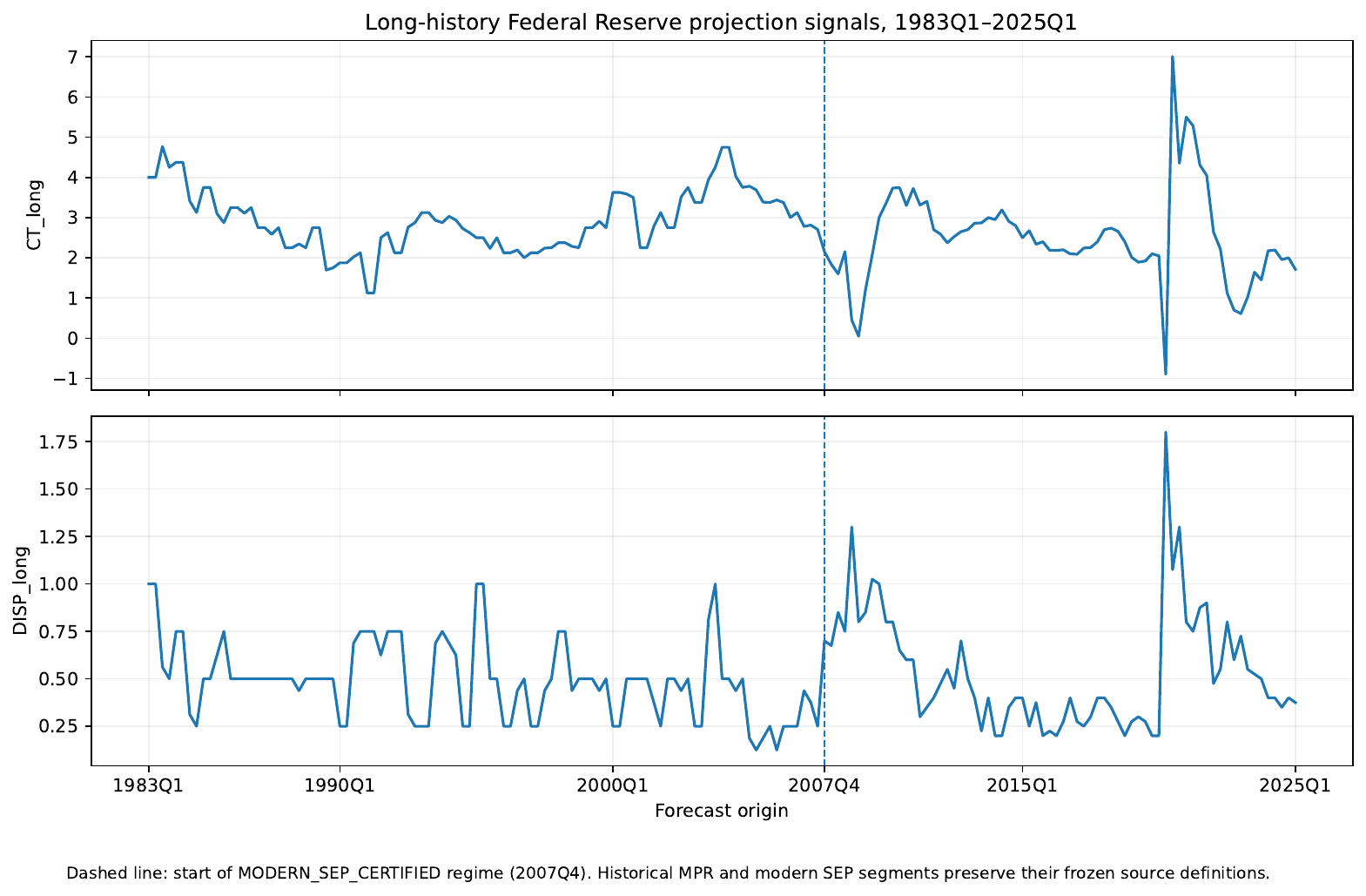}
\caption{Central tendency and dispersion of the Committee's growth projections, 1983--2025. Monetary Policy Reports to 2007Q3, SEP from 2007Q4; quarterly series on the paper's clock.}
\label{fig:ctdisp}
\end{figure}

\paragraph{Dispersion.}
The Committee's declared disagreement is the width of the central tendency, $w_{r,Y}=\mathrm{CT}^{\mathrm{high}}_{r,Y}-\mathrm{CT}^{\mathrm{low}}_{r,Y}$. For each origin the measure is the average of the widths for the calendar years covered by the four target quarters, taken from the round assigned to that origin. No interpolation or backfilling is used. Over the 70 test origins the measure ranges from 0.20 to 1.80 percentage points, with a mean of 0.54.
Table~\ref{tab:ctdisp-desc} and Figure~\ref{fig:ctdisp} describe the two quarterly series by publication.

\begin{table}[!htbp]
\centering
\caption{The Committee's declared outlook and disagreement, by publication}
\label{tab:ctdisp-desc}
\small
\resizebox{\textwidth}{!}{%
\begin{tabular}{llrrrrr}
\toprule
Publication & Measure & $N$ & Mean & S.d. & Min. & Max. \\
\midrule
Monetary Policy Report, 1983Q1--2007Q3 & Central tendency & 99 & 2.92 & 0.75 & 1.13 & 4.77 \\
Monetary Policy Report, 1983Q1--2007Q3 & Dispersion & 99 & 0.49 & 0.21 & 0.13 & 1.00 \\
SEP, 2007Q4--2025Q1 & Central tendency & 70 & 2.49 & 1.19 & $-0.90$ & 7.01 \\
SEP, 2007Q4--2025Q1 & Dispersion & 70 & 0.54 & 0.31 & 0.20 & 1.80 \\
\bottomrule
\end{tabular}}
\begin{minipage}{0.9\textwidth}\footnotesize\vspace{3pt}
Notes: Quarterly series on the paper's clock, in percentage points; the dispersion is the width of the central tendency. Descriptive, not a re-estimation.
\end{minipage}
\end{table}

\paragraph{Dating and treatment.}
Each round is dated by its publication date and assigned to origins on the same conservative clock as the Minutes-based measures; between rounds, the latest published round is used. Both measures enter in levels, unpenalized and without standardization, and are set to zero before 1983Q1, so that every model in a comparison is estimated on the same observations; the zeros are an availability device, not observations of the Committee's outlook. At the first forecast origin the models are estimated on 129 quarters, 96 of them with the measures observed. The decision to enter the measures in levels was recorded before any result with them was examined.

\paragraph{Comparisons.}
Two comparisons were fixed before any forecast with the long series was generated: Hidden Dissent added to the public panel containing the central tendency, in the center, and to the panel with financial conditions containing the dispersion, in the lower block. The two remaining pairings were added once those were known. Holm adjustment is applied within each pair (Figure~\ref*{fig:sep} of the paper; Table~\ref{tab:sep-full}). Table~\ref{tab:sep-conc} shows how the four loss differentials are distributed over time. In the center the gain is spread out: the five most favorable targets account for about 30 percent of the cumulated improvement, and excluding the financial crisis and the pandemic leaves the mean differential close to its full-sample value. In the lower block it is concentrated: five targets account for more than half, and excluding the two episodes reduces the mean differential by more than half, without changing its sign. In the estimation of the center comparison with the central tendency, the solver's authorized fallback from summed to averaged loss was used in 51 of 1,960 fits, 25 for the benchmark and 26 for the model with Hidden Dissent, at dates spread over the sample; the other comparisons used none.

\begin{table}[!t]
\centering\small
\caption{Hidden Dissent and the published projections: how the gain is distributed over time}\label{tab:sep-conc}
\resizebox{\textwidth}{!}{%
\begin{tabular}{llccccc}
\toprule
 & & \multicolumn{3}{c}{Mean loss differential $\bar d$} & \multicolumn{2}{c}{Share of improvement} \\
\cmidrule(lr){3-5}\cmidrule(lr){6-7}
Block & Projection measure & All 70 & Excl.\ crisis & Excl.\ both & Top 5 & Top 10 \\
\midrule
Center (\Mtwo{}) & Central tendency & $-0.0577$ & $-0.0676$ & $-0.0507$ & 30.5 & 50.8 \\
                 & Dispersion$^{\S}$ & $-0.0526$ & $-0.0547$ & $-0.0415$ & 30.2 & 50.2 \\
Lower (\Mthree{}) & Dispersion        & $-0.0395$ & $-0.0195$ & $-0.0143$ & 57.2 & 73.5 \\
                  & Central tendency$^{\S}$ & $-0.0321$ & $-0.0180$ & $-0.0163$ & 53.6 & 72.7 \\
\bottomrule
\end{tabular}}
\begin{minipage}{0.92\textwidth}\footnotesize\vspace{4pt}
Notes: $\bar d$ is the mean, over targets, of the block loss of the model with Hidden Dissent minus that of the benchmark containing the projection measure; negative values favor Hidden Dissent. ``Crisis'': targets 2008Q4--2010Q4; ``both'': also the pandemic targets 2020Q1--2021Q4. The exclusions are descriptive: no model is re-estimated. Excluding only the pandemic targets, $\bar d$ is $-0.0419$, $-0.0410$, $-0.0377$ and $-0.0325$ in the order of the rows. Share of improvement: percentage of the cumulated improvement over the targets at which Hidden Dissent has the lower loss accounted for by the five and ten most favorable. $^{\S}$Added after the two pre-specified comparisons were known.
\end{minipage}
\end{table}

\paragraph{Benchmarks.}
Every comparison with the projection measures holds the estimation history and the block penalty fixed across its two models. All are estimated on the common history from 1974Q4 (129 observations at the first origin), whereas the main forecast sets start in 1973Q1 (136). The benchmarks of these comparisons are not the bare \Mtwo{} and \Mthree{} of Table~\ref*{tab:overview} but \Mtwo{} with the central tendency and \Mthree{} with the dispersion, and Hidden Dissent is compared within each pair at the same history and penalty: against them it improves the center by 6.3 percent and the lower block by 7.0 percent, against 5.6 and 4.5 percent against the main forecast sets without the projection; the difference is the benchmark, not the first origin or the training size. In the lower block every comparison uses the pre-specified penalty. In the center, the comparison with the central tendency uses the pre-specified penalty, and its benchmark without the measure was estimated for this purpose; the comparisons of Hidden Dissent with the dispersion in the benchmark use $\lambda_L=0.562$, $\lambda_C=5.62$ and $\lambda_U=100$ in both arms (an earlier branch with $\lambda_U=0.316$ was superseded and is not used). The common-history public panel without the dispersion exists in the upper block only at $\lambda_U=0.316$; the dispersion's own whole-distribution and upper-block rows in Table~\ref{tab:sep-full} therefore compare forecasts estimated at different upper-block penalties, are reported as descriptive, and are omitted from Table~\ref*{tab:overview}. Table~\ref{tab:sep-full} reports every comparison.

\begin{table}[!t]
\centering\small
\caption{Hidden Dissent and the Committee's published projections: full inference}\label{tab:sep-full}
\resizebox{\textwidth}{!}{%
\begin{tabular}{llccccc}
\toprule
Comparison & Block & Gain (\%) & $\bar d$ & $p$ & 95\% moving-block interval & Wins \\
\midrule
\multicolumn{7}{l}{\emph{Panel A. Hidden Dissent added to the benchmark}} \\
\Mtwo{}, no projections & Center & 6.31 & $-0.0464$ & -- & $[-0.0724,\ -0.0201]$ & 50/70 \\
\Mtwo{} + central tendency (C1) & Center & 7.60 & $-0.0577$ & 0.0002 & $[-0.0850,\ -0.0292]$ & 51/70 \\
\Mtwo{} + dispersion$^{\S}$ & Center & 7.03 & $-0.0526$ & 0.0002 & $[-0.0780,\ -0.0269]$ & 54/70 \\
\Mtwo{} + dispersion$^{\S}$ & All 19 & 4.69 & $-0.0311$ & 0.0001 & $[-0.0458,\ -0.0168]$ & 56/70 \\
\Mtwo{} + dispersion$^{\S}$ & Upper & 1.89 & $-0.0109$ & 0.197 & --$^{\P}$ & 56/70 \\
\Mtwo{} + dispersion$^{\S}$ & Lower & 2.10 & $-0.0126$ & 0.127 & $[-0.0360,\ +0.0040]$ & 32/70 \\
\Mthree{}, no projections & Lower & 6.99 & $-0.0311$ & -- & $[-0.0522,\ -0.0042]$ & -- \\
\Mthree{} + central tendency$^{\S}$ (C4) & Lower & 7.28 & $-0.0321$ & 0.045 & $[-0.0510,\ -0.0036]$ & 42/70 \\
\Mthree{} + dispersion (C2) & Lower & 7.70 & $-0.0395$ & 0.035 & $[-0.0694,\ -0.0075]$ & 48/70 \\
\addlinespace
\multicolumn{7}{l}{\emph{Panel B. Projection measure added to the benchmark}} \\
Central tendency, \Mtwo{} & Center & $-1.86$ & $+0.0138$ & 0.164 & $[-0.0018,\ +0.0348]$ & 29/70 \\
Dispersion, \Mtwo{} & Center & $-1.88$ & $+0.0138$ & 0.244 & $[-0.0062,\ +0.0361]$ & -- \\
Dispersion, \Mtwo{}$^{\ddagger}$ & All 19 & $-6.01$ & $+0.0376$ & 0.157 & $[-0.0157,\ +0.0829]$ & 13/70 \\
Dispersion, \Mtwo{}$^{\ddagger}$ & Upper & $-6.34$ & $+0.0343$ & 0.625 & $[-0.1241,\ +0.1241]$ & 10/70 \\
Dispersion, \Mtwo{} & Lower & $-16.25$ & $+0.0838$ & 0.117 & $[+0.0017,\ +0.1838]$ & -- \\
Central tendency, \Mthree{} & Lower & 0.82 & $-0.0037$ & 0.884 & $[-0.0401,\ +0.0508]$ & 33/70 \\
Dispersion, \Mthree{} & Lower & $-15.19$ & $+0.0676$ & 0.195 & $[-0.0093,\ +0.1468]$ & 32/70 \\
\bottomrule
\end{tabular}}
\begin{minipage}{0.92\textwidth}\footnotesize\vspace{4pt}
Notes: 70 targets, 2008Q4--2026Q1; every model estimated on the public history from 1974Q4; projection measures in levels, zero before 1983Q1. $\bar d$ is the loss of the larger model minus that of the benchmark, averaged over the quantiles of the block. $p$: in Panel A, one-sided HLN for an improvement; in Panel B, two-sided HLN. Moving-block bootstrap with $L=4$ and 10,000 replications. Wins: targets at which the larger model has the lower loss. C1 and C2 are the two comparisons fixed in advance (Holm 0.0003 and 0.035); $^{\S}$added once they were known (Holm across C3 and C4, 0.0003 and 0.045). Penalties as described in the text. Without the pandemic targets (67), the dispersion's whole-distribution loss change is $-6.60$ percent ($p=0.177$). $^{\ddagger}$Descriptive: the benchmark's upper-block forecasts carry $\lambda_U=0.316$ and the dispersion's $\lambda_U=100$, so these two rows are not equal-penalty comparisons and do not appear in Table~\ref*{tab:overview}. $^{\P}$The upper-block comparison of the final forecast sets was evaluated with HLN inference only; its HAC interval is $[-0.0345,\ +0.0127]$. The three sensitivities of the projection series declared before the comparisons were run (the February horizon rule, a common publication frequency, and demeaning within each publication regime) were not executed.
\end{minipage}
\end{table}

\paragraph{The comparison with the SEP alone.}
Restricted to the SEP, the projection series begin in 2007Q4, and the coefficient on each measure is estimated at the first origins from very few observations. Table~\ref{tab:sep-oa} reports the comparisons under that restriction. Over the 70 targets, adding the central outlook raises the benchmark's loss by 42.8 percent, a deterioration the data cannot distinguish from zero and that five targets in 2010--11 largely account for; from the origins of 2012Q1 it disappears. Over the 45 targets from 2015Q1, with every model estimated on the long public history, Hidden Dissent's central gain falls from 6.1 to 0.6 percent when the SEP central outlook is in the benchmark, while its lower-block gain conditional on the SEP dispersion is 3.7 percent ($p=0.026$). With the coefficient estimated on so short a history of projections, these comparisons are imprecise; the long series are the paper's measure. Over the same 45 targets, replacing the SEP-only series by the long series changes the fitted distribution, but in no block distinguishably from zero (two-sided $p$ between 0.14 and 0.82). A further comparison that also truncated the estimation history of every model to the SEP period, leaving about 22 quarters at the first origin, is in Panel D.

\begin{table}[!t]
\centering\small
\caption{Hidden Dissent and the Committee's projections, SEP alone}\label{tab:sep-oa}
\resizebox{\textwidth}{!}{%
\begin{tabular}{llccc}
\toprule
 & & Without the & + central & + \\
Comparison & Block & SEP & outlook & dispersion \\
\midrule
\multicolumn{5}{l}{\emph{Panel A. 45 targets from 2015Q1: gain from adding Hidden Dissent}} \\
\Mtwo{} & All 19 & 5.88 (0.046) & 1.53 (0.318) & 3.03 (0.225) \\
\Mthree{} & Lower & 5.12 (0.006) & 2.59 (0.150) & 3.67 (0.026) \\
\Mtwo{} & Center & 6.11 (0.009) & 0.62 (0.770) & 3.54 (0.298) \\
\Mtwo{} & Upper & 7.00 (0.278) & 3.67 (0.224) & 3.94 (0.260) \\
\addlinespace
\multicolumn{5}{l}{\emph{Panel B. 45 targets from 2015Q1: gain from adding the SEP measure}} \\
\Mtwo{} & All 19 & -- & 2.55 (0.712) & $-4.27$ (0.552) \\
\Mtwo{} & Center & -- & 8.19 (0.271) & 1.68 (n.s.) \\
\Mthree{} & Lower & -- & 2.46 (0.419) & $-0.99$ (0.785) \\
\addlinespace
\multicolumn{5}{l}{\emph{Panel C. Central outlook, full test sample}} \\
SEP added, \Mtwo{} & All 19 & -- & $-42.76$ (n.s.) & \\
SEP added, \Mthree{} & Lower & -- & $-13.76$ (n.s.) & \\
HD added, 70 targets & All 19 / Lower & -- & 0.62 / 5.11 (n.s.) & \\
HD added, origins from 2012Q1 & All 19 / Lower & -- & 1.75 / 1.77 (0.21 / 0.31) & \\
\addlinespace
\multicolumn{5}{l}{\emph{Panel D. Dispersion, estimation history truncated to 2007Q4, 45 targets}} \\
SEP added & All 19 / Lower & -- & & $-9.93$ / $-15.82$ \\
HD added & All 19 / Lower & -- & & $-4.66$ / 2.08 \\
\bottomrule
\end{tabular}}
\begin{minipage}{0.92\textwidth}\footnotesize\vspace{4pt}
Notes: Percentage reduction in block pinball loss; two-sided HLN $p$-values in parentheses; n.s.: not distinguishable from zero at the five percent level. Panels A and B: every model estimated on the public history from 1974Q4, the SEP measures in levels from 2007Q4; in Panel A the benchmark of each column contains the measure named in its heading. In Panel D none of the four entries is distinguishable from zero under HLN; the moving-block bootstrap places the lower-block deterioration from adding dispersion below zero ($p$ of 0.031, 0.042 and 0.049 at $L=4$, 6 and 8).
\end{minipage}
\end{table}

\subsection{Construction of the deliberative signals}\label{app:prompts}

This subsection records how each of the seven Minutes-based measures is built, and reproduces in full the extraction instructions under which the two measures produced for this paper before the family was evaluated---the balance of risks and the insurance motive---were coded; the instructions for attributed heterogeneity, produced afterwards and also the source of the Committee-level balance of risks, are recorded at the end of the subsection. The instructions are given verbatim so that the classifications can be reproduced and audited; nothing in them was changed after the first evaluation of any signal. For the balance of risks two instruction sets exist, one at paragraph level, which produced the earlier version of the measure, and one at document level, whose rules the Committee-level measure now used in the paper applies to the attributed-view extraction; the appendix says which is which at each step.

\paragraph{Balance of risks: concept.}
The measure summarizes whether the Committee sees the risks around its outlook for real activity as weighted to the upside, roughly balanced, or weighted to the downside. It is anchored on the question the Summary of Economic Projections puts to participants (``indicate your judgment of the risk weighting around your projections''), and it is distinct from the direction of the forecast itself: the Committee can expect growth to slow and still see the risks to that expectation as tilted to the upside. The instructions enforce that distinction with worked examples, and they code only explicit risk-balance language.

\paragraph{Balance of risks: the Committee-level measure used in the paper.}
The measure is built from the attributed-view extraction described below for attributed heterogeneity, applying the rules of the document-level instructions in \ref{app:instr-bor-doc} to its fields. For each meeting, the eligible views are those about growth, typed as risk assessments, attributed to the Committee rather than to the staff, and carrying an explicit direction; views conditional on a policy action are excluded. Among them, only the views at the highest quantifier tier present are used: if they share one direction the meeting takes the value $+1$ (upside), $0$ (balanced) or $-1$ (downside); if they point in different directions the meeting is \textsc{divided}, which takes the value $0$ and is flagged separately from \textsc{balanced}; if there is no eligible view the meeting is \textsc{not discussed} and missing. The quantifier tier selects the governing views and is never used as a weight. The quarterly series is the mean over the meetings of the quarter with a value. At the 70 test origins, 66 use the preceding quarter and four (2010Q2, 2012Q3, 2020Q4, 2021Q2) use the most recent earlier quarter with a value, in each case two quarters back; no origin lacks a value.

\paragraph{Balance of risks: why the measure was replaced, and what the evidence was.}
The paper's first version of the measure was the paragraph net share described next. The two versions agree at 0.69 (Pearson) on 112 common quarters and share their sign at 79 percent of them. The decision between them was taken on measurement grounds alone and recorded, with its date, before any forecast with the new version was generated; the evidence did not all point one way, and we report it as it stood. The new version follows the construct as the instructions define it---the Committee's voice only, a split resolved by the dominant quantifier, balanced only when the Committee says so, and absence treated as absence---which the count does not. On the 24 meetings at which the two versions disagree in direction, a blind recoding of the original text by a language model, working from a dossier without either series and with categories fixed in advance, sided with the new version in 15 cases and with the old in 2 (7 ties; exact binomial $p=0.002$ among the 17 that discriminate); because that recoding uses the categories of the new protocol, it is evidence of fidelity to the protocol rather than of accuracy in an external sense. The within-quarter reliability of the two is the same (0.49 against 0.49 on 68 common quarters). Against the most independent benchmark available, the risk weighting reported by participants in the Summary of Economic Projections, both versions agree strongly and the old one slightly more (0.79 against 0.74 on 61 common meetings; the difference has a paired-bootstrap interval of $[-0.16,\ +0.06]$). The old version also agrees more with the same classification applied to the transcripts (0.70 against 0.47), but that comparison shares the old instructions and is a consistency check across corpora, not an independent validation. The new version covers fewer meetings, because it records no assessment where the Committee states none. No blind recoding of the 24 meetings by a human coder was carried out. Once chosen, the new version was not revisited in the light of any forecasting result.

\paragraph{Balance of risks: the earlier paragraph-count version.}
The earlier series is built from a paragraph-level classification of each meeting's Minutes under the instructions reproduced in \ref{app:instr-bor-par}. Every paragraph is coded, for growth and for inflation separately, as stating upside, balanced or downside risks, or as stating no balance of risks for that variable; the coding records whose view the paragraph reports (Committee, staff, other) and a verbatim quote of the decisive language. Let $N^{up}_m$, $N^{bal}_m$ and $N^{down}_m$ be the numbers of paragraphs at meeting $m$ coded upside, balanced and downside for growth. The meeting-level score is the net share of directionally coded paragraphs,
\[
\mathrm{BoR}_m=\frac{N^{up}_m-N^{down}_m}{N^{up}_m+N^{bal}_m+N^{down}_m},
\]
which equals $-1$ when every coded paragraph points to downside risks and $+1$ when every one points to upside risks. A value of zero can arise from balanced codings or from offsetting upside and downside codings; the score does not distinguish the two, and it is therefore not the same object as the document-level category \textsc{balanced} described below. A meeting with no coded growth-risk paragraph is missing, not balanced. The quarterly series is the equal-weight mean of the meeting-level scores over the meetings of the quarter for which a score is observed (one meeting in 7 quarters, two in 116, three in 7); the two test origins without a value are bridged by the latest quarter strictly before the forecast quarter, a rule declared before the runs. On all 130 quarters of the series so constructed, the quarterly values reproduce the earlier forecasting variable exactly. Its results are those reported in earlier versions of this paper: against the public panel, 1.640 percent on the grid, 1.519 without the pandemic quarters, $-1.823$, 3.651 and $-0.004$ in the three blocks, with 12 of 19 quantiles improved; against the real-time panel with financial conditions, 0.585 percent in the lower block ($p=0.31$), with a paired differential against Hidden Dissent of $-0.015$ (two-sided $p=0.015$). Meetings average about five coded growth-risk paragraphs.

\paragraph{Balance of risks: the document-level classification.}
A second classification, under the instructions in \ref{app:instr-bor-doc}, asks for one assessment per meeting of the Committee's collective balance of risks, with how widely the view was held (the verbatim quantifier and an intensity grade), a citation for every positive classification, a negative-search report for every \textsc{not discussed}, and a separate record of any staff assessment. Where the Minutes report a split among participants, this classification resolves it by the Committee's own quantifier ordering (all, most, many, several, some, a few, a couple, one), and it returns \textsc{balanced} only when the document itself characterizes the risks as balanced, never because opposing views coexist. It is a different aggregator of the same text, not a second reading of the same object: across the 92 meetings it classifies as \textsc{balanced}, the paragraph net share averages about $-0.37$, and a document classified \textsc{upside} can carry a negative paragraph net share. The document-level classification is the protocol the Committee-level measure applies.

\paragraph{Balance of risks: convergence of the earlier version across pipelines.}
An independent extraction of the Minutes carried out later under a different instruction set, which decomposes each paragraph into attributed views with a risk direction, allows the paragraph net share to be reconstructed from a second pipeline without reference to the first. On the common sample the two agree at 0.75 (Pearson) and 0.73 (Spearman) at quarterly frequency, at 0.75 and 0.70 at meeting frequency, with the same sign at 80 percent of meetings; only 6 of 159 meetings at which both are non-zero carry opposite signs. A tier-resolved aggregation of the second extraction, in the spirit of the document-level classification, agrees with the paper's series at about 0.70. We report these as convergent validation of the construct across pipelines, not as alternative estimators, and no series in the paper was selected or altered on the basis of them. 

\paragraph{Insurance motive.}
The measure follows the insurance-reasoning construct of \citet{CHMX2023}: a policy action is insurance-motivated when a specific adverse scenario that is regarded as unlikely, rather than as the baseline, directly motivates a concrete policy action or signal. The instructions in \ref{app:instr-ins} apply that two-part test to every paragraph of a meeting's Minutes, in either direction (easing as a hedge against a low-probability downside, tightening as a hedge against a low-probability upside), with two variants that the literal test would miss (entrenchment of inflation; deepening of a crisis already under way) and five exclusions applied strictly: insurance discussed without motivating action, which is coded as a future contingency rather than as insurance; loss-function or asymmetric-risk reasoning, including any balance-of-risks statement; responses to adverse events that have already occurred; insurance about the operating framework rather than the stance; and general vigilance. Each coded paragraph records whose reasoning it is (Committee, staff or other), the verbatim quantifier, the scenario, the target of the reasoning (the current action, formal communication, or a future move only) and a short verbatim quote; a meeting-level summary is then formed from the Committee reasoning that carries the strongest quantifier, with a justification and a confidence grade. Staff reasoning is recorded and never enters the meeting-level summary. The quarterly series is the mean over the meetings of the quarter of the meeting-level intensity of insurance-motivated reasoning.

\paragraph{Attributed heterogeneity.}
The measure records whether the Minutes partition the room. Every paragraph of the participants' discussion is read for the views it attributes and to whom: a view attributed to a subset of the Committee (``a few'', ``some'', ``several'', ``many'', ``most'' participants, or ``others'' set against one of these) is a partition; a view attributed to the Committee or to participants collectively, or an unattributed institutional assessment, is not. A paragraph is \emph{partitioned} when it attributes views to more than one subset; it is \emph{resolved} when the attribution can be read from the text, and unresolved otherwise. Unresolved paragraphs are excluded from both numerator and denominator rather than counted as unpartitioned, so that an extraction failure is never read as agreement. The meeting-level measure is the share $P_m/O_m$ of resolved paragraphs that are partitioned; a meeting with no resolved paragraph is missing rather than zero. The quarterly series is the mean over the meetings of the quarter, dated by release under the $t-1$ rule, and standardized on the expanding training window like every other signal. The construct, the quantifier tiers, the resolved/unresolved rule and the aggregation were fixed in a dated record before any loss involving the series was computed. On the common sample of 128 quarters the series correlates with the \HDlong{} index at $-0.13$ (Pearson) and $-0.05$ (Spearman), and with the financial-conditions index at $0.04$ and $-0.04$; the two readings of division are therefore close to orthogonal in levels, and nothing in the paper selects between them on that basis.

The extraction and the measure are separated. A language model reads one paragraph of the published Minutes at a time and returns, under a strict schema with no free fields, the views the paragraph attributes: for each, the evidence quote, the attribution phrase, the quantifier, the topic, the type of view, its stance and risk direction, whether it is conditional or an alternative scenario, and its relation to other views in the paragraph. It computes no score. The measure is then built by deterministic rules from those fields: a paragraph counts as partitioned when it attributes a view to a small or indefinite subset of the Committee (``a few'', ``some'', ``several'', or ``others'' set against one of these); views attributed to a large but not dominant subset (``many'') enter only an expanded variant reported for robustness. Each paragraph is then observed as partitioned, observed as not partitioned, or unobserved---semantically unresolved, or not extracted---and unobserved paragraphs are excluded from both numerator and denominator. Of the 4{,}006 paragraphs of the corpus, 1{,}909 are partitioned, 1{,}666 are observed as not partitioned, 287 are semantically unresolved and 144 were not extracted, so that 10.8 percent are treated as missing rather than as agreement. The extraction was applied to every paragraph of the corpus without selection---4{,}006 paragraphs from 260 meetings---through the OpenAI Responses API, model \texttt{gpt-5.6-terra} (the provider's public identifier for the model family released in July 2026; the API does not expose a dated revision), with structured output under the strict schema, one published paragraph per call, low reasoning effort and a limit of 16{,}384 output tokens. The instructions confine the model to factual extraction: they prohibit it from inferring hidden dissent, scoring disagreement, or judging predictive relevance or implications for the growth distribution, so that no step of the measure is delegated to the model. The instruction set, the output schema, the validator and the runner are archived with checksums in the replication package, and the end-to-end path from paragraph to quarterly series was verified before any forecast was generated. The companion measure of explicit directional disagreement is built from the same fields, counting the paragraphs in which subsets are attributed opposite directions on the same question. The production instruction set (B29E) is reproduced in \ref{app:instr-ah}.

\paragraph{Salience, quantifier discount and verbosity.}
The remaining measures are built from a large-language-model classification of the Minutes by Hansen and co-authors---the Minutes-based counterpart of the transcript-based measurement of \citet{CHMX2023}---that records, for each meeting, the Committee's balance of risks for growth as stated and as weighted by the quantifiers the Minutes attach to it (``a few'', ``several'', ``many'', ``most'' participants), the number of paragraphs of the participants' discussion in which the outlook for growth is spoken to, the number of paragraphs of the discussion, and the length of the document. Salience is the ratio of the first paragraph count to the second. The quantifier discount is the absolute difference between the stated and the quantifier-weighted balance of risks; it is zero at meetings at which the Minutes attach no quantifier to the assessment, which is the case in 41 of the 70 test quarters, and undefined at two test origins, bridged by the latest quarter strictly before the forecast quarter. Verbosity is the natural logarithm of the word count. All are quarterly means over the meetings of the quarter, which reproduces the source's own quarterly aggregation exactly, and none was modified after its first evaluation. The instructions under which that classification was produced are the source's and are not reproduced here.

\subsubsection{Extraction instructions: balance of risks, paragraph level (source of the series)}\label{app:instr-bor-par}
\begin{lstlisting}
You are analyzing the MINUTES of an FOMC meeting, paragraph by paragraph.

MEETING: {meeting}

The document below has been split into numbered paragraphs.

<document_paragraphs>
{numbered_paragraphs}
</document_paragraphs>

TASK: For EVERY numbered paragraph, record whether that paragraph states a BALANCE OF
RISKS around the outlook, for growth and for inflation, and whose view it reports.
Then give one document-level assessment using the same rules as the document-level
coding, so the two can be compared.

CONTEXT: The Summary of Economic Projections asks FOMC members to "indicate your
judgment of the risk weighting around your projections." That is what we capture.

CRITICAL DISTINCTION - Forecast vs Balance of Risks:
- A FORECAST is the central expectation ("inflation will decline to 2%").
- BALANCE OF RISKS describes where probability mass lies RELATIVE TO that forecast.
- Code only explicit risk-balance language: risks "tilted", "weighted", "skewed" or
  "slanted"; asymmetric concern; scenarios worse/better than baseline; "risks are to
  the upside/downside". A forecast direction alone is NOT a balance of risks.

PER-PARAGRAPH FIELDS:
- "growth_risk" / "inflation_risk": UPSIDE / BALANCED / DOWNSIDE / NONE
  Use NONE when the paragraph states no balance of risks for that variable. Most
  paragraphs are NONE for both; that is expected and correct.
- "agent": whose view the paragraph reports -
  COMMITTEE (members / participants / the Committee collectively),
  STAFF (the staff outlook or the staff's own risk assessment),
  OTHER (market commentary, desk operations, procedural text).
- "quote": at most 15 words of the decisive risk language, verbatim; "NA" when both
  risk fields are NONE. Never reproduce a whole sentence when a fragment will do.

DOCUMENT-LEVEL FIELDS (the Committee's collective view only, staff excluded):
- "doc_growth_risk" / "doc_inflation_risk": UPSIDE / BALANCED / DOWNSIDE / NOT_DISCUSSED
- If the document reports a split among participants, follow the quantifier carrying
  the majority: all > most > many > several = a number > some > a few > a couple > one.
  Classify as BALANCED only if the document itself characterizes the risks as roughly
  balanced, NOT merely because opposing views coexist.
- Staff-only risk assessments never fill the document-level Committee fields.

Respond ONLY with valid JSON, no markdown fences:
{
  "paragraphs": [
    {"id": 1, "growth_risk": "NONE", "inflation_risk": "NONE", "agent": "OTHER", "quote": "NA"}
  ],
  "doc_growth_risk": "UPSIDE/BALANCED/DOWNSIDE/NOT_DISCUSSED",
  "doc_inflation_risk": "UPSIDE/BALANCED/DOWNSIDE/NOT_DISCUSSED",
  "confidence": "high/medium/low"
}

IMPORTANT:
- Return exactly one entry per numbered paragraph, in order, with matching ids.
- Quotes never exceed 15 words. A short decisive fragment is more auditable than a
  long one, and long verbatim reproduction is not permitted.
\end{lstlisting}

\subsubsection{Extraction instructions: balance of risks, document level (complementary)}\label{app:instr-bor-doc}
\begin{lstlisting}
You are analyzing an FOMC policy document (meeting MINUTES, the post-meeting STATEMENT, or the pre-1993 RECORD OF POLICY ACTIONS).

MEETING: {meeting}
DOCUMENT TYPE: {doc_type}

<document_text>
{document_text}
</document_text>

TASK: Extract the COMMITTEE's assessment of the BALANCE OF RISKS around its economic outlook, and separately record any STAFF risk assessment.

CONTEXT: The Summary of Economic Projections (SEP) asks FOMC members to "indicate your judgment of the
risk weighting around your projections." This is what we are trying to capture from the document.

WHOSE VIEW TO CODE:
- The primary fields (growth_risk, inflation_risk) code ONLY the view of the Committee / "members" /
  "participants" collectively. NEVER fill these fields with a staff-only assessment. If the Committee
  does not explicitly discuss the balance of risks for a variable, that field is NOT_DISCUSSED -- this
  is valid information, not a failure.
- Risk assessments attributed to the staff (e.g. "the staff regarded the risks to the GDP forecast as
  skewed to the downside") go ONLY in the separate staff_* fields. The staff assessment must meet the
  same bar of explicit risk-balance language; a staff forecast revision is NOT a risk assessment.
  Statements contain no staff views, so staff_* fields are NOT_DISCUSSED for statements.
- In MINUTES, prioritize the section "Participants' views on current conditions and the economic
  outlook" (or its historical equivalent) for the Committee fields; the staff outlook section feeds
  only the staff_* fields.
- If the document reports a split among participants, follow the quantifier that carries the majority
  using the FOMC's conventional ordering:
      all > most > many > several = a number > some > a few > a couple > one
  Classify as BALANCED only if the document itself characterizes the risks as roughly
  balanced/symmetric, NOT merely because opposing views coexist.

CRITICAL DISTINCTION - Forecast vs Balance of Risks:
- A FORECAST is the central expectation (e.g., "inflation will decline to 2%")
- BALANCE OF RISKS describes where probability mass lies RELATIVE TO that forecast
- The Committee can forecast FALLING inflation but still see UPSIDE RISKS (i.e., inflation might not fall as much as expected)
- The Committee can forecast RISING growth but still see DOWNSIDE RISKS (i.e., growth might disappoint relative to forecast)

DO NOT classify based on forecast direction. Only classify based on explicit statements about:
- Risks being "tilted", "weighted", "skewed" or "slanted" to one side
- Asymmetric uncertainty (more concern about one outcome than the other)
- Scenarios where outcomes could be worse/better than baseline
- Language like "risks are to the upside/downside"

DOCUMENT-TYPE-SPECIFIC RULES:

STATEMENT:
- Statements exist only from February 1994 onward, and only became systematic for every meeting from
  May 1999. 1994-1998 statements are terse and rarely discuss risks: NOT_DISCUSSED is the expected
  and correct output there.
- 1999-2003 dual-mandate boilerplate, map as follows:
  - "risks are weighted mainly toward conditions that may generate economic weakness" -> growth DOWNSIDE
  - "risks are weighted mainly toward conditions that may generate heightened inflation pressures" -> inflation UPSIDE
  - "the risks are balanced with respect to prospects for both goals" -> growth BALANCED and inflation BALANCED
  - If the sentence names only one of the two goals, the other side stays NOT_DISCUSSED unless
    discussed elsewhere in the text.
- Later statements use "the risks to the outlook ... are roughly balanced" or "the risks ... are
  weighted to the downside"; code them directly. If the phrase refers to "the economic outlook"
  without naming inflation, code growth only and leave inflation NOT_DISCUSSED unless inflation risks
  are discussed elsewhere in the text.
- Statements speak with a single institutional voice: set the intensity fields to "institutional".

MINUTES:
- Modern minutes begin in 1993. Apply the quantifier rule above for splits among participants.

RECORD_OF_POLICY_ACTIONS (pre-1993 equivalent of minutes):
- Language conventions differ from modern minutes; apply the same explicit-risk-language bar.
  Committee views may appear as "the members generally agreed that ..." formulations.

For GROWTH/OUTPUT risks, classify as:
- UPSIDE: The Committee explicitly states risks are weighted toward growth EXCEEDING the forecast
- BALANCED: The Committee states risks are roughly symmetric around the forecast
- DOWNSIDE: The Committee explicitly states risks are weighted toward growth FALLING SHORT of the forecast
- NOT_DISCUSSED: The Committee does not explicitly discuss the balance of risks to growth

For INFLATION risks, classify as:
- UPSIDE: The Committee explicitly states risks are weighted toward inflation EXCEEDING the forecast
- BALANCED: The Committee states risks are roughly symmetric around the forecast
- DOWNSIDE: The Committee explicitly states risks are weighted toward inflation FALLING SHORT of the forecast
- NOT_DISCUSSED: The Committee does not explicitly discuss the balance of risks to inflation

INTENSITY (Committee fields only; captures how widely the view was held):
- "unanimous": all members/participants, "members generally", uniform language with no dissenting quantifier
- "strong_majority": most / nearly all, with at most isolated dissent (a couple / one)
- "majority": many / a majority, with a visible minority holding another view (several / some)
- "divided": competing quantifiers of similar rank, or the document emphasizes disagreement
- "institutional": statements (single collective voice)
- "NA": when the corresponding risk field is NOT_DISCUSSED

NEGATIVE CLASSIFICATIONS (NOT_DISCUSSED) -- required reasoning format:
A NOT_DISCUSSED answer is a claim about the whole document and must be as auditable as a
positive one. Its reasoning field must be a NEGATIVE-SEARCH REPORT with two parts:
1. State that no explicit risk-balance language by the relevant agent was found.
2. QUOTE the nearest-miss passage -- the closest candidate you considered -- and say in one
   clause why it does not qualify (forecast direction / current-conditions assessment /
   staff-only view / refers to the other variable / uncertainty level without direction).
Example of a valid negative reasoning:
  "No explicit Committee risk-balance language on growth. Nearest passage: 'the expansion
   was likely to moderate in coming quarters' -- a forecast, not a risk assessment."
Only exception: if the document contains no passage even loosely related to the variable's
outlook (e.g. terse 1994-1998 statements), write exactly that ("document contains no
discussion of growth outlook or risks") -- this is the only case where no quote is required.
NEVER output a bare "NA", empty string, or generic boilerplate in any reasoning field.

QUOTE LENGTH (applies to every quotation in every reasoning field):
Quote at most 25 words. When the relevant passage is longer, quote only the decisive
fragment and mark the omission with "...". Never reproduce a full paragraph or a run of
consecutive sentences verbatim. A short decisive fragment is more auditable than a long one.

EXAMPLES OF WHAT IS NOT BALANCE OF RISKS:
- "Inflation should decline" -> This is a FORECAST, not a risk assessment. Classify as NOT_DISCUSSED.
- "Growth will slow" -> This is a FORECAST, not a risk assessment. Classify as NOT_DISCUSSED.
- "The economy is weak" -> This is a current assessment, not a risk assessment. Classify as NOT_DISCUSSED.
- "The staff revised down its projection for real GDP" -> Staff FORECAST revision, not a risk assessment.

EXAMPLES OF BALANCE OF RISKS:
- "Risks to inflation are tilted to the upside" -> inflation UPSIDE
- "Most participants saw more downside risk to growth than upside" -> growth DOWNSIDE, intensity strong_majority
- "The risks seem roughly balanced" -> BALANCED
- "Participants worried that inflation could be higher than projected" -> inflation UPSIDE
- "The staff continued to view the risks to the GDP forecast as skewed to the downside" -> staff_growth_risk DOWNSIDE (Committee field unaffected)

Respond ONLY with valid JSON, no markdown fences, no additional text:
{
    "growth_risk": "UPSIDE/BALANCED/DOWNSIDE/NOT_DISCUSSED",
    "growth_risk_intensity": "unanimous/strong_majority/majority/divided/institutional/NA",
    "growth_risk_quantifier": "verbatim quantifier phrase used, or NA",
    "growth_risk_reasoning": "if classified: DIRECT QUOTE of the risk language; if NOT_DISCUSSED: negative-search report with nearest-miss quote (see NEGATIVE CLASSIFICATIONS)",
    "inflation_risk": "UPSIDE/BALANCED/DOWNSIDE/NOT_DISCUSSED",
    "inflation_risk_intensity": "unanimous/strong_majority/majority/divided/institutional/NA",
    "inflation_risk_quantifier": "verbatim quantifier phrase used, or NA",
    "inflation_risk_reasoning": "if classified: DIRECT QUOTE of the risk language; if NOT_DISCUSSED: negative-search report with nearest-miss quote (see NEGATIVE CLASSIFICATIONS)",
    "staff_growth_risk": "UPSIDE/BALANCED/DOWNSIDE/NOT_DISCUSSED",
    "staff_growth_risk_reasoning": "if classified: DIRECT QUOTE; if NOT_DISCUSSED in minutes: one line naming what the staff section covers instead; for statements: 'statements contain no staff sections'",
    "staff_inflation_risk": "UPSIDE/BALANCED/DOWNSIDE/NOT_DISCUSSED",
    "staff_inflation_risk_reasoning": "if classified: DIRECT QUOTE; if NOT_DISCUSSED in minutes: one line naming what the staff section covers instead; for statements: 'statements contain no staff sections'",
    "confidence": "high/medium/low"
}

IMPORTANT:
- Only classify UPSIDE/BALANCED/DOWNSIDE if the relevant agent EXPLICITLY discusses risk balance
- If only a forecast direction is stated without risks around it, classify as NOT_DISCUSSED
- Committee fields and staff fields must NEVER be mixed: absence of a Committee view is NOT_DISCUSSED
  even when a staff view is present
- Reasoning fields are ALWAYS substantive: direct quote of risk language for classified
  answers; negative-search report with nearest-miss quote for NOT_DISCUSSED. A bare "NA"
  is a protocol violation in every reasoning field
- "confidence" refers ONLY to classification certainty given the text, never to which agent was coded
\end{lstlisting}

\subsubsection{Extraction instructions: insurance motive}\label{app:instr-ins}
\begin{lstlisting}
You are analyzing the MINUTES of an FOMC meeting, paragraph by paragraph, to locate
INSURANCE REASONING about monetary policy.

MEETING: {meeting}

The document below has been split into numbered paragraphs.

<document_paragraphs>
{numbered_paragraphs}
</document_paragraphs>

TASK: For EVERY numbered paragraph, record whether that paragraph reports insurance
reasoning, whose reasoning it is, and how widely it was held. Then summarise the
meeting.

## WHAT INSURANCE REASONING IS

Insurance reasoning argues for a policy action - the rate decision, the balance-sheet
action, the directive tilt, the statement language, or forward guidance - as a HEDGE
against a SPECIFIC low-probability but highly adverse scenario. The baseline outlook
alone would not justify the action; what motivates it is an outcome considered
unlikely but very costly if it materialised.

The test has TWO parts and BOTH must hold in the same paragraph (or in a paragraph
that explicitly continues the argument of the one before it):
1. a SPECIFIC adverse scenario is identified and treated as UNLIKELY - not the
   baseline, not the most likely outcome.
   WORD TEST: if the text calls the risk "material", "significant", "substantial",
   "considerable", "elevated", "likely" or "increased", it is describing the baseline
   outlook, NOT an unlikely tail, and part 1 FAILS. Language that passes part 1 marks
   the scenario as improbable or contingent: "if", "should X occur", "the possibility
   that", "a small probability", "were the situation to deteriorate further";
2. that scenario DIRECTLY MOTIVATES a concrete policy action or signal.

It runs in both directions: easing, balance-sheet action or dovish guidance as a
hedge against a low-probability downside scenario; tightening or hawkish guidance as
a hedge against a low-probability upside scenario.

Two variants are easy to miss and DO count when both parts of the test hold:
- ENTRENCHMENT: current inflation is not yet alarming, but the paragraph argues for
  acting now because inflation could become embedded in wages or expectations, making
  later removal far more costly.
- CRISIS DEEPENING: an adverse event is already under way, and the paragraph argues
  for action because the situation could turn QUALITATIVELY worse - contagion,
  cascading failures, a deflationary spiral - beyond the already-bad baseline.

## WHAT DOES NOT COUNT - apply these strictly

1. INSURANCE DISCUSSED WITHOUT MOTIVATING ACTION. The concept appears but is tied to
   no policy action or signal. If a specific tail risk is articulated and flagged as
   something that MIGHT require action later, code insurance = "Future_contingency",
   not "Yes".
2. LOSS-FUNCTION OR ASYMMETRIC-RISK LOGIC. Arguments that the costs of error are
   asymmetric, that policy is easier to reverse in one direction, that risks are
   "weighted to the downside", or that one should "err on the side of caution", are
   decision-theoretic claims about BASELINE uncertainty. Comparing plausible forecast
   scenarios is scenario analysis. Neither is insurance.
   NOTE: a balance-of-risks statement is NOT insurance reasoning. This is the single
   most common false positive - do not code a paragraph as insurance merely because
   it says risks are tilted one way.
3. RESPONDING TO WHAT HAS ALREADY HAPPENED. If the adverse event has occurred and the
   paragraph responds to current conditions, that is ordinary policy, not insurance.
   Crisis as the baseline is not a tail risk. Only the CRISIS DEEPENING variant above
   converts such a paragraph into insurance, and only when a specific further
   deterioration is named.
4. INSURANCE ABOUT POLICY IMPLEMENTATION. Repo and open-market operations, reserve
   management, year-end money-market pressures, discount-window mechanics and any
   other plumbing of policy implementation are OUT OF SCOPE, however literally the
   text uses the word insurance or hedge. This exercise is about the macroeconomic
   stance - the rate path, the balance sheet as a stance instrument, the tilt, the
   guidance - not about keeping the operating framework functioning. Code these "No".
5. GENERAL CAUTION OR VIGILANCE. "Remain vigilant", "watch for signs of", "wait and
   see", or listing risks without using them to justify an action.

## WHOSE REASONING

- "COMMITTEE": members or participants, individually or collectively ("a few
  participants argued...", "most members judged...").
- "STAFF": the staff outlook or the staff's own scenario analysis.
- "OTHER": market commentary, desk operations, procedural or factual text.
Staff reasoning is recorded but never enters the Committee-level summary.

## HOW WIDELY HELD

When the paragraph attributes the reasoning to Committee participants, copy the
verbatim quantifier ("a few", "several", "most", "all", "one participant") into
`quantifier`. The FOMC ordering is
    all > most > many > several = a number > some > a few > a couple > one.
When the paragraph attributes the reasoning to the Committee but uses NO quantifier,
do not write "NA": copy the grammatical SUBJECT of the sentence instead - "the
members", "participants", "the Committee", or the named individuals ("Mr. Fisher").
"NA" is only for agent = STAFF or OTHER.
This field is what lets a single set of minutes show how contested the reasoning was,
so record it exactly as written rather than paraphrasing.

## PER-PARAGRAPH FIELDS

- "insurance": "Yes" / "Future_contingency" / "No"
- "direction": "Downside_growth" / "Upside_inflation" / "Other" / "NA"
- "variant": "General" / "Entrenchment" / "Crisis_deepening" / "NA"
- "target": what the reasoning motivates, comma-separated when several apply -
  "Current_action" / "Formal_communication" / "Future_only" / "Unclear" / "NA"
  Current_action is the rate decision or balance-sheet action taken at this meeting.
  Formal_communication is the directive tilt or bias, the wording of the post-meeting
  statement, and forward guidance about the likely path of policy. Minutes discuss
  these in their own passages - the directive adopted, the language agreed, the risk
  assessment to be published - which are usually LATER in the document than the
  economic discussion. Read those passages too: a paragraph in which insurance
  concerns shape the agreed language or the tilt is Formal_communication even if the
  rate itself is unchanged. Use both labels when the same reasoning drives both.
- "agent": "COMMITTEE" / "STAFF" / "OTHER"
- "quantifier": verbatim quantifier phrase, or "NA"
- "scenario": the specific adverse scenario in at most 12 words, or "NA"
- "quote": at most 15 words, verbatim, of the decisive passage; "NA" when
  insurance = "No". Never reproduce a full sentence when a fragment will do.

Keep every field "NA" and `insurance` = "No" for paragraphs that carry no insurance
reasoning. Most paragraphs are "No"; that is expected and correct, and those entries
need no justification.

## MEETING-LEVEL FIELDS

- "doc_insurance": "Yes" / "Future_contingency" / "No" - whether the minutes, taken as
  a whole, report Committee insurance reasoning motivating a policy action or signal.
  Staff-only reasoning does not qualify.
- "doc_direction", "doc_variant", "doc_target": as above, for the Committee reasoning
  that carries the strongest quantifier.
- "doc_justification": two sentences. If "No", name the closest candidate passage and
  say which exclusion rule above disqualifies it.
- "confidence": about the classification only, anchored as follows -
  "high": the text uses insurance/hedge wording explicitly, or names an unlikely
          scenario and the action it motivates in the same passage;
  "medium": the two parts of the test are met but inferred from wording that never
          says insurance, or split across adjacent paragraphs;
  "low": the passage also admits a loss-function reading, or it is unclear whether
          the scenario is treated as unlikely rather than as the baseline.
  A meeting whose evidence is a single borderline passage is not "high".

## BEFORE YOU ANSWER, re-apply the test to every paragraph you were about to mark

For each candidate ask, in order:
  (a) Is the adverse scenario marked as UNLIKELY, or does the text call it material,
      significant, substantial or increased? If the latter, code "No".
  (b) Does that scenario motivate a concrete macroeconomic policy action or signal in
      the same passage? If it motivates only vigilance, monitoring, or a possible
      future move, code "Future_contingency", not "Yes".
  (c) Is the passage about the operating framework - repos, reserves, money-market
      pressure - rather than the stance? If so, code "No".
  (d) Would a careful reader say the passage is really a balance-of-risks statement or
      an asymmetric-loss argument? If so, code "No".
Only paragraphs surviving all four are "Yes".

Respond ONLY with valid JSON, no markdown fences:
{
  "paragraphs": [
    {"id": 1, "insurance": "No", "direction": "NA", "variant": "NA", "target": "NA",
     "agent": "OTHER", "quantifier": "NA", "scenario": "NA", "quote": "NA"}
  ],
  "doc_insurance": "Yes/Future_contingency/No",
  "doc_direction": "Downside_growth/Upside_inflation/Other/NA",
  "doc_variant": "General/Entrenchment/Crisis_deepening/NA",
  "doc_target": "Current_action/Formal_communication/Future_only/Unclear/NA",
  "doc_justification": "...",
  "confidence": "high/medium/low"
}

IMPORTANT:
- Return exactly one entry per numbered paragraph, in order, with matching ids.
- Both parts of the two-part test must hold before coding "Yes".
- A tilted balance of risks is not insurance. A specific unlikely scenario driving a
  concrete action is.
\end{lstlisting}

\subsubsection{Extraction instructions: attributed views (instruction set B29E, production)}\label{app:instr-ah}
\begin{lstlisting}
You are performing structured information extraction from one published
paragraph of Federal Open Market Committee Minutes.

Your task is descriptive.

Extract ONLY economic or monetary-policy views that are explicitly
attributed in the supplied paragraph.

Do not summarize the paragraph.
Do not evaluate the importance of a view.
Do not infer disagreement, dissent, minority status, consensus,
hawkishness, dovishness, predictive relevance, or forecasting value.

MEETING DATE: {meeting_date}
PARAGRAPH ID: {paragraph_id}

<paragraph>
{paragraph_text}
</paragraph>

======================================================================
1. EXTRACTION UNIT
======================================================================

The extraction unit is an ATTRIBUTED VIEW.

A paragraph may contain:

- zero attributed views;
- one attributed view;
- multiple attributed views.

Return one separate view object for each separately attributed
economic or monetary-policy position.

Do NOT collapse multiple attributed positions into one summary.

Example:

"Most participants expected growth to remain solid. Several
participants, however, noted that downside risks to growth had
increased."

This contains TWO attributed views, not one.

======================================================================
2. EXPLICIT ATTRIBUTION IS REQUIRED
======================================================================

A view is eligible only if the paragraph explicitly attributes it to
participants, members, the Committee, or another explicitly named
relevant group.

Examples of attribution language include:

- one participant
- a few participants
- some participants
- several participants
- a number of participants
- many participants
- most participants
- participants generally
- members
- the Committee

Preserve the exact attribution wording in:

    attribution_phrase

Do not infer an individual speaker.
Do not infer an attribution that is not present in the paragraph.

The attribution phrase MUST appear verbatim inside evidence_quote.

======================================================================
3. EVIDENCE QUOTE
======================================================================

Every extracted view MUST contain an evidence_quote.

The evidence_quote must:

- be copied verbatim from the supplied paragraph;
- contain the attribution_phrase;
- contain enough text to support the extracted view;
- be as short as possible while remaining sufficient evidence.

Never paraphrase the evidence_quote.

CONTIGUOUS EVIDENCE RULE

The evidence_quote MUST be one contiguous verbatim span from the source
paragraph.

Never delete words from the middle of a source span in order to create
a shorter topic-specific quotation.

Never concatenate non-contiguous source fragments into one
evidence_quote.

When one coordinated source sentence supports two or more distinct
attributed views, those views MAY and SHOULD share the same complete
contiguous verbatim evidence_quote when necessary to preserve literal
source wording.

Separating views by topic does NOT require separate evidence quotations.
Verbatim grounding takes priority over shortening the quotation.

Never reconstruct wording.
Never quote text that is not in the paragraph.

======================================================================
4. ATTRIBUTION NORMALIZATION
======================================================================

attribution_class must be exactly one of:

PARTICIPANTS
MEMBERS
COMMITTEE
OTHER_ATTRIBUTED_GROUP

quantifier must be exactly one of:

ONE
FEW
SOME
SEVERAL
NUMBER
MANY
MOST
GENERAL
ALL
UNSPECIFIED

Normalize only what is linguistically explicit.

Examples:

"one participant" -> ONE
"a few participants" -> FEW
"some participants" -> SOME
"several participants" -> SEVERAL
"a number of participants" -> NUMBER
"many participants" -> MANY
"most participants" -> MOST
"participants generally" -> GENERAL

Do NOT translate these categories into numerical shares.

Do NOT decide whether a quantifier represents a minority or majority.

======================================================================
5. TOPIC
======================================================================

Assign one primary topic:

GROWTH
INFLATION
LABOR_MARKET
FINANCIAL_CONDITIONS
FINANCIAL_STABILITY
HOUSEHOLDS_CONSUMPTION
BUSINESS_INVESTMENT
HOUSING
EXTERNAL_GLOBAL
FISCAL
MONETARY_POLICY
EXPECTATIONS
OTHER

Use topic_detail only when a short textual clarification is useful.
Otherwise use null.

If a single attributed statement clearly contains distinct views about
different topics, return separate view objects when the views can be
supported by distinct evidence.

======================================================================
6. VIEW TYPE
======================================================================

view_type must be exactly one of:

CURRENT_ASSESSMENT
BASELINE_FORECAST
RISK_ASSESSMENT
ALTERNATIVE_SCENARIO
UNCERTAINTY_ASSESSMENT
POLICY_PREFERENCE
POLICY_CONDITIONALITY
OTHER

Use the most specific category supported by the text.

----------------------------------------------------------------------
CRITICAL DISTINCTION: BASELINE FORECAST VS RISK ASSESSMENT
----------------------------------------------------------------------

A BASELINE_FORECAST is a central expectation.

Examples:

"Several participants expected inflation to decline."
"Most participants anticipated slower growth."

These are forecasts, not risk assessments.

A RISK_ASSESSMENT requires explicit language about:

- upside or downside risks;
- risks being tilted or weighted;
- asymmetric uncertainty;
- outcomes above or below a baseline;
- greater concern about one side of the distribution;
- another explicit statement about the balance or direction of risk.

Do NOT infer risk direction from forecast direction.

For example:

"Inflation is expected to decline"

does NOT imply:

DOWNSIDE inflation risk.

And:

"Growth is expected to slow"

does NOT imply:

DOWNSIDE growth risk.

----------------------------------------------------------------------
ALTERNATIVE SCENARIO
----------------------------------------------------------------------

Use ALTERNATIVE_SCENARIO only when the text explicitly describes a
non-baseline possibility or scenario.

Do not invent an alternative scenario from generic uncertainty.

----------------------------------------------------------------------
UNCERTAINTY ASSESSMENT
----------------------------------------------------------------------

Use UNCERTAINTY_ASSESSMENT when the attributed view concerns the degree,
source, or nature of uncertainty without expressing a directional
balance of risks.

======================================================================
7. STANCE
======================================================================

stance must be exactly one of:

HIGHER
LOWER
STRONGER
WEAKER
TIGHTER
EASIER
MORE_PERSISTENT
LESS_PERSISTENT
UNCHANGED
BALANCED
OTHER
NOT_APPLICABLE

Use stance only when a directional assessment is explicit.

Do not use stance as a substitute for risk_direction.

If none applies, use NOT_APPLICABLE.

======================================================================
8. RISK DIRECTION
======================================================================

risk_direction must be exactly one of:

UPSIDE
DOWNSIDE
BALANCED
TWO_SIDED
UNSPECIFIED
NOT_APPLICABLE

For a RISK_ASSESSMENT:

- UPSIDE means explicitly greater risk of an outcome above the relevant
  baseline or central expectation.
- DOWNSIDE means explicitly greater risk of an outcome below the
  relevant baseline or central expectation.
- BALANCED means explicitly approximately symmetric risks.
- TWO_SIDED means meaningful risks on both sides are explicitly
  discussed without describing them as balanced.
- UNSPECIFIED means risk is explicitly discussed but its direction is
  not stated.

For non-risk views, normally use NOT_APPLICABLE.

======================================================================
9. BASELINE REFERENCE
======================================================================

baseline_reference records an explicit textual baseline against which
the view is expressed.

Examples may include:

- the staff forecast
- participants' central outlook
- the baseline projection
- the expected inflation path

If no baseline is explicit, use null.

Do not invent a baseline.

======================================================================
10. ALTERNATIVE SCENARIO
======================================================================

alternative_scenario records a specific non-baseline scenario that is
explicitly described in the evidence.

If no explicit alternative scenario is present, use null.

Do not convert generic concern or uncertainty into a scenario.

======================================================================
11. CONDITIONALITY
======================================================================

conditionality records an explicit condition, threshold, contingency,
or state-dependent relationship.

Examples include explicit constructions such as:

- if X occurs...
- should inflation remain elevated...
- unless financial conditions improve...
- if labor-market weakness intensifies...

If no such condition is explicit, use null.

Do not invent implicit conditions.

======================================================================
12. POLICY IMPLICATION
======================================================================

policy_implication records a policy consequence or preference only when
the text states it explicitly.

Do not infer policy preferences from economic assessments.

For example:

"Several participants saw upside inflation risks"

does NOT by itself imply a preference for tighter policy.

If no explicit policy implication is stated, use null.

======================================================================
13. POLICY DIRECTION
======================================================================

policy_direction must be exactly one of:

TIGHTER
EASIER
HOLD
COMMUNICATION_HAWKISH
COMMUNICATION_DOVISH
OTHER
UNSPECIFIED
NOT_APPLICABLE

Use a directional category only when explicit policy content supports
it.

If policy is discussed but direction cannot be determined, use
UNSPECIFIED.

If the view contains no policy implication, normally use
NOT_APPLICABLE.

======================================================================
14. RELATION TO OTHER VIEWS
======================================================================

relation_to_other_views must be exactly one of:

AGREES
QUALIFIES
DIFFERS
OPPOSES
ALTERNATIVE
NO_EXPLICIT_RELATION

This field describes only an EXPLICIT textual relationship.

Potential relational language includes:

- while
- whereas
- however
- in contrast
- by contrast
- others
- other participants

But the mere presence of one of these expressions is not sufficient.
The relationship between positions must be explicit.

Do NOT infer:

- minority status;
- majority status;
- dissent;
- hidden dissent;
- degree of disagreement.

If no explicit relation is expressed, use:

NO_EXPLICIT_RELATION

======================================================================
15. CONFIDENCE
======================================================================

confidence refers ONLY to confidence in the extraction/classification:

HIGH
MEDIUM
LOW

It does not measure:

- importance;
- intensity;
- economic significance;
- probability;
- strength of disagreement;
- predictive value.

======================================================================
16. ZERO-VIEW RULE
======================================================================

Return:

"views": []

when the paragraph contains no eligible explicitly attributed economic
or monetary-policy view.

Examples may include:

- procedural material;
- purely factual narration;
- staff analysis not attributed to participants/members/Committee;
- economic statements without explicit attribution;
- text that does not express a view.

Do not manufacture a view merely because the paragraph discusses the
economy.

======================================================================
17. MULTIPLE-VIEW RULE
======================================================================

When separately attributed groups express distinct positions, preserve
them as separate view objects.

Synthetic example:

"Most participants expected growth to remain solid. Several
participants, however, noted that downside risks to growth had
increased."

Correct conceptual decomposition:

VIEW 1
- attribution_phrase: "Most participants"
- quantifier: MOST
- topic: GROWTH
- view_type: BASELINE_FORECAST
- risk_direction: NOT_APPLICABLE
- relation_to_other_views: NO_EXPLICIT_RELATION

VIEW 2
- attribution_phrase: "Several participants"
- quantifier: SEVERAL
- topic: GROWTH
- view_type: RISK_ASSESSMENT
- risk_direction: DOWNSIDE
- relation_to_other_views: DIFFERS

Do not describe VIEW 2 as a minority view.
Do not compute disagreement.

======================================================================
18. PROHIBITED INFERENCES
======================================================================

You must NOT output or infer:

- minority status;
- majority status beyond the literal quantifier normalization;
- dissent;
- hidden dissent;
- disagreement scores;
- consensus scores;
- hawkishness scores;
- dovishness scores;
- importance scores;
- signal strength;
- predictive relevance;
- forecasting implications;
- Growth-at-Risk implications.

Your task ends at factual structured extraction.

======================================================================
19. OUTPUT CONTRACT
======================================================================

Return ONLY one JSON object.

Do not use Markdown.
Do not use code fences.
Do not add commentary before or after the JSON.

The object must have exactly:

{
  "meeting_date": "{meeting_date}",
  "paragraph_id": {paragraph_id},
  "views": [...]
}

Each element of "views" must contain exactly these fields:

view_id
evidence_quote
attribution_phrase
attribution_class
quantifier
topic
topic_detail
view_type
stance
risk_direction
baseline_reference
alternative_scenario
conditionality
policy_implication
policy_direction
relation_to_other_views
confidence

view_id must start at 1 and increase by 1 in textual order.

Use null only where the schema permits null.

Do not add any fields.
\end{lstlisting}

\paragraph{Correlations on the development sample.}
On the development sample 1994Q2--2007Q3 the quantifier discount correlates with \HDlong{} at 0.49, salience at $-0.46$, the balance of risks at 0.43 and verbosity at 0.39. No signal was selected on the basis of these correlations: each entered the family for the dimension of the deliberation it measures, and earned or lost its place through its own out-of-sample contribution.
\section{Properties of the Committee's measures}\label{app:properties}

This appendix describes the nine measures of the Committee used in the paper---the two taken from its published projections and the seven read from the Minutes---in the order of the meeting (Table~\ref*{tab:signals} of the paper): their time-series properties, how they relate to one another, and how they relate to growth and to the public state. Everything here is descriptive. No forecast is made and no comparison of forecast losses is drawn; the in-sample regressions of Table~\ref{tab:prop-growth-reg} describe sign and magnitude and are not a substitute for the out-of-sample results of the paper. Series are shown in their original units (a probability for Hidden Dissent, shares of paragraphs for salience and attributed heterogeneity, a three-valued category for the balance of risks, the logarithm of the word count for verbosity, percentage points for the projections); the standardization under which the Minutes-based measures enter the models (Section~\ref*{sec:protocols} of the paper) does not affect any correlation reported here. Standard errors are HAC (Bartlett kernel, bandwidth 3); intervals are 95 percent moving-block bootstrap intervals with blocks of four quarters, 10,000 replications. All series are on the paper's clock: the value at origin $t$ is what was public at the start of quarter $t$.

\subsection{The series}

\begin{figure}[!t]
\centering
\includegraphics[width=\textwidth]{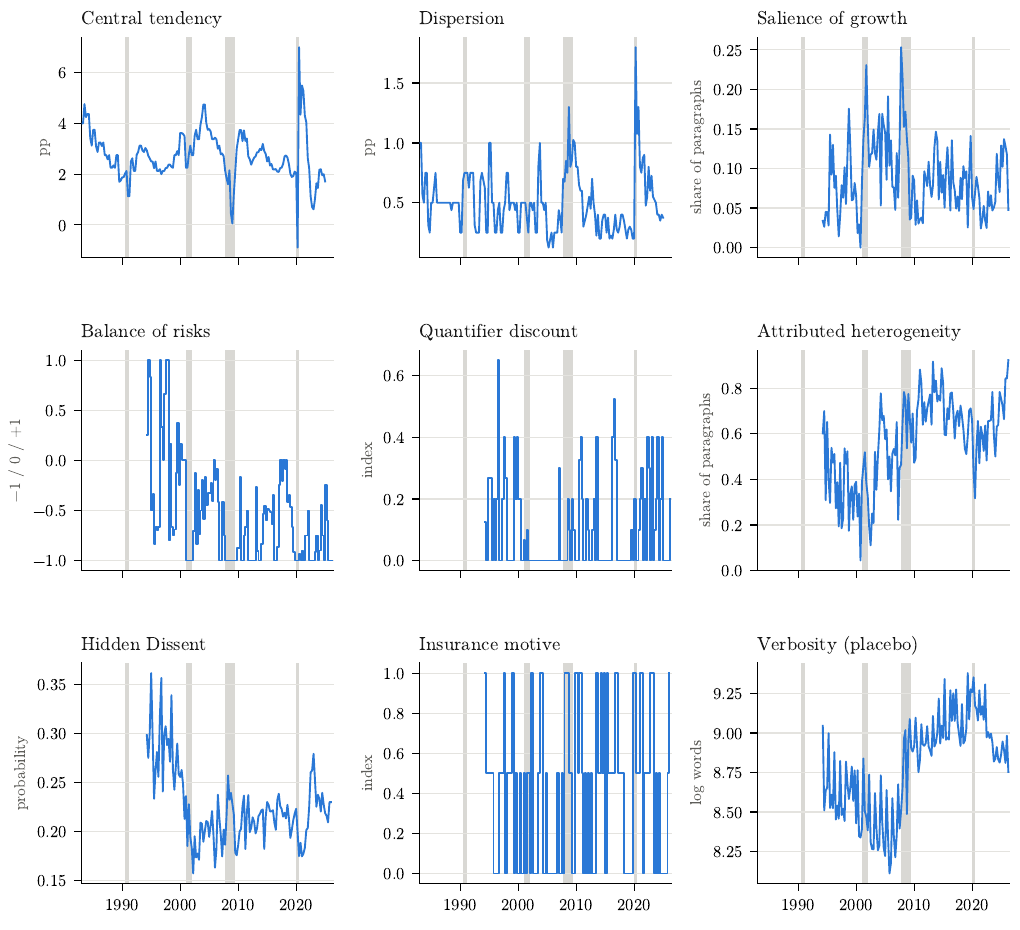}
\caption{The Committee's measures. Quarterly series in original units, on the paper's clock (the value at origin $t$ comes from the meetings of quarter $t-1$ for the Minutes-based measures, and from the latest published round for the projections). Shaded areas are NBER recessions. The central tendency and the dispersion are the midpoint and the width of the central tendency of the growth projections, from the Monetary Policy Reports until 2007 and the SEP since.}\label{fig:prop-series}
\end{figure}

\begin{table}[!t]
\centering\small
\caption{Descriptive statistics of the Committee's measures}\label{tab:prop-stats}
\resizebox{\textwidth}{!}{%
\begin{tabular}{lcccccccccc}
\toprule
 & & & & & & \multicolumn{3}{c}{Autocorrelation} & & \\
\cmidrule(lr){7-9}
Measure & From & $N$ & Mean & S.D. & Min / Max & 1 & 2 & 4 & Share neutral & Unit \\
\midrule
\multicolumn{11}{l}{\emph{Inputs: the published projections}} \\
Central tendency & 1983Q1 & 169 & 2.74 & 0.98 & $-$0.90 / 7.01 & .66 & .59 & .27 & -- & pp \\
Dispersion & 1983Q1 & 169 & 0.51 & 0.25 & 0.13 / 1.80 & .59 & .36 & .22 & -- & pp \\
\multicolumn{11}{l}{\emph{Discussion: what it is about and what the Committee says it sees}} \\
Salience of growth & 1994Q2 & 129 & 0.089 & 0.047 & 0 / 0.25 & .53 & .36 & .19 & -- & share \\
Balance of risks & 1994Q2 & 129 & $-$0.55 & 0.49 & $-$1 / 1 & .63 & .48 & .27 & .32 & $-1/0/{+}1$ \\
\multicolumn{11}{l}{\emph{Discussion: how divided the room is}} \\
Quantifier discount & 1994Q2 & 121 & 0.095 & 0.147 & 0 / 0.65 & .21 & .04 & .04 & .63 & index \\
Attributed heterogeneity & 1994Q2 & 129 & 0.565 & 0.189 & 0.04 / 0.92 & .67 & .63 & .48 & -- & share \\
Hidden Dissent & 1994Q2 & 128 & 0.226 & 0.039 & 0.16 / 0.36 & .74 & .63 & .60 & -- & probability \\
\multicolumn{11}{l}{\emph{Decision and record}} \\
Insurance motive & 1994Q2 & 128 & 0.41 & 0.38 & 0 / 1 & .28 & .21 & $-$.02 & .40 & index \\
Verbosity (placebo) & 1994Q2 & 129 & 8.80 & 0.31 & 8.11 / 9.38 & .73 & .72 & .89 & -- & log words \\
\bottomrule
\end{tabular}}
\begin{minipage}{0.95\textwidth}\footnotesize\vspace{4pt}
Notes: Quarterly series in original units over each measure's full sample (through 2025Q1 for the projections and 2026Q1--Q2 for the Minutes-based measures); the Supplementary Material data files give the same statistics over the test origins 2007Q4--2025Q1. Share neutral: fraction of quarters at which the measure takes its neutral value (balanced, for the balance of risks; zero, for the quantifier discount and the insurance motive). Verbosity is the logarithm of the number of words in the Minutes (the mean word count is 6,925). The projections combine 50 publication rounds of the Monetary Policy Report (1983--2007) and 68 rounds of the SEP (2007--2025); Supplementary Material~\ref*{app:data} describes the splice.
\end{minipage}
\end{table}

Three features of Table~\ref{tab:prop-stats} and Figure~\ref{fig:prop-series} matter for reading the paper. First, the measures differ widely in persistence. Hidden Dissent, attributed heterogeneity and verbosity are slow-moving (first-order autocorrelations of 0.67--0.74, and still 0.5--0.9 at four quarters), the published projections and the balance of risks are moderately persistent (0.59--0.66), while the quantifier discount and the insurance motive are close to white noise (0.21 and 0.28) because they are near-discrete: the quantifier discount is zero in 63 percent of quarters and the insurance motive in 40 percent. A measure that is zero most of the time can only carry information in the quarters when it is not, which is one reason the quantifier discount does not replicate Hidden Dissent in the paper even though both are read as disagreement. Second, Hidden Dissent is a probability that moves within a narrow band (0.16--0.36 over the full sample, 0.17--0.28 over the test origins); its level fell from the mid-1990s to the early 2000s and has fluctuated around 0.22 since (Figure~\ref{fig:prop-ty} shows the longer history). Third, the published projections have two regimes. The Monetary Policy Report years (1983--2007) show a central tendency with a mean of 2.9 percent and a standard deviation of 0.75, revised twice a year; the SEP years are more volatile (standard deviation 1.19, dominated by 2020--2021, when the central tendency ranged from $-$0.9 to 7.0 percent) and more frequently revised. Section~\ref{sec:prop-diag} returns to this.

\subsection{How the measures relate to one another}

\begin{figure}[!t]
\centering
\includegraphics[width=0.8\textwidth]{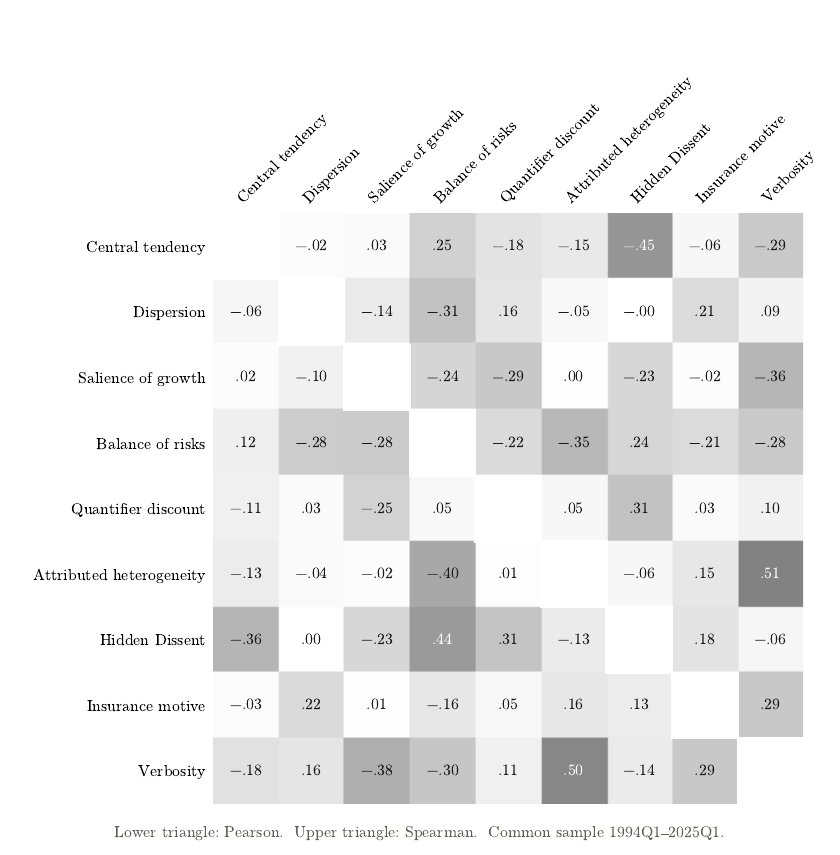}
\caption{Correlations among the Committee's measures. Levels, common sample 1994--2025. Lower triangle: Pearson; upper triangle: Spearman. Darker cells are larger absolute correlations. Table~\ref{tab:prop-corr} gives the HAC standard errors and the same matrix in first differences.}\label{fig:prop-corr}
\end{figure}

\begin{table}[!t]
\centering\footnotesize
\caption{Correlations among the Committee's measures: levels and first differences}\label{tab:prop-corr}
\setlength{\tabcolsep}{4pt}
\begin{tabular}{lcccccccc}
\toprule
 & CT & DISP & SAL & BoR & QD & AH & HD & INS \\
\midrule
\multicolumn{9}{l}{\emph{Panel A. Levels}} \\
Dispersion (DISP) & $-$.06 &  &  &  &  &  &  &  \\
 & (.17) &  &  &  &  &  &  &  \\
Salience of growth (SAL) & .02 & $-$.10 &  &  &  &  &  &  \\
 & (.11) & (.11) &  &  &  &  &  &  \\
Balance of risks (BoR) & .12 & $-$.28 & $-$.28 &  &  &  &  &  \\
 & (.12) & (.10) & (.10) &  &  &  &  &  \\
Quantifier discount (QD) & $-$.11 & .03 & $-$.25 & .05 &  &  &  &  \\
 & (.08) & (.08) & (.09) & (.16) &  &  &  &  \\
Attributed heterogeneity (AH) & $-$.13 & $-$.04 & $-$.02 & $-$.40 & .01 &  &  &  \\
 & (.09) & (.09) & (.12) & (.12) & (.12) &  &  &  \\
Hidden Dissent (HD) & $-$.36 & .00 & $-$.23 & .44 & .31 & $-$.13 &  &  \\
 & (.12) & (.12) & (.09) & (.13) & (.09) & (.13) &  &  \\
Insurance motive (INS) & $-$.03 & .22 & .01 & $-$.16 & .05 & .16 & .13 &  \\
 & (.12) & (.08) & (.09) & (.08) & (.11) & (.09) & (.09) &  \\
Verbosity & $-$.18 & .16 & $-$.38 & $-$.30 & .11 & .50 & $-$.14 & .29 \\
 & (.14) & (.15) & (.10) & (.10) & (.10) & (.11) & (.13) & (.09) \\
\addlinespace
\multicolumn{9}{l}{\emph{Panel B. First differences}} \\
Dispersion (DISP) & $-$.46 &  &  &  &  &  &  &  \\
 & (.22) &  &  &  &  &  &  &  \\
Salience of growth (SAL) & $-$.03 & $-$.10 &  &  &  &  &  &  \\
 & (.05) & (.07) &  &  &  &  &  &  \\
Balance of risks (BoR) & .08 & $-$.09 & $-$.24 &  &  &  &  &  \\
 & (.04) & (.09) & (.10) &  &  &  &  &  \\
Quantifier discount (QD) & .06 & .00 & $-$.04 & $-$.03 &  &  &  &  \\
 & (.04) & (.09) & (.09) & (.19) &  &  &  &  \\
Attributed heterogeneity (AH) & $-$.05 & $-$.02 & .01 & $-$.35 & $-$.02 &  &  &  \\
 & (.06) & (.11) & (.10) & (.10) & (.11) &  &  &  \\
Hidden Dissent (HD) & $-$.12 & .16 & $-$.03 & $-$.01 & .14 & .24 &  &  \\
 & (.07) & (.10) & (.08) & (.13) & (.07) & (.08) &  &  \\
Insurance motive (INS) & .08 & $-$.14 & .10 & $-$.16 & $-$.05 & $-$.04 & .13 &  \\
 & (.05) & (.08) & (.10) & (.06) & (.11) & (.07) & (.08) &  \\
Verbosity & $-$.05 & .01 & $-$.42 & $-$.03 & .01 & .16 & $-$.09 & .06 \\
 & (.11) & (.11) & (.08) & (.10) & (.09) & (.10) & (.08) & (.11) \\
\bottomrule
\end{tabular}
\begin{minipage}{0.95\textwidth}\footnotesize\vspace{4pt}
Notes: Pearson correlations, HAC standard errors in parentheses. Common sample 1994--2025; $N$ between 118 and 125 in levels and between 112 and 124 in differences, depending on the pair. CT is the central tendency. Spearman correlations and moving-block intervals are in the data files; the Spearman matrix is in the upper triangle of Figure~\ref{fig:prop-corr}.
\end{minipage}
\end{table}

The measures are, for the most part, weakly correlated: of the 36 pairs in levels, five have an absolute correlation above 0.35 and twenty-two are below 0.15. Three readings follow. The three measures of division in the room do not measure the same thing. Hidden Dissent correlates at 0.31 with the quantifier discount and at $-$0.13 with attributed heterogeneity, and the latter two are uncorrelated with each other (0.01); in first differences the picture is the same (0.14, 0.24, $-$0.02). The paper's finding that Hidden Dissent carries information that the other two do not is therefore not a matter of three noisy readings of one latent variable, of which one happens to be the cleanest; the three pick up different aspects of how the discussion is recorded. The declared and the hidden assessments are related, and with the expected signs. Hidden Dissent is higher when the Committee's central tendency is lower ($-$0.36, Spearman $-$0.45) and when the balance of risks is tilted to the downside (0.44), and it is unrelated to the published dispersion (0.00): the dispersion of the participants' point projections and the disagreement a reader infers from the Minutes are, in levels, orthogonal. Finally, verbosity is not neutral. It correlates at 0.50 with attributed heterogeneity and at $-$0.38 with salience: longer Minutes attribute more views to more participants and devote a smaller share of their paragraphs to growth. This is why verbosity is kept as a placebo throughout the paper rather than discarded as uninformative; it is correlated with the record, not with the economy (Table~\ref{tab:prop-growth}). The two published measures, nearly uncorrelated in levels ($-$0.06), move in opposite directions quarter to quarter ($-$0.46): when the Committee marks its growth projection down, the range of its participants' projections widens.

\subsection{How the measures relate to growth}

\begin{figure}[!t]
\centering
\includegraphics[width=\textwidth]{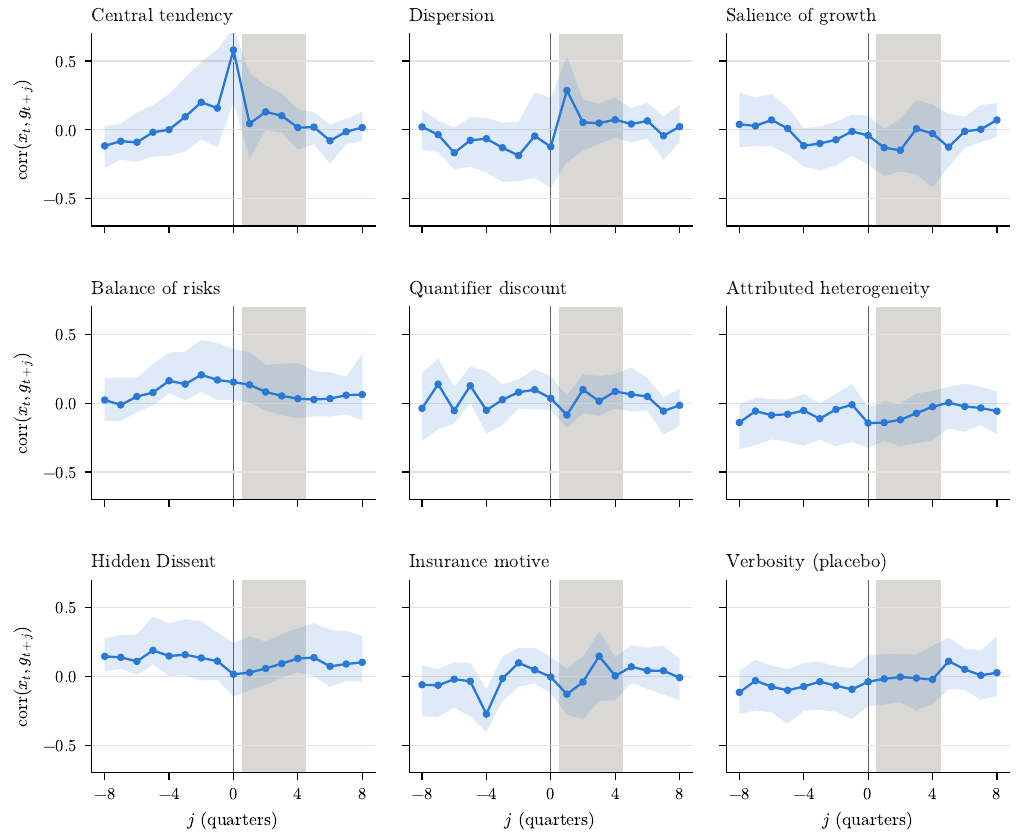}
\caption{Cross-correlations with growth. Correlation of each measure at origin $t$ with annualized real GDP growth in quarter $t+j$, for leads and lags of up to eight quarters; bands are 95 percent moving-block bootstrap intervals ($L=4$). Positive $j$ is future growth; the shaded band marks $j=1,\dots,4$, the four quarters the paper's target averages. Projections from 1983Q1; Minutes-based measures from 1994Q2.}\label{fig:prop-xcorr}
\end{figure}

\begin{table}[!t]
\centering\small
\caption{Correlations of the Committee's measures with growth}\label{tab:prop-growth}
\resizebox{\textwidth}{!}{%
\begin{tabular}{lcccccccc}
\toprule
 & \multicolumn{2}{c}{Past} & Current & \multicolumn{3}{c}{Future} & & SPF error \\
\cmidrule(lr){2-3}\cmidrule(lr){4-4}\cmidrule(lr){5-7}\cmidrule(lr){9-9}
Measure & $g_{t-4}$ & $g_{t-1}$ & $g_t$ & $g_{t+1}$ & $g_{t+4}$ & $y_{t+4}$ & & $y_{t+4}-\mathrm{SPF}_t$ \\
\midrule
Central tendency & $-$.04 & .15 & .58 & .05 & .02 & .16 & & $-$.12 \\
 & (.10) & (.16) & (.23) & (.15) & (.06) & (.15) & & (.12) \\
Dispersion & $-$.09 & $-$.05 & $-$.12 & .29 & .07 & .25 & & .18 \\
 & (.08) & (.13) & (.15) & (.25) & (.05) & (.17) & & (.12) \\
Salience of growth & $-$.12 & $-$.01 & $-$.04 & $-$.13 & $-$.03 & $-$.17 & & $-$.20 \\
 & (.07) & (.06) & (.07) & (.07) & (.13) & (.18) & & (.18) \\
Balance of risks & .16 & .17 & .15 & .13 & .03 & .17 & & .16 \\
 & (.05) & (.06) & (.06) & (.06) & (.07) & (.12) & & (.12) \\
Quantifier discount & $-$.05 & .10 & .04 & $-$.09 & .09 & .06 & & .16 \\
 & (.12) & (.06) & (.04) & (.08) & (.06) & (.07) & & (.07) \\
Attributed heterogeneity & $-$.05 & $-$.01 & $-$.14 & $-$.14 & $-$.03 & $-$.21 & & $-$.14 \\
 & (.06) & (.09) & (.07) & (.09) & (.06) & (.11) & & (.13) \\
Hidden Dissent & .15 & .11 & .02 & .03 & .13 & .18 & & .38 \\
 & (.07) & (.07) & (.08) & (.07) & (.05) & (.11) & & (.10) \\
Insurance motive & $-$.27 & .05 & $-$.00 & $-$.13 & .00 & $-$.01 & & .01 \\
 & (.15) & (.04) & (.05) & (.10) & (.06) & (.12) & & (.12) \\
Verbosity (placebo) & $-$.08 & $-$.10 & $-$.06 & $-$.00 & $-$.04 & $-$.02 & & $-$.00 \\
 & (.09) & (.08) & (.10) & (.10) & (.12) & (.17) & & (.15) \\
\bottomrule
\end{tabular}}
\begin{minipage}{0.95\textwidth}\footnotesize\vspace{4pt}
Notes: Pearson correlations in levels, HAC standard errors (Bartlett, bandwidth 3) in parentheses; $g$ is annualized quarterly real GDP growth, final vintage; $y_{t+4}$ is the paper's target, the average of $g_{t+1},\dots,g_{t+4}$; the last column is the target minus the SPF median forecast of it made at $t$, so a positive correlation means the measure anticipates the direction in which professional forecasters will turn out to have erred. Samples: 1983Q1--2025Q1 for the projections ($N=169$); 1994Q2--2025Q1 for the Minutes-based measures ($N=118$--129 depending on the column; 124 for $y_{t+4}$). Moving-block intervals are in the data files.
\end{minipage}
\end{table}

\begin{table}[!t]
\centering\small
\caption{In-sample regressions of future growth on each measure}\label{tab:prop-growth-reg}
\resizebox{\textwidth}{!}{%
\begin{tabular}{lccccccccc}
\toprule
 & \multicolumn{3}{c}{Alone} & \multicolumn{3}{c}{With $g_t$ and NFCI} & \multicolumn{3}{c}{+ four FRED-MD components} \\
\cmidrule(lr){2-4}\cmidrule(lr){5-7}\cmidrule(lr){8-10}
Measure & $\beta$ & (s.e.) & $R^2$ & $\beta$ & (s.e.) & $\Delta R^2$ & $\beta$ & (s.e.) & $\Delta R^2$ \\
\midrule
Central tendency & .16 & (.15) & .02 & $-$.04 & (.14) & .00 & .06 & (.17) & .00 \\
Dispersion & .25 & (.17) & .06 & .37 & (.15) & .12 & .32 & (.12) & .08 \\
Salience of growth & $-$.17 & (.18) & .03 & $-$.13 & (.14) & .02 & $-$.15 & (.14) & .02 \\
Balance of risks & .17 & (.12) & .03 & .11 & (.10) & .01 & .17 & (.08) & .03 \\
Quantifier discount & .06 & (.07) & .00 & .08 & (.07) & .01 & .12 & (.07) & .01 \\
Attributed heterogeneity & $-$.21 & (.11) & .04 & $-$.23 & (.11) & .05 & $-$.23 & (.09) & .05 \\
Hidden Dissent & .18 & (.11) & .03 & .19 & (.11) & .04 & .22 & (.08) & .05 \\
Insurance motive & $-$.01 & (.12) & .00 & .02 & (.10) & .00 & .03 & (.10) & .00 \\
Verbosity (placebo) & $-$.02 & (.17) & .00 & $-$.03 & (.16) & .00 & $-$.06 & (.14) & .00 \\
\bottomrule
\end{tabular}}
\begin{minipage}{0.95\textwidth}\footnotesize\vspace{4pt}
Notes: OLS of $y_{t+4}$ on each measure, with the measure and the dependent variable standardized, so $\beta$ is the change in the target, in standard deviations, for a one-standard-deviation change in the measure. HAC standard errors (Bartlett, bandwidth 3). $\Delta R^2$ is the increase in $R^2$ from adding the measure to the controls; the controls alone explain 0.15 of the variance of the target in the second specification and 0.22 in the third. The FRED-MD components are the first principal component of each of the four FRED-MD families described in Section~\ref*{sec:public} of the paper. Sample 1994Q2--2025Q1 ($N=124$) in the second and third specifications and for the Minutes-based measures in the first; 1983Q1--2025Q1 ($N=169$) for the projections alone. In-sample and overlapping; the paper's claims rest on the out-of-sample comparisons.
\end{minipage}
\end{table}

\begin{figure}[!t]
\centering
\includegraphics[width=0.7\textwidth]{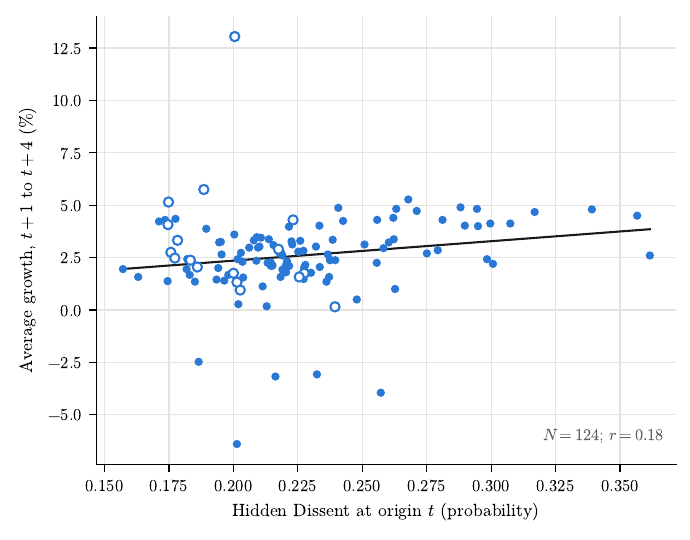}
\caption{Hidden Dissent and future growth. Average growth over the next four quarters against the index at the origin, 1994Q2--2025Q1; the line is the in-sample regression of Table~\ref{tab:prop-growth-reg}, first specification. Hollow markers are the origins of the financial crisis (2008Q4--2010Q4) and the pandemic (2020Q1--2021Q4).}\label{fig:prop-hd-scatter}
\end{figure}

Table~\ref{tab:prop-growth} and Figure~\ref{fig:prop-xcorr} separate what each measure says about the past and present of growth from what it says about its future. The central tendency is, as it should be, a reading of the present: it correlates at 0.58 with current-quarter growth and at 0.16 with the paper's target, and the cross-correlogram peaks sharply at $j=0$ and falls to zero within two quarters. Hidden Dissent is the mirror image. It is unrelated to current growth (0.02) and weakly related to past growth, and its correlation with future growth rises with the horizon, from 0.03 at one quarter to 0.13 at four and 0.18 with the four-quarter average: higher disagreement in the room precedes higher growth. The sign agrees with the coefficient of the index in the paper's central quantile regressions, which is positive in every fit (Section~\ref*{sec:consensus} of the paper), but it is not the point of this appendix; what matters is that the association is with the future, not the present, so that it is not a restatement of the state of the economy. The strongest entry in the table is Hidden Dissent against the error of the professional forecasters: 0.38 (HAC standard error 0.10, moving-block interval $[0.15, 0.60]$). When the room disagreed more, the SPF median made at the same time turned out to be too low. No other measure comes close on this column; the dispersion reaches 0.18 and the balance of risks 0.16, with intervals that include zero, and the placebo is 0.00. Figure~\ref{fig:prop-hd-scatter} shows that the relation with the target is not a product of the two crisis episodes: the hollow markers sit on both sides of the line.

The published dispersion is the other measure with a forward-looking correlation (0.25 with the target, 0.29 with next-quarter growth), and the only one whose association strengthens when the public state is controlled for (Table~\ref{tab:prop-growth-reg}: 0.25 alone, 0.37 with $g_t$ and the NFCI, 0.32 with the FRED-MD components). Attributed heterogeneity has the opposite sign ($-$0.21, stable across specifications): Minutes that attribute views to more distinct groups of participants precede lower growth. Hidden Dissent's coefficient is stable at 0.18--0.22 across the three specifications, with an incremental $R^2$ of 0.03--0.05; it is small, as every single-variable in-sample association with a noisy four-quarter target is, and it does not shrink when the public state is added, which is the property that matters for a measure meant to complement a benchmark rather than to replace it. The salience of growth, the quantifier discount, the insurance motive and verbosity have no in-sample relation with the target in any specification.

\subsection{How the measures relate to the public state}

Table~\ref{tab:declared} summarizes, for the Committee's declared outlook, its declared disagreement and Hidden Dissent, the correlations with growth, the public state and public uncertainty that Section~\ref*{sec:consensus} of the paper reads.

\begin{table}[!htbp]
\centering\footnotesize
\caption{The declared outlook, the declared disagreement and the disagreement absent from the vote against growth and the public state}\label{tab:declared}
\resizebox{\textwidth}{!}{%
\begin{tabular}{lccccccccc}
\toprule
 & \multicolumn{3}{c}{Growth} & \multicolumn{3}{c}{Public state} & \multicolumn{3}{c}{Public uncertainty} \\
\cmidrule(lr){2-4}\cmidrule(lr){5-7}\cmidrule(lr){8-10}
Measure & $g_t$ & $y_{t+4}$ & $y_{t+4}-\mathrm{SPF}_t$ & Activity & NFCI & SPF median & SPF dispersion & EPU & VIX \\
\midrule
Central tendency & .58 & .16 & $-$.12 & .62 & $-$.44 & .64 & $-$.12 & $-$.15 & $-$.21 \\
 & (.23) & (.15) & (.12) & (.22) & (.24) & (.19) & (.18) & (.16) & (.18) \\
Dispersion & $-$.12 & .25 & .18 & $-$.28 & .41 & .20 & .58 & .51 & .42 \\
 & (.15) & (.17) & (.12) & (.19) & (.24) & (.20) & (.24) & (.18) & (.14) \\
Hidden Dissent & .02 & .18 & .38 & .06 & .01 & $-$.40 & $-$.12 & $-$.15 & $-$.01 \\
 & (.08) & (.11) & (.10) & (.07) & (.08) & (.12) & (.09) & (.08) & (.11) \\
\bottomrule
\end{tabular}}
\begin{minipage}{0.97\textwidth}\footnotesize\vspace{4pt}
Notes: Pearson correlations in levels, HAC standard errors in parentheses. $y_{t+4}$ is the target; $\mathrm{SPF}_t$ the consensus forecast of it at the origin; Activity the first principal component of the real-activity block of the public panel; SPF dispersion the interquartile range of the individual forecasts. Projections 1983Q1--2025Q1, Hidden Dissent 1994Q2--2025Q1. The moving-block interval of Hidden Dissent against the consensus error is $[0.15, 0.60]$. The rest of this section gives all nine measures.
\end{minipage}
\end{table}

\begin{table}[!t]
\centering\small
\caption{Correlations of the Committee's measures with the public state and with external measures of uncertainty}\label{tab:prop-public}
\resizebox{\textwidth}{!}{%
\begin{tabular}{lcccccccccc}
\toprule
 & & \multicolumn{4}{c}{FRED-MD first components} & \multicolumn{2}{c}{SPF} & \multicolumn{3}{c}{Uncertainty} \\
\cmidrule(lr){3-6}\cmidrule(lr){7-8}\cmidrule(lr){9-11}
Measure & NFCI & Activity & Prices & Credit, rates & Markets & Median & Dispersion & EPU & GPR & VIX \\
\midrule
Central tendency & $-$.44 & .62 & .24 & $-$.01 & .11 & .64 & $-$.12 & $-$.15 & .06 & $-$.21 \\
 & (.24) & (.22) & (.13) & (.13) & (.12) & (.19) & (.18) & (.16) & (.10) & (.18) \\
Dispersion & .41 & $-$.28 & $-$.06 & $-$.21 & $-$.01 & .20 & .58 & .51 & $-$.19 & .42 \\
 & (.24) & (.19) & (.11) & (.08) & (.13) & (.20) & (.24) & (.18) & (.07) & (.14) \\
Salience of growth & .14 & $-$.09 & $-$.03 & $-$.31 & $-$.12 & .01 & $-$.08 & .04 & .33 & .09 \\
 & (.15) & (.08) & (.08) & (.11) & (.12) & (.11) & (.09) & (.10) & (.18) & (.11) \\
Balance of risks & $-$.27 & .25 & $-$.03 & .24 & $-$.01 & .08 & $-$.20 & $-$.34 & $-$.18 & $-$.18 \\
 & (.13) & (.07) & (.05) & (.11) & (.09) & (.12) & (.09) & (.09) & (.10) & (.12) \\
Quantifier discount & .03 & .07 & .10 & .22 & .07 & $-$.20 & $-$.03 & $-$.04 & $-$.13 & .02 \\
 & (.05) & (.04) & (.07) & (.09) & (.07) & (.08) & (.08) & (.08) & (.08) & (.08) \\
Attributed heterogeneity & .04 & $-$.09 & .02 & .01 & .11 & $-$.23 & $-$.15 & .21 & .04 & $-$.26 \\
 & (.10) & (.06) & (.06) & (.12) & (.09) & (.10) & (.09) & (.13) & (.16) & (.09) \\
Hidden Dissent & .01 & .06 & $-$.06 & .24 & .01 & $-$.40 & $-$.12 & $-$.15 & $-$.29 & $-$.01 \\
 & (.08) & (.07) & (.06) & (.11) & (.09) & (.12) & (.09) & (.08) & (.11) & (.11) \\
Insurance motive & .09 & $-$.01 & .05 & $-$.07 & $-$.02 & $-$.04 & .10 & .06 & $-$.06 & .06 \\
 & (.09) & (.05) & (.07) & (.10) & (.08) & (.11) & (.06) & (.11) & (.09) & (.09) \\
Verbosity (placebo) & .03 & $-$.11 & .11 & .04 & .10 & $-$.06 & .22 & .28 & $-$.09 & $-$.05 \\
 & (.07) & (.13) & (.09) & (.13) & (.10) & (.17) & (.19) & (.12) & (.13) & (.14) \\
\bottomrule
\end{tabular}}
\begin{minipage}{0.95\textwidth}\footnotesize\vspace{4pt}
Notes: Pearson correlations in levels, HAC standard errors in parentheses. The FRED-MD columns are the first principal component of each of the four FRED-MD families described in Section~\ref*{sec:public} of the paper, computed on the final vintage. SPF median: the Survey of Professional Forecasters' median forecast of growth over the next four quarters; SPF dispersion: the interquartile range of the individual forecasts at the same horizon, both from the Federal Reserve Bank of Philadelphia. EPU: \citet{BakerBloomDavis2016}; GPR: \citet{CaldaraIacoviello2022}; VIX: quarterly average. Samples: 1983Q1--2025Q1 for the projections against the FRED-MD and SPF columns ($N=169$), otherwise from the start of each external series (NFCI, EPU and GPR from 1994, the VIX from 1990); Minutes-based measures 1994Q2--2025Q1 ($N=120$--129).
\end{minipage}
\end{table}

Table~\ref{tab:prop-public} asks how much of each measure is already in the public state the paper conditions on. The answer separates the published projections from the Minutes-based measures. The central tendency is largely public information: it correlates at 0.62 with the activity component of FRED-MD, at 0.64 with the SPF median and at $-$0.44 with the NFCI. The published dispersion is a reading of public uncertainty: 0.58 with the dispersion of the professional forecasters, 0.51 with the EPU index, 0.42 with the VIX and 0.41 with the NFCI. These are the two measures that, in the paper, add nothing to a benchmark that already contains the public state (Section~\ref*{sec:robustness} of the paper, Figure~\ref*{fig:sep}), and this table says why. The Minutes-based measures are different. Hidden Dissent is uncorrelated with the NFCI (0.01), the activity component (0.06), the VIX ($-$0.01) and the SPF dispersion ($-$0.12), and it is negatively related to the SPF median ($-$0.40) and to geopolitical risk ($-$0.29): the room disagrees more when the consensus outlook is weak, not when measured uncertainty is high. Its only positive association with the public state is with the credit-and-rates component (0.24). This is the pattern \citet{TsangYang2025} report for their index against the standard uncertainty measures, and it is the sense in which the disagreement the paper exploits is not a repackaging of something already observed. Attributed heterogeneity and the quantifier discount are likewise close to orthogonal to the public state (no correlation above 0.26 in absolute value). The balance of risks and the salience of growth sit in between: the former follows activity (0.25), financial conditions ($-$0.27) and policy uncertainty ($-$0.34); the latter rises with geopolitical risk (0.33). Verbosity, finally, correlates with none of the activity or financial variables, and modestly with the EPU (0.28).

\subsection{Two diagnostics}\label{sec:prop-diag}

\paragraph{The projections across publication regimes.}
The long projection series splices two documents. Table~\ref{tab:prop-regimes} compares them. In the Monetary Policy Report years the central tendency was less volatile and only moderately related to current growth (0.29); in the SEP years it tracks current growth closely (0.68) and is unrelated to the target (0.01). The dispersion has the opposite profile: in both regimes it is weakly or negatively related to current growth and positively to the target (0.32 and 0.26). The paper's long-series results (Section~\ref*{sec:robustness}) are obtained on the SEP years, where the central tendency is most nearly a function of the current state; the MPR years enter only as estimation history.

\begin{table}[!h]
\centering\small
\caption{The published projections in the Monetary Policy Report and the SEP years}\label{tab:prop-regimes}
\begin{tabular}{llcccrr}
\toprule
Measure & Regime & $N$ & Mean & S.D. & $r(g_t)$ & $r(y_{t+4})$ \\
\midrule
Central tendency & MPR, 1983Q1--2007Q3 & 99 & 2.92 & 0.75 & .29 (.16) & .25 (.18) \\
 & SEP, 2007Q4--2025Q1 & 70 & 2.49 & 1.19 & .68 (.27) & .01 (.17) \\
Dispersion & MPR, 1983Q1--2007Q3 & 99 & 0.49 & 0.21 & .11 (.15) & .32 (.15) \\
 & SEP, 2007Q4--2025Q1 & 70 & 0.54 & 0.31 & $-$.20 (.18) & .26 (.28) \\
\bottomrule
\end{tabular}
\begin{minipage}{0.9\textwidth}\footnotesize\vspace{4pt}
Notes: Quarterly series in percentage points; Pearson correlations with HAC standard errors in parentheses. The SEP years coincide with the paper's test origins.
\end{minipage}
\end{table}

\begin{figure}[!htb]
\centering
\includegraphics[width=\textwidth]{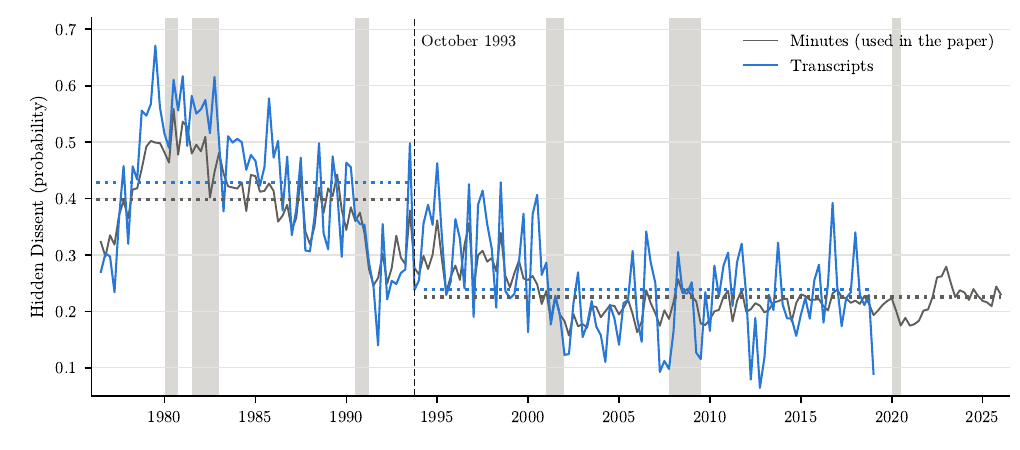}
\caption{The two Hidden Dissent indices. Quarterly means of the meeting-level indices of \citet{TsangYang2025} computed from the transcripts (available through 2018) and from the Minutes (before 1993, from the Record of Policy Actions and the Minutes of Actions). The vertical line marks October 1993, when the Committee decided to release its transcripts with a five-year lag; the dotted horizontal segments are the means of each index before 1993Q3 and after 1994Q1 (Table~\ref{tab:prop-1993-level}). Shaded areas are NBER recessions.}\label{fig:prop-ty}
\end{figure}

\paragraph{Hidden Dissent in the transcripts and in the Minutes.}
\citet{TsangYang2025} compute their index both from the transcripts, which are released after five years, and from the Minutes, the latter with a second model trained to predict the transcript index from the Minutes text; both series are distributed on the authors' data site (\url{https://www.digitecon.org/hidden-dissent-index}). The paper uses the Minutes index, which is public three weeks after the meeting. Figure~\ref{fig:prop-ty} plots the two. Over their common sample, 1976Q2--2018Q4, the two indices correlate at 0.85 by meeting and 0.90 by quarter. The correlation is lower after the FOMC learned, in October 1993, that its transcripts would be published: 0.81 by meeting before 1993Q4 and 0.66 after (0.72 by quarter), and 0.45 over 2008--2018. The disagreement that reaches the Minutes has become a noisier reading of the disagreement in the room since the Committee has known that the room would eventually be read---consistent with the more guarded deliberation that \citet{HansenMcMahonPrat2018} document after the change---and the paper's results are obtained with the noisier of the two instruments. The figure also shows the long decline in the level of both indices from the early 1980s to the early 2000s, which the paper's sample does not see; the paper's inference does not depend on the level, since the index enters standardized over the history available at each origin.

\subsection{Hidden Dissent before and after the transparency change of 1993}\label{sec:prop-1993}

In October 1993 the FOMC learned that its transcripts, until then internal, would be published with a five-year lag. \citet{HansenMcMahonPrat2018} show that deliberation changed: members, and in particular the less experienced, became both more disciplined (more data, more quantitative references) and more conformist (less willing to dissent in the policy round), and conclude that the first effect dominated. The two indices of \citet{TsangYang2025} allow the change to be described from the record rather than from the room, and allow the question the paper cannot ask on its own sample: whether what the index says about future growth changed when the Committee knew it would be read. Everything here is descriptive and in sample; no forecast is made. Two limitations are declared and not corrected: the transcript index is available with a five-year lag, and the Tsang--Yang model is trained on the votes of 1976--2018, so the index of each year incorporates information from the whole period, which affects both regimes alike. A third applies to the Minutes index before 1993, which is computed on the Record of Policy Actions and the Minutes of Actions, a shorter document than the Minutes that replaced them.

The regimes are origins up to 1993Q3 (meetings before the announcement) and from 1994Q2 (the two origins around the announcement are excluded); the post period runs to 2019Q1 for the transcript index and to 2026Q1 for the Minutes index. Moving the cut to 1994Q1 changes nothing reported below.

\begin{table}[!t]
\centering\small
\caption{The two indices before and after 1993}\label{tab:prop-1993-level}
\resizebox{\textwidth}{!}{%
\begin{tabular}{llcccccc}
\toprule
Index & Regime & $N$ & Mean & S.D. & ACF(1) & ACF(4) & Var.\ ratio post/pre \\
\midrule
Transcripts & 1976Q3--1993Q3 & 69 & .429 & .118 & .68 & .60 & \\
 & 1994Q2--2019Q1 & 100 & .238 & .087 & .39 & .15 & .55 [.35, .93] \\
Minutes & 1976Q3--1993Q3 & 69 & .399 & .074 & .83 & .69 & \\
 & 1994Q2--2019Q1 & 100 & .227 & .041 & .73 & .65 & .31 [.15, .64] \\
 & 1994Q2--2026Q1 & 128 & .226 & .039 & .74 & .60 & .27 [.14, .57] \\
\addlinespace
\multicolumn{8}{l}{\emph{Minutes index on transcript index}} \\
 & 1976Q3--1993Q3 & 69 & \multicolumn{5}{l}{Pearson .87 (.07); slope .55 (.04); $R^2$ .76} \\
 & 1994Q2--2019Q1 & 100 & \multicolumn{5}{l}{Pearson .72 (.10); slope .34 (.05); $R^2$ .52} \\
\bottomrule
\end{tabular}}
\begin{minipage}{0.95\textwidth}\footnotesize\vspace{4pt}
Notes: Quarterly means of the meeting-level indices on the paper's clock, original units (probability). The fall in the mean is significant in both indices (HAC $z$ of $-7.0$ and $-9.7$). Variance ratios with 95 percent moving-block bootstrap intervals ($L=4$, $B=10{,}000$). The slope is the OLS coefficient of the Minutes index on the transcript index, HAC standard errors in parentheses.
\end{minipage}
\end{table}

\paragraph{What changed in the index.}
Expressed disagreement fell by almost half (Table~\ref{tab:prop-1993-level}, Figure~\ref{fig:prop-ty}): the transcript index from a mean of 0.43 to 0.24, the Minutes index from 0.40 to 0.23. Its variance fell by more in the Minutes (ratio 0.31) than in the transcripts (0.55), and the Minutes index became a weaker function of the transcript index (slope 0.55 to 0.34, $R^2$ 0.76 to 0.52). Less is disagreed in the room since the Committee has known it would be read, and the record passes on a smaller fraction of what is disagreed. This is the conformity that \citet{HansenMcMahonPrat2018} document, seen from the document.

\begin{table}[!t]
\centering\small
\caption{The index and future growth before and after 1993}\label{tab:prop-1993-content}
\resizebox{\textwidth}{!}{%
\begin{tabular}{llcccccc}
\toprule
 & & & \multicolumn{2}{c}{Correlation with} & \multicolumn{3}{c}{Coefficient on $y_{t+4}$ (per s.d.)} \\
\cmidrule(lr){4-5}\cmidrule(lr){6-8}
Index & Regime & $N$ & $y_{t+4}$ & SPF error & Alone & $+\,g_t$, NFCI & $+\,g_t$, NFCI, SPF \\
\midrule
Transcripts & before 1993Q3 & 69 & $-$.28 (.17) & $-$.18 (.18) & $-$.78 (.48) & $-$.09 (.44) & .00 (.33) \\
 & after 1994Q1 & 100 & .18 (.11) & .24 (.12) & .45 (.30) & .35 (.25) & .19 (.25) \\
Minutes & before 1993Q3 & 69 & $-$.30 (.18) & $-$.18 (.21) & $-$.94 (.57) & $-$.05 (.59) & .08 (.44) \\
 & after 1994Q1 & 124 & .18 (.11) & .38 (.10) & .91 (.53) & .94 (.55) & 1.74 (.56) \\
\addlinespace
\multicolumn{8}{l}{\emph{Stability of the coefficient ($+\,g_t$, NFCI specification)}} \\
Transcripts & \multicolumn{7}{l}{Chow at 1993Q3/1994Q1: change $+$.29 (.49), $p=.55$; Andrews sup-Wald 5.8, bootstrap $p=.33$} \\
Minutes & \multicolumn{7}{l}{Chow at 1993Q3/1994Q1: change $+$.67 (.80), $p=.40$; Andrews sup-Wald 4.7, bootstrap $p=.50$} \\
\addlinespace
\multicolumn{8}{l}{\emph{Quantile regressions after 1994Q1, Minutes index ($+\,g_t$, NFCI); coefficient and moving-block 95 percent interval}} \\
 & \multicolumn{7}{l}{$\tau=.10$: .53 [.26, 2.30]; $\tau=.25$: 1.01 [.18, 2.11]; $\tau=.50$: 1.33 [.14, 2.12]; $\tau=.75$: 1.12 [$-$.03, 2.22]; $\tau=.90$: .54 [$-$.68, 2.24]} \\
 & \multicolumn{7}{l}{Before 1993Q3, no quantile coefficient of either index is distinguishable from zero.} \\
\addlinespace
\multicolumn{8}{l}{\emph{Both indices together, 1994Q2--2019Q1 ($+\,g_t$, NFCI)}} \\
 & \multicolumn{7}{l}{Transcripts alone .35 (.25); Minutes alone 1.07 (.42); together: transcripts $-$.31 (.35), Minutes 1.41 (.56)} \\
\bottomrule
\end{tabular}}
\begin{minipage}{0.95\textwidth}\footnotesize\vspace{4pt}
Notes: In-sample OLS of $y_{t+4}$ on the standardized index (standardized over its full coverage) and the controls shown; HAC standard errors (Bartlett, bandwidth 3) in parentheses. SPF error is $y_{t+4}$ minus the consensus forecast of it at the origin. The Chow test allows a change in the intercept and tests the change in the index slope; the Andrews test searches the central 70 percent of the sample, with a moving-block residual bootstrap ($B=999$). Quantile regression intervals from a moving-block bootstrap ($L=4$, $B=999$). Over the paper's own origins with transcripts available (2008Q4--2019Q1, 42 observations) the in-sample mean coefficient is zero for both indices; the paper's results concern the distribution, out of sample, against a 123-series panel, and Table~\ref{tab:prop-growth-reg} gives the in-sample association over 1994--2025.
\end{minipage}
\end{table}

\begin{figure}[!t]
\centering
\includegraphics[width=\textwidth]{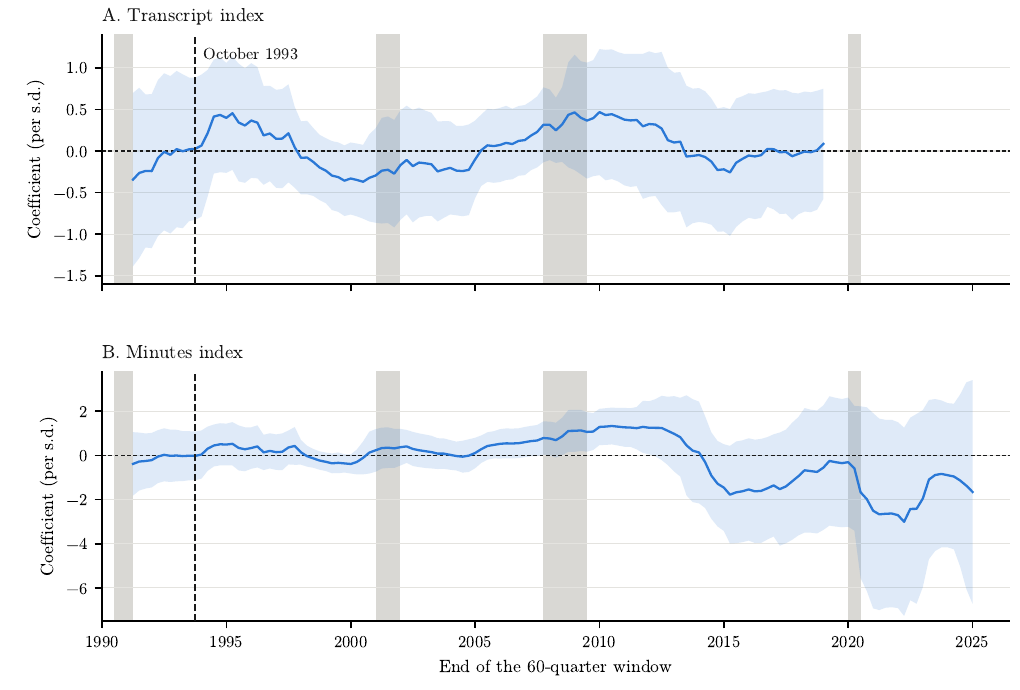}
\caption{Rolling coefficient of each index on future growth. OLS coefficient of the standardized index in a regression of $y_{t+4}$ on the index, current growth and the NFCI, over 60-quarter windows ending at the date shown; bands are 95 percent HAC intervals. The vertical line marks October 1993; shaded areas are NBER recessions.}\label{fig:prop-1993-rolling}
\end{figure}

\paragraph{What did not change: the content.}
Table~\ref{tab:prop-1993-content} asks whether the disagreement that remained says less about the economy. It does not. Before 1993 neither index is related to future growth: the correlations with $y_{t+4}$ are negative ($-0.28$, $-0.30$) and the coefficients conditional on current growth and financial conditions are zero ($-0.09$, $-0.05$), at the mean and at every quantile. After 1993 the relation is positive---0.35 for the transcript index, 0.94 for the Minutes index and 1.74 once the consensus is controlled for---and in the Minutes index it is distinguishable from zero in the lower half of the distribution ($\tau=0.10$, $0.25$, $0.50$). The formal tests do not reject a stable coefficient, because the pre-1993 estimates are imprecise; the Andrews test finds no break at any date. What can be said is that the change of 1993 reduced how much disagreement is expressed, and did not reduce what the expressed disagreement says about future growth: in the point estimates it went from nothing to something. This is the net reading of \citet{HansenMcMahonPrat2018}---discipline dominating conformity---with a macroeconomic outcome as the yardstick, and it places the paper's result in the transparency regime, where it belongs: a 60-quarter rolling coefficient (Figure~\ref{fig:prop-1993-rolling}) is zero in windows ending before 1994 and positive in those ending between 2005 and 2013, for both indices; in the Minutes index it turns negative and very imprecise in windows that contain the pandemic.

\paragraph{The room and the record.}
After 1993 the content is in the Minutes index rather than in the transcript index: alone, 1.07 (HAC 0.42) against 0.35 (0.25); together, the Minutes index keeps its coefficient (1.41) and the transcript index does not ($-0.31$). The disagreement that reaches the record is the part of the disagreement in the room that is informative about growth---the opposite of the suspicion that the Minutes, being a noisier reading of the room, would carry less of its content. The paper's instrument is the noisier of the two in the sense of Table~\ref{tab:prop-1993-level}, and the more informative in the sense of this one.

\subsection{The Beige Book regional dispersion}\label{sec:prop-beige}

The regional dispersion used in Section~\ref*{sec:consensus} of the paper measures how heterogeneous the information brought to the meeting is before it is discussed. Its source is the Beige Book: two weeks before each FOMC meeting, each of the twelve Reserve Banks summarizes the anecdotal information of its district, and the twelve sections are written independently of one another. We use every public edition from the first (July 1, 1983) to November 2025: 341 editions and 4,092 district sections, taken from the archive of the Federal Reserve Bank of Minneapolis (two editions of 2016, absent from that archive, from the Board's own pages). Each district section is coded, by a language model under a fixed written instruction, on a five-point scale for the pace of activity in the district over the period the edition covers ($-2$ clear contraction, $-1$ slight decline or weakening, $0$ flat or mixed, $+1$ slight or modest growth, $+2$ moderate or strong growth), anchored in the Beige Book's own vocabulary, with a verbatim quote supporting the code; a code is always returned, and sections that do not characterize the district as a whole (8 of 4,092) are coded zero and flagged. The regional dispersion of an edition is the standard deviation of the twelve codes; the regional mean is their average. Editions are averaged within the quarter and dated on the paper's clock (the editions published in quarter $t-1$ enter at origin $t$); the series starts at 1983Q4 and is zero before, under the available-only treatment of the Minutes-based measures. A first version of the instruction, which asked for the change in activity relative to the previous edition, left 36 percent of sections uncoded because the text rarely makes that comparison; it was replaced, before any forecast was made, by the version described here, and the replacement is recorded.

The coding was validated in three ways. A human reader (one of the authors), blind to the model's codes, coded the opening paragraph of 24 sections drawn from a random sample of 40 (two per district where available); the quadratically weighted kappa between the model's code of the full section and the human code of its opening paragraph is 0.80 (bootstrap 95 percent interval $[0.48, 0.93]$) for the activity tone. The model's own repeatability, on 205 sections coded twice, is 0.95; on the 40-section sample coded in two separate runs, 0.99. The coding ran as a single batch (4,377 calls, none failed), at temperature zero with a fixed output schema, and the raw responses are archived with the instruction and the corpus hashes.

\begin{table}[!t]
\centering\small
\caption{The Beige Book series}\label{tab:prop-beige}
\resizebox{\textwidth}{!}{%
\begin{tabular}{lcccccc}
\toprule
Series & $N$ & Mean & S.D. & ACF(1) & ACF(2) & ACF(4) \\
\midrule
Regional mean (tone of activity) & 170 & 0.80 & 0.67 & .71 & .57 & .30 \\
Regional dispersion & 170 & 0.64 & 0.16 & .31 & .12 & .22 \\
Regional dispersion, flagged districts excluded & 170 & 0.64 & 0.16 & .31 & .12 & .22 \\
Share of districts off the median sign & 170 & 0.22 & 0.19 & .63 & .50 & .36 \\
\addlinespace
\multicolumn{7}{l}{\emph{Correlations of the regional dispersion}} \\
 & \multicolumn{6}{l}{Hidden Dissent: .10 (1983Q4--1993Q4, $N=41$), .12 (1994Q1--2025Q1, $N=126$)} \\
 & \multicolumn{6}{l}{SPF dispersion: .25 (1983Q4--1993Q4), $-$.08 (1994Q1--2025Q1)} \\
\bottomrule
\end{tabular}}
\begin{minipage}{0.92\textwidth}\footnotesize\vspace{4pt}
Notes: Quarterly series on the paper's clock, 1983Q4--2026Q1. Tone of activity coded $-2$ to $+2$ by district; the regional dispersion is the cross-district standard deviation, averaged over the editions of the quarter. The series of the declared outlook for the coming months, coded in the same instruction, are in the data files.
\end{minipage}
\end{table}

The regional dispersion is weakly persistent (first-order autocorrelation 0.31) and nearly uncorrelated with Hidden Dissent (0.12 after 1994) and with the dispersion of professional forecasters ($-0.08$): it measures something the paper's other series do not.

\clearpage
\section{Real-time reconstruction of the public panel}\label{app:realtime}

This appendix documents the exercise reported in Section~\ref*{sec:realtime}: the design, the verification carried out before any loss was computed, the full inference under real-time data, and the paired stability diagnostic between the two information sets.

\subsection*{Arm 1: against the public panel}

\paragraph{Design and execution.}
The public benchmark is rebuilt from the historical vintages of FRED-MD, taking at each forecast origin the vintage, the transformation codes and the series definitions in force at the time, under a release-lag rule fixed before any result was seen, with no backfilling from later vintages and no imputation. Both versions of the benchmark are restricted, origin by origin, to the same predictors, so that the only difference between them is the vintage of the data. The target, the horizon, the 70 origins, the 19 quantiles, the block penalties ($\lambda_L=0.562$, $\lambda_C=100$, $\lambda_U=100$, in both arms of every comparison), the available-only transformation, the estimation code and the signal are those of the main specification. The design is two information sets $\times$ two models (public benchmark, benchmark plus signal) $\times$ 19 quantiles, that is 76 estimation jobs of 70 forecasts each, 5{,}320 forecasts in all.

The exercise was executed blind. All 76 jobs completed with no failures, and before the first examination of any loss the whole set of forecasts was structurally verified: the forecast clock, the realized-target identity and the penalty identity were checked both between the benchmark and the augmented model and between the two information sets; all predictions were finite; no penalty was selected and no predictor availability was chosen after the fact; and no loss, gain or performance figure was displayed at any point before that verification was complete. Equality of predictions between the two information sets was deliberately not required, since the two sets differ by construction.

\paragraph{Inference.}
The procedure is the one fixed for the main results and is applied here unchanged. Within each block, the pinball losses are first averaged across the quantiles of the block for each target, and the inference is then carried out on the resulting series of 70 target-level differentials $d_{bt}=L_{bt}(\text{benchmark}+\HD{})-L_{bt}(\text{benchmark})$, so that quantiles are not treated as independent observations. Negative $\bar d$ favors the signal. Standard errors are Bartlett HAC of bandwidth 3 with the Harvey--Leybourne--Newbold correction at $h=4$ and 69 degrees of freedom, and the reported $p$-value is one-sided for a loss reduction. The moving-block bootstrap uses $B=10{,}000$ resamples, block lengths $L=4$ (primary), 6 and 8; it is a percentile and sign diagnostic and is not null-recentered, so its sign probabilities are reported as descriptive robustness statistics and not as a second hypothesis test.

\begin{table}[!t]
\centering
\caption{Real-time arm: full inference and moving-block diagnostics}
\label{tab:rt-detail}
\small
\resizebox{\textwidth}{!}{%
\begin{tabular}{l d{2.3} d{2.6} d{1.6} d{2.3} d{1.4} l d{1.4}}
\toprule
Block & \multicolumn{1}{c}{Gain (\%)} & \multicolumn{1}{c}{$\bar d$} & \multicolumn{1}{c}{HAC SE} & \multicolumn{1}{c}{HLN $t$} & \multicolumn{1}{c}{$p_{1s}$} & \multicolumn{1}{c}{MBB 95\%, $L=4$} & \multicolumn{1}{c}{Sign $p$} \\
\midrule
All 19 quantiles  &  3.789 & -0.023608 & 0.009513 & -2.358 & 0.0106 & $[-0.0411,\ -0.0041]$ & 0.0174 \\
Lower (Q05--Q25)  & -1.338 &  0.006806 & 0.012101 &  0.534 & 0.7026 & $[-0.0068,\ +0.0355]$ & 0.2582 \\
Center (Q30--Q70) &  5.559 & -0.040690 & 0.013936 & -2.774 & 0.0036 & $[-0.0687,\ -0.0136]$ & 0.0030 \\
Upper (Q75--Q95)  &  4.298 & -0.023276 & 0.007795 & -2.837 & 0.0030 & $[-0.0417,\ -0.0109]$ & 0.0000 \\
\bottomrule
\end{tabular}}
\begin{minipage}{0.93\textwidth}\footnotesize\vspace{4pt}
Notes: Targets won by the augmented model: 52 of 70 on the grid, 29 in the lower block, 48 in the center, 59 in the upper block. Two-sided HLN $p$-values are 0.0212 on the grid, 0.5948 in the lower block, 0.0071 in the center and 0.0060 in the upper block. At $L=6$ and $L=8$ the percentile intervals are $[-0.0422,-0.0052]$ and $[-0.0426,-0.0078]$ on the grid, $[-0.0699,-0.0152]$ and $[-0.0709,-0.0191]$ in the center, $[-0.0421,-0.0114]$ and $[-0.0422,-0.0113]$ in the upper block; the lower-block interval contains zero at every length.
\end{minipage}
\end{table}

\paragraph{Stability between the two information sets.}
Comparing percentage gains across the two arms is not the right object, because the two benchmarks have different mean losses. We therefore compare the loss differentials themselves, target by target: writing $\delta_{bt}=d^{\,\mathrm{RT}}_{bt}-d^{\,\mathrm{final}}_{bt}$, a positive $\delta$ means that the advantage of the signal weakens under real-time data. This is a stability diagnostic, not an equivalence test; no equivalence margin was pre-specified.

\begin{table}[!t]
\centering
\caption{Paired stability of the signal's loss differential between information sets}
\label{tab:rt-stability}
\small
\resizebox{\textwidth}{!}{%
\begin{tabular}{l d{2.6} d{2.6} d{2.6} d{1.6} d{2.3} d{1.4} c}
\toprule
Block & \multicolumn{1}{c}{$\bar d$, final} & \multicolumn{1}{c}{$\bar d$, RT} & \multicolumn{1}{c}{$\delta$} & \multicolumn{1}{c}{HAC SE} & \multicolumn{1}{c}{HLN $t$} & \multicolumn{1}{c}{$p_{2s}$} & MBB, $L=4$ \\
\midrule
All 19 quantiles  & -0.025589 & -0.023608 &  0.001981 & 0.001080 &  1.742 & 0.0860 & above zero \\
Lower (Q05--Q25)  & -0.000853 &  0.006806 &  0.007660 & 0.004151 &  1.753 & 0.0841 & above zero \\
Center (Q30--Q70) & -0.040614 & -0.040690 & -0.000076 & 0.000057 & -1.270 & 0.2082 & spans zero \\
Upper (Q75--Q95)  & -0.023281 & -0.023276 &  0.000005 & 0.000023 &  0.203 & 0.8401 & spans zero \\
\bottomrule
\end{tabular}}
\begin{minipage}{0.93\textwidth}\footnotesize\vspace{4pt}
Notes: $\delta=\bar d^{\,\mathrm{RT}}-\bar d^{\,\mathrm{final}}$; positive values mean the signal's advantage weakens under real-time data. The last column summarizes the moving-block percentile interval at $L=4$, with the same conclusion at $L=6$ and $L=8$: the grid and lower-block intervals lie above zero, the center and upper intervals contain it. The HAC statistics are more conservative than the bootstrap for the grid and the lower block, which is why the text describes the attenuation rather than assigning it a significance label. In the center the change is 0.19 percent of the magnitude of the final-vintage differential, and in the upper block 0.02 percent.
\end{minipage}
\end{table}

\subsection*{Arm 2: against the public panel with financial conditions}

\paragraph{Design and execution.}
The second arm repeats the exercise against the benchmark that also contains the canonical financial-conditions index, which is the benchmark under which the paper's lower-region contribution is measured. The index is placed on the same real-time clock as the rest of the information set: at each origin it takes the value constructible from the weekly releases available up to that origin's cutoff, rather than its current published history, and it enters unpenalized in both members of every pair. The signal is the challenger in one member of the pair only; the penalties are selected on the benchmark without the signal and imposed unchanged on the augmented model; the target, the horizon, the 70 origins, the 19 quantiles and the availability rule are those of the main specification. Execution was blind: 76 jobs, 5{,}320 forecasts, no failures, and the same structural verification as the first arm---presence of the index in both members of each pair, the signal in the challenger only, the realized-target identity, the penalty identity and the equality of targets between the two information sets---completed before any loss was computed.

\begin{table}[!t]
\centering
\caption{Real-time arm against the panel with financial conditions: full inference}
\label{tab:rt-detail-m3}
\small
\resizebox{\textwidth}{!}{%
\begin{tabular}{l d{2.3} d{2.6} d{1.4} l l}
\toprule
Block & \multicolumn{1}{c}{Gain (\%)} & \multicolumn{1}{c}{$\bar d$} & \multicolumn{1}{c}{$p_{1s}$} & \multicolumn{1}{c}{95\% HAC interval} & \multicolumn{1}{c}{MBB 95\%, $L=4$} \\
\midrule
All 19 quantiles  &  5.855 & -0.036299 & 0.0000 & $[-0.0493,\ -0.0233]$ & $[-0.0500,\ -0.0235]$ \\
Lower (Q05--Q25)  &  4.123 & -0.017769 & 0.0084 & $[-0.0316,\ -0.0039]$ & $[-0.0259,\ -0.0030]$ \\
Center (Q30--Q70) &  8.337 & -0.063380 & 0.0000 & $[-0.0899,\ -0.0369]$ & $[-0.0939,\ -0.0395]$ \\
Upper (Q75--Q95)  &  1.094 & -0.006085 & 0.2258 & $[-0.0211,\ +0.0090]$ & $[-0.0195,\ +0.0104]$ \\
\bottomrule
\end{tabular}}
\begin{minipage}{0.93\textwidth}\footnotesize\vspace{4pt}
Notes: the benchmark is the matched public panel plus the financial-conditions index on the real-time clock; the challenger adds the deliberative signal. Inference as in Table~\ref{tab:rt-detail}. In the lower block the HLN statistic is $-2.393$, the two-sided $p$-value 0.0167, and the moving-block sign probability 0.0108; the percentile intervals remain below zero at $L=6$ and $L=8$. The matched final-vintage arm gives gains of 6.036, 5.099, 8.334 and 1.109 percent for the grid, lower, center and upper blocks respectively.
\end{minipage}
\end{table}

\begin{table}[!t]
\centering
\caption{Paired stability between information sets, panel with financial conditions}
\label{tab:rt-stability-m3}
\small
\begin{tabular}{l d{2.3} d{2.3} d{2.6} d{1.4} l}
\toprule
Block & \multicolumn{1}{c}{Gain, final} & \multicolumn{1}{c}{Gain, RT} & \multicolumn{1}{c}{$\delta$} & \multicolumn{1}{c}{$p_{2s}$} & \multicolumn{1}{c}{95\% HAC interval} \\
\midrule
All 19 quantiles  & 6.036 & 5.855 &  0.001079 & 0.1750 & $[-0.00040,\ +0.00256]$ \\
Lower (Q05--Q25)  & 5.099 & 4.123 &  0.004075 & 0.1755 & $[-0.00153,\ +0.00968]$ \\
Center (Q30--Q70) & 8.334 & 8.337 & -0.000032 & 0.3461 & $[-0.00010,\ +0.00003]$ \\
Upper (Q75--Q95)  & 1.109 & 1.094 &  0.000083 & 0.1781 & $[-0.00003,\ +0.00020]$ \\
\bottomrule
\end{tabular}
\begin{minipage}{0.93\textwidth}\footnotesize\vspace{4pt}
Notes: $\delta=\bar d^{\,\mathrm{RT}}-\bar d^{\,\mathrm{final}}$; positive values mean the signal's advantage weakens under real-time data. Moving-block percentile intervals at $L=4$ contain zero in every block, as do the HAC intervals. The lower-block entry pairs an undetected change with a descriptive fall in the gain of about one percentage point; with seventy overlapping targets the exercise does not have the precision to characterize a change of that size, and we report it as such rather than as evidence of stability.
\end{minipage}
\end{table}

\paragraph{Describing the attenuation.}
In both arms the point estimates move in the same direction, and the tests are not sharp enough to characterize the movement. Against the public panel, the HAC statistics put the grid and lower-block changes at two-sided $p$-values of 0.086 and 0.084, while the moving-block intervals for those two changes lie above zero; the center and upper changes span zero under both procedures. Against the panel with financial conditions, every paired change is undetectable by both procedures (two-sided $p$-values between 0.18 and 0.35, and intervals containing zero in every block), but the lower-block point estimate is about one percentage point of gain below the matched final-vintage arm, and we do not read an undetected difference of that size, in a sample of seventy overlapping targets, as evidence that nothing changed. What the exercise establishes is the positive statement---the contribution is present and established under vintage-specific information against both benchmarks---and not the stronger claim that the two information sets deliver the same thing.

\paragraph{The matched final-vintage benchmark.}
Both arms restrict the predictor set, origin by origin, to what is constructible from the historical vintage, so the final-vintage columns are matched benchmarks rather than the unrestricted specifications of the main text. The restriction is close to innocuous in most places: against the public panel the matched losses reproduce the unrestricted ones to within a thousandth in every block, and against the panel with financial conditions the center and upper gains are 8.334 and 1.109 percent against 8.335 and 1.103 in Table~\ref*{tab:anatomy}. The exception is the lower block of the second arm, where the matched benchmark gives a gain of 5.099 percent against 4.495 in the main specification, and we trace that difference rather than describe it.

At the 20 origins where the matched and the unrestricted predictor sets happen to coincide, the two designs agree exactly: across the 100 lower-block cells at those origins, 20 origins by the five quantiles of the block, the largest absolute difference between the two designs' predictions is zero for the benchmark and for the augmented model alike, and so is the largest absolute difference between their loss differentials. The designs are therefore the same design wherever the information set is the same, which is the check one would want on an exercise of this kind. At the remaining 50 origins the matched set drops the series that cannot be reconstructed consistently from historical vintages; the exclusions are dominated by the price-earnings ratio of the S\&P composite, absent at 50 of the 70 origins, and by the two help-wanted series, absent at 36 origins each. The difference in the lower block is concentrated around the twentieth percentile.

The entire gap between 4.495 and 5.099 is thus the common-support requirement, and the two figures are not competing estimates of the same object: the first is the gain against the full final-vintage benchmark, the second the gain against that benchmark restricted to what is reconstructible in real time. The properly matched real-time comparison is 5.099 against 4.123. What the reconciliation does show substantively is that the lower-block contribution moves by six tenths of a percentage point when a handful of series leave the panel, while the center reproduces to the third decimal; that sensitivity is a property of the lower block and we record it as such.

\subsection*{Arm 3: the family and the external signals under real-time information}

\paragraph{Design and execution.}
The five original other measures of the deliberation and the five public signals from outside the Committee are each added, unpenalized, to the matched real-time public panel, on the 70 targets, at the 19 quantiles, under the pre-specified penalties: 190 jobs and 13{,}300 forecasts for the deliberative measures, 95 jobs and 6{,}650 forecasts for the external signals, all structurally verified before any loss was computed. The effective size of the public block is 119 to 123 predictors and is identical, target by target, between benchmark and challenger. The external-signal jobs recorded 180 solver fallbacks to the mean (38 for forecaster dispersion, 22 for geopolitical risk, 29 for policy uncertainty, 39 for implied volatility and 52 for the oil price); every final forecast is finite, the fallbacks are a computational diagnostic and not a selection rule, and the concentration in the oil-price jobs is one more reason to treat that signal's extreme asymmetry with caution. For the lower block only, the five deliberative measures and the five external signals are then added to the real-time benchmark with financial conditions, 1{,}750 and 1{,}750 further forecasts, verified in the same way; the external signals were pre-specified as a set before any was opened. Attributed heterogeneity, added to the family later, was put through both arms in full---all 19 quantiles against the real-time public panel and against the real-time panel with financial conditions---under the same matched benchmarks; its block gains are in Table~\ref{tab:ah-realtime}, its grid entries are repeated in Table~\ref{tab:rt-family} for completeness, and its lower-block entry against the panel with financial conditions is in Table~\ref*{tab:rt-lower}.

\begin{table}[!t]
\centering
\caption{Competing signals added to the real-time public panel: gains by block}
\label{tab:rt-family}
\small
\begin{tabular}{l d{2.3} d{2.3} d{2.3} d{2.3}}
\toprule
Measure added to real-time \Mtwo{} & \multicolumn{1}{c}{All 19} & \multicolumn{1}{c}{Lower} & \multicolumn{1}{c}{Center} & \multicolumn{1}{c}{Upper} \\
\midrule
\multicolumn{5}{l}{\emph{Deliberative family (FOMC Minutes)}} \\
Attributed heterogeneity &  3.357 & -6.697 &  4.859 &  9.142 \\
Balance of risks      &  1.363 & -0.198 &  2.125 &  0.973 \\
Quantifier discount   & -1.489 & -0.395 & -2.117 & -0.991 \\
Salience of growth    & -1.916 & -2.035 & -1.882 & -1.886 \\
Insurance motive      & -2.846 & -7.367 & -1.734 & -1.304 \\
Verbosity             & -1.098 & -4.012 & -0.268 & -0.383 \\
\midrule
\multicolumn{5}{l}{\emph{Public signals from outside the Committee}} \\
Forecaster dispersion (SPF) & -0.215 & -2.495 & -0.062 &  1.552 \\
Geopolitical risk     &  0.401 &  0.095 &  0.167 &  1.259 \\
Economic policy uncertainty & -3.108 & -1.715 & -2.965 & -4.763 \\
Implied volatility (VIX) &  0.199 &  3.187 & -1.397 &  1.275 \\
Oil price             & -1.222 & -20.026 &  0.104 & 13.211 \\
\bottomrule
\end{tabular}
\begin{minipage}{0.93\textwidth}\footnotesize\vspace{4pt}
Notes: Same construction as Table~\ref*{tab:overview}, with the public panel rebuilt from historical vintages. Every entry is within a few tenths of a percentage point of its final-vintage counterpart. Inference: attributed heterogeneity improves the grid (one-sided $p=0.024$), the center ($0.003$) and the upper block ($<0.001$) and worsens the lower block (interval $[-0.011,\ +0.079]$), as in Table~\ref{tab:ah-realtime}, Panel~A; the balance of risks, in its Committee-level version, reproduces its final-vintage profile to within half a percentage point in every block (the earlier paragraph-count version gave 1.546, $-2.226$, 3.643 and $-0.013$); the insurance motive worsens the grid and the lower block (two-sided $p$ of 0.010 and 0.002, moving-block intervals above zero); no external signal improves the lower block detectably (one-sided $p$ of 0.92, 0.42, 0.85, 0.20 and 0.996 in the order of the table); geopolitical risk in the upper block has one-sided $p=0.035$ and a moving-block interval just below zero, which we read as localized and suggestive; policy uncertainty worsens the grid, center and upper block (two-sided $p$ of 0.001, 0.018 and 0.007); the oil price worsens the lower block and improves the upper (two-sided $p$ of 0.008 and 0.0005), both confirmed by the moving-block intervals.
\end{minipage}
\end{table}

\begin{table}[!t]
\centering
\caption{Paired comparisons with Hidden Dissent on the real-time public panel}
\label{tab:rt-paired-m2}
\small
\resizebox{\textwidth}{!}{%
\begin{tabular}{l d{2.6} d{1.4} d{1.4} c d{2.6} d{1.4} d{1.4}}
\toprule
 & \multicolumn{3}{c}{All 19 quantiles} & & \multicolumn{3}{c}{Center (Q30--Q70)} \\
\cmidrule(lr){2-4}\cmidrule(lr){6-8}
Hidden Dissent against & \multicolumn{1}{c}{$\bar d^{\HD}-\bar d^{s}$} & \multicolumn{1}{c}{$p_{2s}$, HLN} & \multicolumn{1}{c}{$p$, MBB} & & \multicolumn{1}{c}{$\bar d^{\HD}-\bar d^{s}$} & \multicolumn{1}{c}{$p_{2s}$, HLN} & \multicolumn{1}{c}{$p$, MBB} \\
\midrule
Balance of risks (earlier version) & -0.013974 & 0.1406 & 0.1204 & & \multicolumn{1}{c}{--} & \multicolumn{1}{c}{--} & \multicolumn{1}{c}{--} \\
Quantifier discount   & -0.032888 & 0.0032 & 0.0006 & & \multicolumn{1}{c}{--} & \multicolumn{1}{c}{--} & \multicolumn{1}{c}{--} \\
Salience of growth    & -0.035545 & 0.0078 & 0.0052 & & \multicolumn{1}{c}{--} & \multicolumn{1}{c}{--} & \multicolumn{1}{c}{--} \\
Insurance motive      & -0.041339 & 0.0014 & 0.0000 & & \multicolumn{1}{c}{--} & \multicolumn{1}{c}{--} & \multicolumn{1}{c}{--} \\
Verbosity             & -0.030451 & 0.0085 & 0.0040 & & \multicolumn{1}{c}{--} & \multicolumn{1}{c}{--} & \multicolumn{1}{c}{--} \\
\addlinespace
Forecaster dispersion (SPF) & \multicolumn{1}{c}{--} & \multicolumn{1}{c}{--} & \multicolumn{1}{c}{--} & & -0.041143 & 0.0231 & 0.0130 \\
Geopolitical risk     & \multicolumn{1}{c}{--} & \multicolumn{1}{c}{--} & \multicolumn{1}{c}{--} & & -0.039471 & 0.0082 & 0.0038 \\
Economic policy uncertainty & \multicolumn{1}{c}{--} & \multicolumn{1}{c}{--} & \multicolumn{1}{c}{--} & & -0.062391 & 0.0000 & 0.0000 \\
Implied volatility (VIX) & \multicolumn{1}{c}{--} & \multicolumn{1}{c}{--} & \multicolumn{1}{c}{--} & & -0.050915 & 0.0009 & 0.0002 \\
Oil price             & \multicolumn{1}{c}{--} & \multicolumn{1}{c}{--} & \multicolumn{1}{c}{--} & & -0.039930 & 0.6054 & 0.4030 \\
\bottomrule
\end{tabular}}
\begin{minipage}{0.93\textwidth}\footnotesize\vspace{4pt}
Notes: Paired target-level differentials $d^{\HD}_t-d^{s}_t$ on the same real-time public benchmark; negative favors \HD{}. Attributed heterogeneity, added later, was compared with the index directly only on the benchmark where the lower-region claim is made (Table~\ref*{tab:rt-lower}); its paired comparison on the public panel is not reported in this version. Against the public panel the deliberative signal has detectably lower grid loss than four of the five original other Minutes measures, the earlier version of the balance of risks being the exception; the Committee-level version was compared with the index on the benchmark with financial conditions only (Table~\ref*{tab:rt-lower}), and detectably lower center loss than four of the five external signals, the oil price being the exception. In the lower block against the public panel none of the external comparisons is informative: the differentials are $-0.006$, $+0.007$, $-0.002$, $+0.023$ and $-0.095$ in the order of the table, with only the oil-price comparison distinguishable from zero ($p=0.015$), and the point estimate against implied volatility favors implied volatility. That is the comparison the main text carries to the benchmark with financial conditions (Table~\ref*{tab:rt-lower}).
\end{minipage}
\end{table}

\subsection*{Arm 4: current growth from its historical vintages}

\paragraph{Design and execution.}
The fourth arm completes the second. In both the benchmark with financial conditions and the challenger, current growth and its estimation history are taken, at each origin, from the vintage of real GDP (GDPC1) available at that origin, whose last observation is the origin quarter itself; the public panel stays on the real-time clock of Arm~2, the financial-conditions index on its published history, and the realized target $y_{t+4}=\tfrac14\sum_{j=1}^{4}g_{t+j}$ on final data, so that the vintage changes the forecaster's information set and not the variable against which the forecast is scored (verified at every origin). The design covers the comparison on which the lower-region claim rests---the lower block, five quantiles, against the benchmark with financial conditions---with the pre-specified lower-block penalty and no re-selection. It was fixed before any forecast was generated; the 700 forecasts (two models, five quantiles, 70 origins) were verified, with no estimator fallback, before any loss was computed.

\paragraph{Results.}
The benchmark's lower-block loss is 0.4329 and the challenger's 0.4117, a gain of 4.901 percent ($\bar d=-0.021218$, 43 of 70 targets won). Bartlett HAC of bandwidth 3 with the HLN correction gives a statistic of $-2.956$, one-sided $p=0.0021$ (two-sided $0.0043$) and a 95 percent interval of $[-0.0346,\ -0.0079]$. The moving-block bootstrap places the probability of a non-negative mean differential at 0.0014, 0.0005 and 0.0003 for $L=4$, 6 and 8. Against the real-time arm of Table~\ref*{tab:realtime}, Panel~B (4.123 percent), the paired change in the differential is $-0.003449$, with HAC interval $[-0.0109,\ +0.0040]$ and two-sided HLN $p=0.389$; the primary moving-block interval ($L=4$) contains zero, while those at $L=6$ and 8 lie just below it. We read this as stability of the contribution under vintage-specific output data, not as evidence that vintage data raise it. The arm was not run against the public panel alone or for the other blocks, where the paper makes no lower-region claim.

\paragraph{What the exercise establishes, and what it does not.}
It establishes that the incremental content of the deliberative signal against the public panel is not explained by the benchmark's use of subsequently revised macroeconomic data: on the grid, in the center and in the upper block the contribution survives the restriction to vintage-specific information, and in the two blocks where the content sits it is essentially unchanged. It does not establish that the signal contains private information, that it causes anything, or that the two information sets are formally equivalent. The same holds against the benchmark that also contains financial conditions, where the lower-region contribution remains 4.12 percent and established under vintage-specific information. It does not establish that the two information sets are equivalent: in both arms the paired changes point in the same direction, and in the lower block of the second arm an undetected change accompanies a descriptive fall of about one percentage point in the gain. The third arm establishes that, on the benchmark with financial conditions and under real-time information, no other measure of the deliberation, attributed heterogeneity included, and no external proxy for risk that was carried to that benchmark reproduces the lower-region contribution; it does not establish that the deliberative signal has lower loss than each of them: of the ten direct comparisons only three reject equality at the unadjusted level and one after Holm adjustment, and forecaster dispersion has a point estimate of the same size with far greater uncertainty. And the realized outcomes are at their final vintage in every arm, by construction, so that all models are scored against the same realized growth.
\section{Which public information does deliberation complement?}\label{sec:mechanism}

The public panel is partitioned into four groups following the database's own classification: real activity (62 series), prices (23), credit and interest rates (34), and market prices (4). The same design can therefore ask not only whether the deliberative signal helps but \emph{what it is about}. Removing one group from the benchmark and re-evaluating the signal's contribution asks whether that group is a complement to the signal, in which case the contribution falls when the group is absent, or a substitute, in which case it rises. The exercise is run on the benchmark with financial conditions, \Mthree{}, because that is where the paper's lower-region claim is made, and in the lower block. The object is the paired target-level contrast $\Delta_{j,t}=d^{(-j)}_t-d^{\mathrm{full}}_t$, where $d^{(-j)}_t$ is the signal's loss differential on the benchmark without group $j$ and $d^{\mathrm{full}}_t$ its differential on the full benchmark, expressed on the common denominator $L(\Mthree{})$ so that the four groups are comparable; a negative value means the signal is worth less without that group.

One element of the design decides what the exercise can see, and we state it before the result. The block penalty is \emph{inherited}: every reduced benchmark uses the pre-specified lower-block penalty of the full model, so that the shrinkage on each retained coefficient is held fixed while the information set changes, which is the treatment every other challenger in the paper receives. Re-selecting the penalty for each reduced benchmark is not an option here. On the validation window, re-selection sends the penalty of any reduced public block to the upper boundary of the grid, from $0.56$ to $1{,}000$, in 19 of 20 lower-block cases; the retained block is then shrunk to nothing and the four reduced benchmarks become one model, with forecast correlations above $0.99999$ across groups against $0.77$ to $0.89$ with the full benchmark, so that removing 62 series and removing 4 would give the same answer.

\begin{table}[!t]
\centering
\caption{Removing one group of public predictors from \Mthree{}: the signal's lower-block contribution}
\label{tab:groups}
\small
\resizebox{\textwidth}{!}{%
\begin{tabular}{l c d{1.6} d{1.6} d{2.3} d{2.3} l}
\toprule
 & & \multicolumn{3}{c}{Within the reduced benchmark} & \multicolumn{2}{c}{Against the full benchmark} \\
\cmidrule(lr){3-5}\cmidrule(lr){6-7}
Group removed & Series & \multicolumn{1}{c}{$L$(bench.)} & \multicolumn{1}{c}{$L$(bench.${}+{}$signal)} & \multicolumn{1}{c}{Gain (\%)} & \multicolumn{1}{c}{$\Delta_j$ (pp)} & \multicolumn{1}{c}{MBB 95\%, $L=4$} \\
\midrule
None (full \Mthree{})   & 123 & 0.426365 & 0.407200 &  4.495 & \multicolumn{1}{c}{--} & \multicolumn{1}{c}{--} \\
Real activity           &  62 & 0.436777 & 0.440096 & -0.760 & -5.273 & $[-8.859,\ -0.355]$ \\
Prices                  &  23 & 0.434479 & 0.423020 &  2.637 & -1.807 & $[-5.609,\ +1.272]$ \\
Credit and rates        &  34 & 0.479599 & 0.458698 &  4.358 &  0.407 & $[-2.305,\ +4.261]$ \\
Market prices           &   4 & 0.431431 & 0.409526 &  5.077 &  0.643 & $[-0.673,\ +1.676]$ \\
\bottomrule
\end{tabular}}
\begin{minipage}{0.95\textwidth}\footnotesize\vspace{4pt}
Notes: Lower block (Q05--Q25), 70 targets, corrected target, $h=4$. Every reduced benchmark inherits the pre-specified lower-block penalty of the full model; the financial-conditions index and current growth are unpenalized throughout; the signal is the only difference between benchmark and challenger within each row. $\Delta_j$ is the mean of the paired target-level contrast $d^{(-j)}_t-d^{\mathrm{full}}_t$, scaled by $-100/\overline{L(\Mthree{})}$ so that negative values mean the signal is worth less without the group. For real activity the two-sided HLN $p$-value is $0.057$, and $0.227$ after Holm adjustment across the four contrasts; the moving-block intervals at $L=6$ lie below zero and at $L=8$ include it. The three other intervals contain zero at every block length. Table~\ref{tab:placebo} reads each $\Delta_j$ against random removals of the same size.
\end{minipage}
\end{table}

Table~\ref{tab:groups} and Figure~\ref*{fig:groups} of the paper report the result. Within each reduced benchmark the signal's lower-block gain is 5.1 percent without market prices, 4.4 without credit and rates, 2.6 without prices, and $-0.8$ without real activity, against 4.5 on the full benchmark. Against the full benchmark, removing market prices or credit and rates leaves the signal's contribution where it was ($+0.6$ and $+0.4$ points, intervals centered on zero); removing prices lowers it by 1.8 points with an interval that contains zero; and removing real activity lowers it by 5.3 points, with a moving-block interval that excludes zero at $L=4$ and $L=6$ and includes it at $L=8$, a two-sided HLN $p$-value of $0.057$, and no survival of Holm adjustment across the four contrasts. Real activity is also the largest group, half the panel, so the contrast that stands out is the one most exposed to the explanation that removing many predictors, whatever they are, changes the benchmark most. The placebo that separates the two explanations was specified, with its decision rule, before it was run.

\paragraph{Size-matched placebo.}
For each group, 40 random subsets of public predictors of the same size (4, 23, 34 and 62) were drawn without replacement from the 123-predictor universe under a seed fixed in advance, removed from the benchmark with the same inherited penalty, and evaluated on the same 70 targets and five lower-block quantiles---160 subsets and 112{,}000 forecasts, generated and verified before any loss was computed. Each group's $\Delta_j$ is then read against the distribution of what removing an arbitrary set of the same size produces, and the randomization $p$-value counts the draws at least as extreme, with the usual $+1$ correction over 41. Table~\ref{tab:placebo} reports the outcome. Two things are true at once. Removing public information lowers the signal's contribution on average whatever is removed, and by more the more is removed: the placebo mean falls from $-0.2$ points at 4 series to $-1.4$ at 62. The signal is worth more against a richer benchmark, which is what one expects of a measure whose content is a correction to the state the benchmark describes. And the real-activity removal is not explained by that size effect: at $-5.3$ points it lies below every one of the 40 random removals of 62 series (5th percentile $-3.6$, median $-1.6$; lower-tail $p=0.024$, two-sided $0.049$). Prices sit inside their placebo distribution (rank 15 of 40), and market prices and credit and rates sit near its upper edge, which is what their unchanged contributions in Table~\ref{tab:groups} already said.

\begin{table}[!t]
\centering
\caption{Each group's removal against 40 random removals of the same size, and the real-activity contrast under three penalties}
\label{tab:placebo}
\small
\resizebox{\textwidth}{!}{%
\begin{tabular}{l c d{2.3} d{2.3} d{2.3} d{2.3} d{2.3} c d{1.3} d{1.3}}
\toprule
 & & \multicolumn{1}{c}{Observed} & \multicolumn{4}{c}{Placebo distribution of $\Delta$} & & \multicolumn{2}{c}{Randomization $p$} \\
\cmidrule(lr){4-7}\cmidrule(lr){9-10}
 & \multicolumn{1}{c}{Series} & \multicolumn{1}{c}{$\Delta_j$} & \multicolumn{1}{c}{Mean} & \multicolumn{1}{c}{P05} & \multicolumn{1}{c}{P50} & \multicolumn{1}{c}{P95} & Rank & \multicolumn{1}{c}{lower} & \multicolumn{1}{c}{two-sided} \\
\midrule
\multicolumn{10}{l}{\emph{Panel A. The four groups, inherited penalty}} \\
Market prices    &  4 &  0.643 & -0.207 & -1.125 & -0.137 &  0.626 & 38 of 40 & 0.951 & 0.146 \\
Prices           & 23 & -1.807 & -0.899 & -2.125 & -0.827 &  0.689 & 15 of 40 & 0.171 & 0.342 \\
Credit and rates & 34 &  0.407 & -1.174 & -2.705 & -1.307 &  0.326 & 38 of 40 & 0.951 & 0.146 \\
Real activity    & 62 & -5.273 & -1.429 & -3.574 & -1.621 &  0.749 &  0 of 40 & 0.024 & 0.049 \\
\addlinespace
\multicolumn{10}{l}{\emph{Panel B. Real activity against the same 40 random removals of 62 series, by penalty}} \\
Penalty halved   & 62 & -1.480 & -0.530 & -3.047 & -0.307 &  1.604 & 9 of 40 & 0.244 & 0.488 \\
Penalty inherited (Panel A) & 62 & -5.273 & -1.429 & -3.574 & -1.621 &  0.749 &  0 of 40 & 0.024 & 0.049 \\
Penalty doubled  & 62 & -2.983 & -0.490 & -1.520 & -0.527 &  0.479 &  0 of 40 & 0.024 & 0.049 \\
\bottomrule
\end{tabular}}
\begin{minipage}{0.97\textwidth}\footnotesize\vspace{4pt}
Notes: Lower block, 70 targets, benchmark \Mthree{}. $\Delta$ as in Table~\ref{tab:groups}: change in the signal's contribution when the set is removed, on the common denominator $L(\Mthree{})$; negative means the signal is worth less. Placebo: 40 subsets per size drawn without replacement from the 123 public predictors, seed fixed before any forecast, no stratification, the same subset in benchmark and challenger. Rank is the number of draws with a $\Delta$ below the observed one (mid-ranks for ties); randomization $p$-values are $(1+\#\{\Delta_{\text{draw}}\le\Delta_{\text{obs}}\})/41$ and the two-sided analogue. With 40 draws the smallest attainable lower-tail $p$ is $0.024$, so the design cannot deliver Holm significance across the four groups by construction; the test is informative about each group on its own. Panel B repeats the real-activity comparison with the inherited lower-block penalty ($0.562$) halved and doubled, using the same 40 subsets and no new draws; the signal's gain on the full benchmark is 4.50, 3.89 and 3.25 percent under the inherited, halved and doubled penalties. The three penalties were fixed before any loss was computed, and no further penalty, seed or draw was added after the results were seen.
\end{minipage}
\end{table}

\paragraph{What the placebo licenses, and what it does not.}
The decision rule fixed before the placebo was opened required two things for the paper to say that the signal's contribution depends on the composition of the public information set rather than on its size: that the real-activity removal lie below the 5th percentile of its size-matched placebo, and that this survive halving and doubling the inherited penalty. The first is met. The second is met when the penalty is doubled (again below all 40 draws, $p=0.024$) and not when it is halved: with less shrinkage the reduced benchmarks are freer to exploit particular combinations of the 60 series that remain, the placebo distribution widens from a span of 4.3 points to one of 4.7 with a much heavier right tail, and the real-activity removal, still negative at $-1.5$, falls at the 22nd percentile. The pre-specified criterion is therefore not met in full, and we state the result at the level the evidence reaches. The sign is the same under all three penalties: removing real activity lowers the signal's contribution every time. Under the inherited penalty and the stronger one, that fall is larger than any of 40 random removals of the same size. Under the weaker one it is not distinguishable from a random removal. We read this as a directional pattern of dependence on real-activity information whose exceptionality relative to panel size is conditional on the strength of regularization, and we do not select the two penalties under which it holds as the ones that count. No further penalty value, seed, draw or grouping was examined after these results were seen.

The exercise is informative in one further direction. It gives no support to the reading that the deliberative signal duplicates financial-market or credit information: removing those groups changes nothing, the signal continues to add on a benchmark that has neither, and both removals sit at the upper edge of their placebo distributions. Whatever the signal carries, it is not a repackaging of what markets have already priced. What the pattern points to instead is that the signal's content is read \emph{jointly} with the observable state of real activity: the disagreement is informative about the correction to that state, and loses its meaning when the state is not in the model. That is consistent with the reading in Section~\ref*{sec:interpretation}, and it is the form in which the paper holds it---as a reading, not as an identified complementarity.\footnote{An exploratory contrast grouping credit and market prices against real activity and prices, specified after these results were seen, is positive with interval excluding zero at every block length ($+4.1$ points, HAC $[+0.2,+7.9]$). It is not shown in Figure~\ref*{fig:groups} of the paper and is not used as evidence: a grouping chosen after the four estimates were known cannot be evidence about them.}
\section{Robustness: detail}\label{app:robust}

\paragraph{The penalties and the richness of the benchmark.}
The pre-specified penalties shrink the public panel almost entirely in the center and upper blocks (Section~\ref*{sec:protocols}). On the 70 targets the ladder confirms it: in those blocks the public benchmark is indistinguishable from the canonical regression of current growth, and the benchmark with financial conditions from the canonical regression with the index, while in the lower block the panel does its work. It also qualifies the lower-region result. The contribution of the deliberative signal in the lower block appears only when the benchmark contains the public panel and the financial-conditions index together; the index alone improves the lower block of the public panel by a large but imprecise 16.1 percent (one-sided $p=0.244$). The lower-region claim is therefore a claim about the benchmark with financial conditions, which is the one we use for it because it is the richest information set, not because it is the most favorable.

A central gain measured against a benchmark close to the canonical regression could be a gain that any cyclical indicator would deliver. To separate the two we re-estimate the benchmarks at the minimizers of the same validation criterion, $\lambda_C=5.62$ and $\lambda_U=0.316$ ($\lambda_L$ unchanged), and add the signal on top with nothing else changed. At these values the public panel moves the fitted central quantiles about twelve times more than under the pre-specified penalties, and the benchmark with financial conditions improves its central loss by 2.4 percent ($p=0.028$). The deliberative signal's gain does not fall; it rises (Table~\ref{tab:robust-lambda-spf}, Panel~A). Against the public panel it is 6.5 percent in the center ($p=0.0001$) and 5.5 percent on the grid ($p=0.0095$); against the panel with financial conditions it is 8.4 percent in the center ($p<0.0001$) and 8.2 percent on the grid ($p=0.0001$), where these and all $p$-values in this section are two-sided. The lower block is unaffected, since its penalty is the same under both rules. The upper block is where the choice of penalty matters: the point gains rise, to 8.0 percent against the public panel and 10.3 percent against the panel with financial conditions, but neither is established ($p=0.133$ and $0.087$). We therefore state the central result without qualification as to the penalty, and the upper-block result as established under the pre-specified penalties only. The same perturbation applied to attributed heterogeneity removes its content (Panel~A, last two rows): its grid gain against the public panel falls from 3.5 to 1.2 percent (two-sided $p=0.27$), its upper-block gain against the panel with financial conditions from 9.3 to $-0.8$ percent ($p=0.84$), and its grid gain against that panel to $-0.1$ percent. The index and the transparent measure respond in opposite directions to the same change in the benchmark, which is why the paper treats the second as suggestive.

\paragraph{Beyond professional forecasters.}
The first control a reader will ask for is whether the signal adds to what professional forecasters already expect. We take the median forecast of the Survey of Professional Forecasters for average real GDP growth over the four quarters after the origin, built from the survey's quarterly forecasts to match the target, from the latest survey released before the origin, and use it as the benchmark, alone and with current growth, adding the deliberative signal on top (Panel~B). Against the consensus alone the signal improves the grid by 11.0 percent, the center by 12.5 percent and the upper block by 12.8 percent, all established at the two-sided five percent level ($p$ of 0.038, 0.042 and 0.020); with current growth added to the consensus, the central gain is 14.0 percent ($p=0.022$) and the grid gain 10.7 percent ($p=0.063$). The lower block is not improved in either comparison. The consensus benchmark is weaker than the public panel---its central loss is 0.756 against 0.726 for the public panel at the validation-minimizing penalty---so on its own the exercise establishes that the signal's central content is not the consensus forecast, not that it survives a benchmark containing both.

We therefore add the consensus to the public panel at the validation-minimizing penalties, with current growth and the consensus unpenalized, and add the signal on top. The consensus requires all four of its quarterly horizons, and one early survey (1974Q3) lacks the last of them; rather than impute it, every model in this comparison starts in 1974Q4, and the same comparison without the consensus on that common history reproduces the result above (6.3 percent in the center, $p=0.001$; 5.5 percent on the grid, $p=0.008$). The center and the grid were fixed in advance as the cells read. Adding the consensus to the panel lowers the central loss from 0.734 to 0.689, the lowest of any benchmark in the paper, although that improvement is not itself distinguishable from zero ($p=0.52$). Against this benchmark the deliberative signal lowers the central loss by 12.6 percent ($p=0.003$; HAC interval excluding zero; 50 of 70 targets) and the grid loss by 11.5 percent ($p=0.008$); the lower block is unchanged ($+0.6$ percent, $p=0.79$), and the upper-block point gain of 20.7 percent is not established ($p=0.074$). Five of the 5,320 forecasts, all in the central quantiles of the augmented model, required the estimator's pre-specified fallback; excluding those five targets altogether leaves the central gain at 11.7 percent ($p=0.005$).

The signal's gain against this benchmark is larger than against the panel alone, and the deliberative index and the consensus are negatively correlated in levels ($-0.62$ over the test origins, $-0.22$ in first differences), which raises the possibility that the signal matters only once the consensus is partialled out. It does not. We re-estimated the central-quantile regressions of both models, reproducing their forecasts exactly, and recovered the signal's coefficient in each of the 630 fits (nine quantiles, 70 origins): it is positive in every fit with the consensus and in every fit without it. Conditioning on the consensus raises its median magnitude by about half, most at the upper quantiles of the center, and changes which origins carry the largest coefficients. The consensus therefore amplifies a relation that is already present rather than creating it. Because the recursive windows overlap almost entirely, the uniformity of the sign is a description of stability, not a separate piece of statistical evidence; the evidence is the loss differential.

\paragraph{The lower region and the consensus.}
The paper's lower-region claim is made against the public panel with financial conditions, so the same question has to be asked there (Table~\ref{tab:spf-lower}). All models start in 1974Q4, and the lower-block penalty is the same under both rules. On that common history, without the consensus, the deliberative signal lowers the lower-block loss by 7.0 percent (one-sided $p=0.036$, two-sided $0.071$; moving-block intervals below zero at $L=4$, 6 and 8), so the shorter history does not remove the result. Adding the consensus to this benchmark barely changes it: its lower-block loss falls by 1.1 percent, with a two-sided $p$ of 0.88. Conditional on the consensus, the signal's point gain is 5.2 percent, but it is attenuated---the mean loss differential is about a quarter smaller than without the consensus---and less precisely estimated (one-sided $p=0.146$, two-sided $0.292$; no moving-block interval excludes zero). Part of what the signal carries in the lower region is therefore shared with the consensus, and its contribution beyond the consensus is not established in this sample. The conditional differential is also concentrated: the ten targets with the largest absolute differentials account for about 92 percent of the aggregate, and the single largest, the four quarters from 2008Q1, for about half of it, with signs that are mixed within the financial crisis itself. No target is excluded from estimation or inference.

\begin{table}[!t]\centering\small
\caption{The lower region against the professional consensus (lower block, common history from 1974Q4)}\label{tab:spf-lower}
\begin{tabular}{lcccc}
\toprule
Addition & Gain (\%) & $p$, 1-sided & $p$, 2-sided & MBB, $L=4$ \\
\midrule
Hidden Dissent added to \Mthree{}                 & 6.99 & 0.036 & 0.071 & 0.009 \\
Consensus (SPF) added to \Mthree{}                & 1.09 & -- & 0.879 & 0.899 \\
Hidden Dissent added to \Mthree{} + SPF           & 5.18 & 0.146 & 0.292 & 0.469 \\
\bottomrule
\end{tabular}
\begin{minipage}{0.92\textwidth}\footnotesize\vspace{4pt}
Notes: Percentage reduction in lower-block pinball loss from the addition in each row, 70 targets, $h=4$, pre-specified lower-block penalty, all models estimated from 1974Q4 (the first quarter with all four horizons of the SPF forecast). Losses: \Mthree{} 0.4450; \Mthree{}+HD 0.4139; \Mthree{}+SPF 0.4401; \Mthree{}+SPF+HD 0.4173. HAC (Bartlett, bandwidth 3) with the Harvey--Leybourne--Newbold correction at $h=4$; the last column is the two-sided moving-block sign $p$-value. The second row compares two benchmarks and has no one-sided test. The lower-block result of Table~\ref*{tab:anatomy} starts in 1973Q1.
\end{minipage}
\end{table}
The Panel~A $p$-values of Table~\ref{tab:robust-lambda-spf} for Hidden Dissent added to \Mtwo{} (0.0095, 0.0001 and 0.133 over the whole distribution, the center and the upper block) are two-sided HLN values, as the other entries are.

\begin{table}[!t]\centering\small
\caption{The central gain under the validation-minimizing penalties and against the professional consensus}\label{tab:robust-lambda-spf}
\begin{tabular}{lcccc}
\toprule
 & All 19 & Lower & Center & Upper \\
\midrule
\multicolumn{5}{l}{\emph{Panel A. Validation-minimizing penalties}} \\
HD added to \Mtwo{}   & 5.451 & $-0.044$ & 6.51 & 8.0 \\
                   & (0.0095) &  & (0.0001) & (0.133) \\
HD added to \Mthree{} & 8.154 & 4.495 & 8.445 & 10.268 \\
                   & (0.0001) &  & ($<$0.0001) & (0.087) \\
AH added to \Mtwo{} & 1.236 & $-5.863$ & 4.183 & 0.788 \\
                   & (0.269) & (0.210) & (0.0019) & (0.787) \\
AH added to \Mthree{} & $-0.142$ &  & 3.764 & $-0.790$ \\
                   & (0.972) &  & (0.512) & (0.840) \\
\midrule
\multicolumn{5}{l}{\emph{Panel B. Against the SPF median forecast}} \\
SPF                & 10.98 & 6.60 & 12.54 & 12.83 \\
                   & (0.038) & (0.405) & (0.042) & (0.020) \\
Current growth + SPF & 10.70 & 4.80 & 14.03 & 9.83 \\
                   & (0.063) & (0.311) & (0.022) & (0.365) \\
Public panel + SPF & 11.51 & 0.60 & 12.64 & 20.68 \\
                   & (0.008) & (0.785) & (0.003) & (0.074) \\
\bottomrule
\end{tabular}
\begin{minipage}{0.92\textwidth}\footnotesize\vspace{4pt}
Notes: Percentage reduction in block pinball loss from adding the signal, unpenalized; two-sided $p$-values in parentheses, HAC (Bartlett, bandwidth 3) with the Harvey--Leybourne--Newbold correction at $h=4$; one-sided $p$-values for an improvement are half of these; 70 targets, $h=4$. Panel A: $\lambda_C=5.62$, $\lambda_U=0.316$; the lower-block penalty is the same under both rules, so the lower entries, and their inference, are those of Table~\ref*{tab:anatomy}. AH: attributed heterogeneity. Panel B: SPF is the median forecast of average growth over the four quarters after the origin, from the latest survey released before it. ``Public panel + SPF'': \Mtwo{} at the validation-minimizing penalties plus the SPF forecast, all models from 1974Q4; center and all 19 quantiles were the cells fixed in advance. 
\end{minipage}
\end{table}

\paragraph{The staff's own forecast.}
The last comparison of Section~\ref*{sec:consensus} of the paper asks whether the signal's content is the information the Committee has in front of it: the staff forecast of real GDP growth prepared for each meeting (the Greenbook until 2010, the Tealbook since), released by the Federal Reserve Bank of Philadelphia with a five-year lag and available through the meeting of December 2020. The design was fixed and dated before any forecast was made. The staff series, $\mathrm{GB}_t$, is the average of the staff's projections for quarters $t+1,\dots,t+4$ from the last Greenbook of quarter $t-1$ (horizons two to five quarters ahead, complete at every meeting used), the same construction as the SPF consensus of Section~\ref*{sec:consensus}; it enters in levels, unpenalized, with the estimation history from 1974Q4. The four new forecast sets carry $\lambda_L=0.562$, $\lambda_C=100$ and $\lambda_U=1$, the upper-block value the design stated; the reused common-history sets without the staff forecast carry the validation-minimizing center and upper penalties (Supplementary Material~\ref{app:protocol}), so the two tests fixed in advance, whose two arms are new sets, are at equal penalty, while the premise comparisons over the 19 quantiles and the paired differences $\Delta$ compare sets estimated at different penalties and are reported as descriptive; the lower-block premise comparison is at equal penalty. The sample is every origin from 2008Q4 to the last with a published Greenbook, 2021Q1, 50 targets; the 48 origins strictly free of the two public-panel series with partial 2020Q2 coverage, and the 47 outside the pandemic, are reported as sensitivities. All reused benchmarks are scored on the same targets.

Two tests were fixed in advance: Hidden Dissent added to the public panel containing the staff forecast, over the 19 quantiles (T1), and to the panel with financial conditions containing it, over the lower block (T2), with Holm adjustment over the pair; and, before them, a premise, that the staff forecast adds to the benchmark or beats the consensus. The premise fails. The staff forecast is a worse point forecast of the target than the consensus over 2008Q4--2021Q1 (RMSE 2.18 against 1.71; 1.64 against 1.50 through 2020Q1), it does not improve the public panel (6.5 percent over the 19 quantiles, one-sided $p=0.24$, moving-block interval $[-0.16, 0.04]$; $-1.3$ percent on the 48 strict origins) or the lower block of the panel with financial conditions (8.1 percent, $p=0.10$), and the panel with the staff forecast has a 13.6 percent higher loss than the panel with the consensus. The Greenbook of 2008--2021 does not carry, for the distribution of growth four quarters ahead, information beyond a rich public panel, consistent with the evidence that the Federal Reserve's forecasting advantage has narrowed since the 2000s \citep{HoeschRossiSekhposyan2023}.

Table~\ref{tab:greenbook} gives the tests. Hidden Dissent improves the panel containing the staff forecast by 12.7 percent over the whole distribution (39 of 50 targets; HLN one-sided $p=0.013$, Holm $0.025$; HAC interval $[-0.130,-0.014]$, moving-block $[-0.143,-0.035]$), by 13.6 percent in the center and 21.8 percent in the upper block, and leaves the lower block unchanged ($-1.5$ percent); on the panel with financial conditions the lower-block gain is 5.4 percent ($p=0.077$, moving-block $[-0.058, 0.002]$). The whole-distribution result holds on the 48 strict origins (10.9 percent, $p=0.011$), outside the pandemic (13.3 percent, $p=0.023$) and under the validation-minimizing center penalty (13.4 percent, $p=0.016$). The paired difference between the signal's gain with and without the staff forecast in the benchmark, descriptive because the two benchmarks carry different penalties, is negative, that is, the gain is larger with it (whole distribution: 6.6 to 12.7 percent, $\Delta=-0.032$, HAC s.e.\ 0.016, moving-block $[-0.069,-0.008]$; lower block: 1.6 to 5.4 percent, $\Delta=-0.016$, interval includes zero). The target of 2021Q2, forecast from 2020Q2, is the most influential single observation in T1, accounting for about a third of the mean differential; the sensitivities without it are those reported. Hidden Dissent correlates at 0.24 with the staff forecast's error on the target through 2020Q1 and at 0.11 once the pandemic origins are included.

Under the reading rules fixed in advance, the failure of the premise means this exercise cannot separate information from deliberation in general; what it establishes is that the signal's content is not the staff forecast, and that, as with the consensus, conditioning on the forecast sharpens the relation rather than removing it. The exercise is one of mechanism with data published five years after the fact; it changes no number in the paper's main results.

\begin{table}[!t]
\centering\small
\caption{Hidden Dissent and the staff's forecast (Greenbook/Tealbook), 50 targets 2008Q4--2021Q1}\label{tab:greenbook}
\resizebox{\textwidth}{!}{%
\begin{tabular}{llcccccc}
\toprule
Comparison & Block & Gain (\%) & $\bar d$ & $p$ (HLN) & HAC 95\% & MBB 95\% ($L=4$) & W/L \\
\midrule
\multicolumn{8}{l}{\emph{Premise: the staff forecast added to the benchmark}} \\
GB $\mid$ \Mtwo{} & All 19 & 6.51 & $-0.0394$ & 0.24 & $[-0.143, 0.064]$ & $[-0.159, 0.035]$ & 22/28 \\
GB $\mid$ \Mthree{} & Lower & 8.09 & $-0.0385$ & 0.10 & $[-0.094, 0.017]$ & $[-0.088, 0.019]$ & 27/23 \\
\addlinespace
\multicolumn{8}{l}{\emph{Tests fixed in advance: Hidden Dissent added to the benchmark containing the staff forecast}} \\
T1: HD $\mid$ \Mtwo{}+GB & All 19 & 12.71 & $-0.0719$ & 0.013 (Holm 0.025) & $[-0.130, -0.014]$ & $[-0.143, -0.035]$ & 39/11 \\
\quad & Center & 13.59 & $-0.0935$ & -- & -- & -- & -- \\
\quad & Upper & 21.84 & $-0.1108$ & -- & -- & -- & -- \\
\quad & Lower & $-1.46$ & $0.0059$ & -- & -- & -- & -- \\
T2: HD $\mid$ \Mthree{}+GB & Lower & 5.44 & $-0.0238$ & 0.077 (Holm 0.077) & $[-0.055, 0.007]$ & $[-0.058, 0.002]$ & 21/29 \\
\addlinespace
\multicolumn{8}{l}{\emph{Paired difference: gain with the staff forecast in the benchmark minus gain without it ($\Delta<0$: larger with it)}} \\
$\Delta$ T1 & All 19 & 6.55 $\to$ 12.71 & $-0.0323$ & 0.033 & $[-0.064, -0.000]$ & $[-0.069, -0.008]$ & -- \\
$\Delta$ T2 & Lower & 1.57 $\to$ 5.44 & $-0.0163$ & 0.12 & $[-0.042, 0.010]$ & $[-0.044, 0.007]$ & -- \\
\addlinespace
\multicolumn{8}{l}{\emph{Sensitivities}} \\
T1, 48 strict origins & All 19 & 10.88 & $-0.0458$ & 0.011 & $[-0.082, -0.010]$ & $[-0.087, -0.024]$ & -- \\
T1, excluding the pandemic (47) & All 19 & 13.28 & $-0.0661$ & 0.023 & $[-0.126, -0.006]$ & $[-0.139, -0.028]$ & -- \\
T1, re-tuned center penalty & Center & 13.37 & $-0.0915$ & 0.016 & $[-0.169, -0.014]$ & $[-0.177, -0.032]$ & -- \\
T2, 48 strict origins & Lower & 4.04 & $-0.0161$ & 0.17 & $[-0.047, 0.015]$ & $[-0.052, 0.008]$ & -- \\
T2, excluding the pandemic (47) & Lower & 6.17 & $-0.0201$ & 0.13 & $[-0.052, 0.012]$ & $[-0.056, 0.006]$ & -- \\
\bottomrule
\end{tabular}}
\begin{minipage}{0.97\textwidth}\footnotesize\vspace{4pt}
Notes: Percentage reduction in pinball loss; $\bar d$ is the mean loss differential (negative favors the model with the added variable); $p$-values are one-sided HLN with HAC variance (Bartlett, bandwidth 3), Holm-adjusted over \{T1, T2\} where indicated; moving-block bootstrap with $B=10{,}000$. New forecast sets at $\lambda=0.562/100/1$, the re-tuned center row at $\lambda_C=5.62$; the premise row over the 19 quantiles and the $\Delta$ rows compare them with reused sets at the validation-minimizing center and upper penalties and are descriptive. GB enters in levels, unpenalized; estimation history from 1974Q4; the public-panel series CP3Mx and COMPAPFFx are excluded at the 2020Q2--Q3 origins exactly as in the paper's main forecast sets. W/L: targets with lower/higher loss for the augmented model.
\end{minipage}
\end{table}

\paragraph{The information brought to the meeting.}
The last comparison of Section~\ref*{sec:consensus} of the paper adds the Beige Book regional dispersion (Supplementary Material~\ref*{sec:prop-beige}) to the benchmarks, as a measure of how heterogeneous the information carried into the room is before deliberation. The design was fixed and dated before any forecast was made, with the series built and validated first; the tests are the signal's own content (T0: regional dispersion added to the public panel, 19 quantiles, and to the panel with financial conditions, lower block), Hidden Dissent added to the benchmark containing it (T1, 19 quantiles; T2, lower block; Holm over the pair), the paired difference between Hidden Dissent's gain with and without the regional dispersion in the benchmark ($\Delta$), and the paired comparison of the two signals. All new forecast sets on the history from 1974Q4 with $\lambda_L=0.562$, $\lambda_C=100$ and $\lambda_U=1$, the upper-block value the design stated, 70 targets; no existing forecast set was re-estimated. T1 and T2, whose two arms are new sets, are at equal penalty; T0 against the panel with financial conditions compares sets with the same lower-block penalty and history; T0 against the public panel over the 19 quantiles and the paired difference $\Delta$ of T1 compare the new sets with sets estimated at other upper-block or center penalties (Supplementary Material~\ref{app:protocol}) and are descriptive. Table~\ref{tab:beigebook} gives the results.

The regional dispersion has no content of its own: $-1.1$ percent in the lower block of the panel with financial conditions, with a moving-block interval including zero, and, descriptively, $-1.2$ percent over the whole distribution against the public panel, where the 5 of 19 quantiles it improves are all in the upper block, at a different upper-block penalty. Neither the regional mean (0.9 percent, 10 of 19) nor the share of districts off the median sign ($-0.4$ percent) does better. Hidden Dissent added to the benchmark containing the regional dispersion improves the whole distribution by 5.2 percent (57 of 70 targets; HLN one-sided $p=0.0002$, Holm $0.0003$; moving-block $[-0.048,-0.017]$), the center by 6.7 percent and the upper block by 6.1 percent, against 5.5, 6.3 and 6.1 percent without it, a descriptive comparison with sets at other penalties under which the paired difference is zero ($\Delta=0.001$, $p=0.86$). In the lower block it improves the panel with financial conditions by 4.9 percent ($p=0.043$, Holm $0.043$; moving-block $[-0.035,-0.001]$) against 7.0 percent without it, and the paired difference is positive and at the margin ($\Delta=0.009$, one-sided $p=0.058$, moving-block $[0.000, 0.018]$), spread over the five lower quantiles. Excluding the financial crisis the lower-block gain falls to 3.2 percent, and excluding the crisis and the pandemic to 3.7 percent; five targets account for three-quarters of it, the forecast of growth over 2008 alone for a third. Over the whole distribution the gain is 7.3 percent without the two episodes. Against the regional dispersion directly, Hidden Dissent has the lower loss by 6.6 percent over the whole distribution and 8.0 percent in the lower block, with moving-block intervals below zero.

Under the reading rules fixed in advance this is the first case: the heterogeneity of the information brought to the meeting does not contain what the disagreement that leaves it contains, over the whole distribution without qualification and in the lower region with the same qualification that the consensus comparison carries. The exercise describes information content, not a mechanism.

\begin{table}[!t]
\centering\small
\caption{Hidden Dissent and the Beige Book regional dispersion, 70 targets 2008Q4--2026Q1}\label{tab:beigebook}
\resizebox{\textwidth}{!}{%
\begin{tabular}{llcccccc}
\toprule
Comparison & Block & Gain (\%) & $\bar d$ & $p$ (HLN) & HAC 95\% & MBB 95\% ($L=4$) & W/L \\
\midrule
\multicolumn{8}{l}{\emph{T0: the regional dispersion added to the benchmark}} \\
RD $\mid$ \Mtwo{} & All 19 & $-1.21$ & $0.0076$ & 0.63 & -- & $[-0.042, 0.035]$ & -- \\
RD $\mid$ \Mthree{} & Lower & $-1.08$ & $0.0048$ & 0.92 & $[-0.001, 0.011]$ & $[-0.001, 0.012]$ & 30/40 \\
\addlinespace
\multicolumn{8}{l}{\emph{T1, T2: Hidden Dissent added to the benchmark containing the regional dispersion}} \\
T1: HD $\mid$ \Mtwo{}+RD & All 19 & 5.21 & $-0.0330$ & 0.0002 (Holm 0.0003) & $[-0.049, -0.017]$ & $[-0.048, -0.017]$ & 57/13 \\
\quad & Center & 6.69 & $-0.0503$ & 0.0003 & $[-0.077, -0.024]$ & $[-0.077, -0.025]$ & 56/14 \\
\quad & Upper & 6.07 & $-0.0320$ & 0.0001 & $[-0.047, -0.017]$ & $[-0.048, -0.019]$ & 60/10 \\
\quad & Lower & 0.53 & $-0.0028$ & 0.37 & $[-0.019, 0.013]$ & $[-0.009, 0.015]$ & 32/38 \\
T2: HD $\mid$ \Mthree{}+RD & Lower & 4.89 & $-0.0220$ & 0.043 (Holm 0.043) & $[-0.046, 0.002]$ & $[-0.035, -0.001]$ & 41/29 \\
\addlinespace
\multicolumn{8}{l}{\emph{Paired difference: gain with the regional dispersion in the benchmark minus gain without it ($\Delta>0$: smaller with it)}} \\
$\Delta$ T1 & All 19 & 5.46 $\to$ 5.21 & $0.0012$ & 0.43 & -- & $[-0.010, 0.015]$ & -- \\
$\Delta$ T2 & Lower & 6.99 $\to$ 4.89 & $0.0091$ & 0.058 & -- & $[0.000, 0.018]$ & -- \\
\addlinespace
\multicolumn{8}{l}{\emph{Paired comparison: Hidden Dissent against the regional dispersion on the same benchmark}} \\
HD $-$ RD $\mid$ \Mtwo{} & All 19 & 6.59 & $-0.0417$ & -- & -- & $[-0.064, -0.009]$ & -- \\
HD $-$ RD $\mid$ \Mthree{} & Lower & 7.98 & $-0.0359$ & 0.048$^{\dagger}$ & $[-0.070, -0.002]$ & $[-0.060, -0.009]$ & 44/26 \\
\addlinespace
\multicolumn{8}{l}{\emph{Sensitivities}} \\
T1, excluding crisis and pandemic (58) & All 19 & 7.26 & $-0.0339$ & -- & -- & -- & -- \\
T2, excluding the financial crisis (61) & Lower & 3.21 & $-0.0117$ & -- & -- & -- & -- \\
T2, excluding crisis and pandemic (58) & Lower & 3.67 & $-0.0092$ & -- & -- & -- & -- \\
T1, re-tuned center penalty & Center & 6.45 & $-0.0479$ & -- & -- & -- & -- \\
Regional mean $\mid$ \Mtwo{} & All 19 & 0.89 & $-0.0056$ & -- & -- & -- & -- \\
Share off the median sign $\mid$ \Mtwo{} & All 19 & $-0.40$ & $0.0025$ & -- & -- & -- & -- \\
\bottomrule
\end{tabular}}
\begin{minipage}{0.97\textwidth}\footnotesize\vspace{4pt}
Notes: RD is the Beige Book regional dispersion, entering standardized on its available observations and unpenalized; new forecast sets at $\lambda=0.562/100/1$ (the lower-block value is the rounded form of the main sets' 0.5623), the re-tuned center row at $\lambda_C=5.62$; history from 1974Q4. The T0 row against \Mtwo{} and the $\Delta$ T1 row compare the new sets with reused sets at other penalties and are descriptive. Percentage reduction in pinball loss; $\bar d$ the mean loss differential (negative favors the model with the added variable); $p$-values one-sided HLN with HAC variance (Bartlett, bandwidth 3), $^{\dagger}$ two-sided; moving-block bootstrap with $B=10{,}000$. W/L: targets with lower/higher loss for the augmented model. The paired difference $\Delta$ in the lower block is positive at every one of the five quantiles (from 0.004 at $\tau=0.20$ to 0.014 at $\tau=0.25$).
\end{minipage}
\end{table}

\subsection{Hidden Dissent on each rung of the benchmark ladder}\label{sec:ladder}

Table~\ref{tab:hd-ladder} adds Hidden Dissent, unpenalized, to the four benchmarks of Section~\ref*{sec:lineage} of the paper, from the canonical regression on current growth to the public panel with financial conditions; Section~\ref*{sec:penalty} of the paper reads it.

\begin{table}[!htbp]
\centering\small
\caption{Hidden Dissent added to each rung of the benchmark ladder}\label{tab:hd-ladder}
\begin{tabular}{lcccc}
\toprule
Benchmark & All 19 & Lower & Center & Upper \\
\midrule
\Mzero{}: current growth (no penalty)            & 4.245 & 1.180 & 5.459 & 4.309 \\
                                                  & (0.003) & (0.305) & (0.003) & (0.003) \\
\Mone{}: + financial conditions (no penalty)      & 4.984 & $-0.324$ & 8.349 & 1.159 \\
                                                  & ($<$0.001) & (0.596) & ($<$0.001) & (0.214) \\
\Mtwo{}: current growth + public panel            & 4.066 & $-0.044$ & 5.555 & 4.303 \\
                                                  & (0.007) & (0.507) & (0.003) & (0.002) \\
\Mthree{}: + financial conditions                 & 5.927 & 4.495 & 8.335 & 1.103 \\
                                                  & ($<$0.001) & (0.006) & ($<$0.001) & (0.224) \\
\bottomrule
\end{tabular}
\begin{minipage}{0.9\textwidth}\footnotesize\vspace{4pt}
Notes: Percentage reduction in block pinball loss from adding Hidden Dissent, unpenalized; one-sided $p$-values in parentheses. \Mzero{} and \Mone{} are the canonical regressions of \citet{ABG2019}, estimated without any penalty. Benchmark losses (all 19, lower, center, upper): \Mzero{} 0.630, 0.533, 0.732, 0.542; \Mone{} 0.630, 0.467, 0.761, 0.557; \Mtwo{} 0.623, 0.508, 0.732, 0.541; \Mthree{} 0.619, 0.426, 0.760, 0.556. The comparisons against \Mzero{} and \Mone{} were fixed before their inference was computed but added after the main results were known.
\end{minipage}
\end{table}

\subsection{Formal dissents}\label{sec:formaldissents}

Section~\ref*{sec:interpretation} of the paper reads the index as disagreement that a committee deciding by consensus does not put in its vote. This subsection measures that premise: it asks whether the formal dissents of the FOMC carry the content of the index. The exercise was added after the design, at the suggestion of a reader, and is descriptive; it enters no pre-specified family and no adjustment for multiple testing.

The signal is the share of dissenting votes among the voting members at each scheduled meeting, from the record of FOMC dissents maintained at the Federal Reserve Bank of St.\ Louis \citep{ThorntonWheelock2014}, dated and standardized exactly as Hidden Dissent: the value at an origin is the mean over the meetings of the previous quarter, from 1994, standardized on its own history up to the origin. A dissent is recorded in 36 percent of the meetings between 1994 and 2025, and the series is zero at 31 of the 70 origins. Its correlation with Hidden Dissent is 0.23 in levels over the evaluation sample and 0.47 in first differences: the two move together from quarter to quarter but do not measure the same thing. The forecasts use the main penalties, with the signals unpenalized, and are compared with the forecasts of the main specification on the same seventy targets.

\begin{table}[!htbp]
\centering\small
\caption{Formal dissents and Hidden Dissent, 70 targets 2008Q4--2026Q1}\label{tab:formaldissents}
\begin{tabular}{lcccc}
\toprule
 & All 19 & Lower & Center & Upper \\
\midrule
\multicolumn{5}{l}{\emph{Do the votes add to the benchmark?}} \\
Share of dissenting votes, added to \Mtwo{}       & $-3.318$ & $-8.756$ & $-1.778$ & $-1.961$ \\
                                                  & (0.013) & (0.009) & (0.334) & ($<$0.001) \\
Share of dissenting votes, added to \Mthree{}     & $-1.141$ & 1.709 & $-2.104$ & $-0.957$ \\
                                                  & (0.406) & (0.424) & (0.355) & (0.478) \\[3pt]
\multicolumn{5}{l}{\emph{Does Hidden Dissent keep its gain once the votes are in the model?}} \\
Hidden Dissent, added to \Mtwo{} + votes          & 4.416 & $-2.860$ & 7.444 & 4.349 \\
                                                  & (0.024) & (0.326) & (0.019) & (0.008) \\
Hidden Dissent, added to \Mthree{} + votes        & 7.537 & 5.440 & 11.198 & $-0.004$ \\
                                                  & ($<$0.001) & (0.076) & ($<$0.001) & (0.999) \\[3pt]
\multicolumn{5}{l}{\emph{Other definitions of the vote, added to \Mtwo{}}} \\
Number of dissenting votes                        & $-2.934$ & $-8.144$ & $-1.658$ & $-1.146$ \\
Any dissent at the meeting                        & $-2.511$ & $-6.280$ & $-1.479$ & $-1.483$ \\
Share of dissenting votes, history from 1973      & $-0.583$ & $-0.208$ & $-0.718$ & $-0.605$ \\
\bottomrule
\end{tabular}
\begin{minipage}{0.9\textwidth}\footnotesize\vspace{4pt}
Notes: Percentage reduction in block pinball loss; negative values mean that the added signal worsens the forecast. Two-sided $p$-values (HAC, bandwidth 3, with the HLN correction) in parentheses. Main penalties in both arms of every comparison, signals unpenalized. Added after the design; descriptive.
\end{minipage}
\end{table}

The votes do not improve the forecast. Against the public panel the share of dissenting votes worsens it by 3.3 percent over the whole distribution and by 8.8 percent in the lower block, and the moving-block intervals exclude zero; five targets, three of them with origins in 2008, account for 63 percent of that deterioration, which is what an unpenalized regressor that is zero at almost half of the origins would be expected to produce. Against the panel with financial conditions no entry is distinguishable from zero, and the two other definitions of the vote give the same answer. With the votes in the model, Hidden Dissent keeps its gain over the whole distribution and in the center against both benchmarks (4.4 and 7.5 percent over the whole distribution, against 4.1 and 5.9 in the main specification; 43 and 51 of the 70 targets), with moving-block intervals that exclude zero at block lengths of four, six and eight. In the lower block against the panel with financial conditions the gain is of the size of the main specification's (5.4 against 4.5 percent) but less precisely estimated: the interval reaches zero. At the FOMC, on this evidence, the vote does not carry what the recorded disagreement carries.
\section{When the information arrives, and whether it runs through policy}\label{app:timing}

This appendix reports the two exercises summarized in Section~\ref*{sec:interpretation} of the paper. Both were fixed in a dated design before any regression was run; neither re-estimates or touches a sealed forecast set. Hidden Dissent is used at the meeting level, read from the Minutes of each scheduled meeting from February 1994 to December 2025 (255 meetings; unscheduled meetings, conference calls and notation votes excluded). The \emph{surprise} in the index at meeting $m$ is the residual of an AR(1) with constant estimated recursively on all previous scheduled meetings, with the first prediction after twenty meetings (August 1996); its standard deviation over 2005--2025 is 0.021.

\subsection{Event study: Minutes release versus decision}\label{app:timing-a}

Since January 2005 the Minutes are released three weeks after the meeting, at 2:00 pm ET; the primary sample is the 167 scheduled meetings of 2005--2025. For each meeting the response is the close-to-close change in the two-year and ten-year Treasury yields (\texttt{DGS2}, \texttt{DGS10}) and the five-year breakeven inflation rate (\texttt{T5YIE}) on the day of the Minutes release, on the day of the decision, and on a placebo day, the same weekday one week before the Minutes release. Each is regressed on the surprise in the index, the announcement surprise $u_m$ (the change in the two-year yield on the day of the decision), and, in the controlled specification, dummies for a scheduled release of the employment report, the CPI or GDP on the event day, built from the ALFRED vintage dates of \texttt{PAYEMS}, \texttt{CPIAUCSL} and \texttt{GDPC1}. Standard errors are HAC (Bartlett, lag 3); intervals are from the moving-block bootstrap ($L=4$, $B=10{,}000$, seed 20260903); the Holm adjustment is over the three Minutes-day responses. The pre-specified test is a positive coefficient on the two-year yield on the Minutes day.

\begin{table}[!htbp]
\centering
\caption{Market response to the surprise in Hidden Dissent, by window}
\label{tab:timing-a}
\small
\begin{minipage}{\textwidth}\centering
\begin{tabular}{llrrrrrr}
\toprule
Response & Window & $\beta$ & HAC s.e. & $p$ & Holm & MBB 95\% & bp per s.d. \\
\midrule
2-year  & Minutes  & 0.133 & 0.215 & 0.27$^{a}$ & 1.00 & $[-0.30,\ 0.53]$ & 0.3 \\
2-year  & Decision$^{b}$ & 0.297 & 0.325 & 0.36 & -- & $[-0.36,\ 0.92]$ & 0.6 \\
2-year  & Placebo  & $-0.117$ & 0.231 & 0.61 & -- & $[-0.57,\ 0.32]$ & $-0.2$ \\
10-year & Minutes  & 0.231 & 0.168 & 0.17 & 0.51 & $[-0.11,\ 0.55]$ & 0.5 \\
10-year & Decision & 0.001 & 0.206 & 1.00 & -- & $[-0.44,\ 0.39]$ & 0.0 \\
10-year & Placebo  & 0.009 & 0.224 & 0.97 & -- & $[-0.44,\ 0.44]$ & 0.0 \\
Breakeven & Minutes & $-0.024$ & 0.139 & 0.86 & 1.00 & $[-0.28,\ 0.29]$ & $-0.1$ \\
Breakeven & Decision & 0.256 & 0.195 & 0.19 & -- & $[-0.12,\ 0.65]$ & 0.5 \\
Breakeven & Placebo & $-0.081$ & 0.147 & 0.58 & -- & $[-0.34,\ 0.24]$ & $-0.2$ \\
\bottomrule
\end{tabular}
\par\vspace{4pt}\begin{flushleft}\footnotesize
Coefficient on the AR(1) surprise in Hidden Dissent in a regression of the daily change (percentage points) on the surprise, the announcement surprise and the release dummies; $N=167$ scheduled meetings, 2005--2025. $p$ is two-sided HAC unless marked. The last column converts the coefficient to basis points per one standard deviation of the surprise (0.021). Without the release dummies the coefficients differ by less than 0.01. $^{a}$One-sided $p$ for the pre-specified test ($\beta>0$). $^{b}$Added after the opening of the results: in the pre-specified regression the announcement surprise is the dependent variable in this window, so the cell is degenerate; the row reported omits it. The pre-specified contrast between the Minutes and decision coefficients for the two-year yield is 0.13 ($p=0.55$). The full set of cells, the first-difference surprise, the 1994--2004 sample with its release dates, and the VIX are in the replication package.
\end{flushleft}
\end{minipage}
\end{table}

No response is distinguishable from zero in any window, and the magnitudes are negligible: a one-standard-deviation surprise in the index moves the two-year yield by less than one basis point on either day. The information the index carries is public once the Minutes are released, and it is not priced.

\subsection{The path of policy}\label{app:timing-b}

Two regressions ask whether the index predicts policy beyond what the term structure implies, and whether its association with growth survives conditioning on the policy that was realized. The first regresses the change in the effective federal funds rate between quarter $t$ and $t+h$, $h=1,\dots,4$, on the standardized quarterly index (the paper's series, from the Minutes of quarter $t-1$), the level of the two-year yield, the ten-minus-two-year slope, current growth and the NFCI, over origins 1994Q1--2025Q1, with HAC standard errors at lag $h$ and the Holm adjustment over the four horizons. The second is a regression of mean growth over the four quarters after each of the paper's 70 targets on the index, current growth, the NFCI and the SPF consensus, with and without the realized change in the funds rate from the origin to the end of the target window; it is a descriptive decomposition, not a forecast, and the change in the index's coefficient is bootstrapped with paired moving blocks.

\begin{table}[!htbp]
\centering
\caption{Hidden Dissent and the path of policy}
\label{tab:timing-b}
\small
\begin{minipage}{\textwidth}\centering
\resizebox{\textwidth}{!}{%
\begin{tabular}{lrrrrrr}
\toprule
 & Coefficient & HAC s.e. & $p$ & Holm & MBB 95\% & $N$ \\
\midrule
\multicolumn{7}{l}{\emph{A. Change in the funds rate from $t$ to $t+h$ on the index, conditional on the term structure}} \\
$h=1$ & 0.255 & 0.144 & 0.078 & 0.27 & $[-0.07,\ 0.60]$ & 124 \\
$h=2$ & 0.541 & 0.308 & 0.080 & 0.27 & $[-0.07,\ 1.20]$ & 123 \\
$h=3$ & 0.799 & 0.435 & 0.066 & 0.27 & $[-0.04,\ 1.64]$ & 122 \\
$h=4$ & 0.978 & 0.549 & 0.075 & 0.27 & $[-0.06,\ 2.02]$ & 121 \\
\multicolumn{7}{l}{\emph{B. Mean growth over the target window on the index, 70 targets}} \\
Without the realized change in the funds rate & 3.260 & 1.595 & 0.041 & -- & $[-1.19,\ 5.61]$ & 70 \\
With the realized change in the funds rate & 2.247 & 1.771 & 0.204 & -- & $[-2.32,\ 4.94]$ & 70 \\
Change in the coefficient & $-1.013$ & -- & -- & -- & $[-3.15,\ 0.57]$ & 70 \\
\bottomrule
\end{tabular}}
\par\vspace{4pt}\begin{flushleft}\footnotesize
Panel A: coefficient on the standardized index, in percentage points of the funds rate per standard deviation, controlling for the two-year yield, the ten-minus-two-year slope, current growth and the NFCI; origins 1994Q1--2025Q1. The Wu--Xia shadow rate, available to February 2022, gives coefficients of 0.24 to 0.75 with $p$ between 0.15 and 0.17. Panel B: coefficient on the index in a regression of mean growth over the four quarters after the target origin on the index, current growth, the NFCI and the SPF consensus; HAC lag 3; the last row is the paired moving-block interval for the change. The coefficient falls by 31 percent, less than the half fixed in advance as the threshold for reading the association as running through policy.
\end{flushleft}
\end{minipage}
\end{table}

The index precedes higher policy rates at every horizon, with coefficients that do not survive the adjustment for multiple testing, and it precedes higher growth; a channel from disagreement to growth through policy would require the opposite sign on one of the two. Conditioning on the realized change in the funds rate removes a third of the association with growth, which is what one expects if policy responds to the growth the index anticipates.
\section{The record and the room: speaker-level evidence}\label{app:speakers}

This appendix reports the exercises summarized in the paragraphs ``Two kinds of disagreement'', ``The record, not the room'' and ``Not a matter of who is in the room'' of Section~\ref*{sec:interpretation}. All were fixed in dated designs before any forecast set was opened; none re-estimates or touches a sealed forecast set of the paper.

\subsection{Data}\label{app:speakers-data}

\citet{TsangYang2025} score each speaker's statements at each FOMC meeting with their model of dissent language, trained on the transcripts, which are public with a five-year lag. The authors made their speaker-level scores available to us for this paper: 8,021 speaker--meeting observations over 370 meetings from 1976 to 2018. The scores are private and are not redistributed; every number in this appendix is an aggregate. We classify each observation as a voting member, a non-voting participant (a governor or Reserve Bank president without a vote at that meeting) or a member of the staff, using the official Board membership list and the Reserve Banks' chronologies of presidents, with homonyms resolved by term dates; 80 rows (1.00 percent), truncated names and interim presidents, are left as staff by rule. Thirteen conference calls are excluded, leaving 357 scheduled meetings. From the scores we build four meeting-level series, the mean score of the voters ($HD_V$), of the non-voting participants ($HD_{NV}$), of all participants ($HD_P$) and of the staff ($HD_S$), and aggregate them to quarters by meeting date and dated as the paper's index. The transcript index the authors publish is the mean over voters: $HD_V$ reproduces it exactly ($r=1.000$ by meeting). Over 1994--2018 the quarterly correlation of $HD_P$ with the Minutes index of the paper is 0.70 ($HD_V$ 0.72, $HD_S$ 0.51). The three rules fixed in the design for proceeding to the forecasting exercise (ambiguous rows below 2 percent, correlation with the Minutes index of at least 0.5, and at least 36 of the 41 targets of 2008Q4--2018Q4 covered by all four series) were met (1.00 percent, 0.70, 41 of 41).

\subsection{Transcript-based disagreement in the forecasting exercise}\label{app:speakers-forecast}

Each of the four series is added, standardized expansively as the index is, to the public benchmark M2 (19 quantiles) and to M3 (lower block), over the 41 targets with origins 2008Q4--2018Q4 that the transcripts cover, with the estimator, penalties and tests of the paper. Table~\ref{tab:speakers-forecast} reports the reduction in out-of-sample pinball loss relative to the benchmark; the yardstick is the Minutes index of the paper over the same 41 targets, computed from the paper's sealed forecasts. The two pre-specified primary tests, participants against M2 and voters against M2, were conjunctive over the whole distribution and the center, with Holm within each family.

\begin{table}[!htbp]
\centering
\caption{Transcript-based disagreement added to the public benchmark, 41 targets 2008Q4--2018Q4}
\label{tab:speakers-forecast}
\small
\begin{minipage}{\textwidth}\centering
\resizebox{\textwidth}{!}{%
\begin{tabular}{lrrr}
\toprule
Signal added & All 19, vs M2 & Center, vs M2 & Lower block, vs M3 \\
\midrule
Minutes index (paper), same targets & $+3.82$ & $+6.32$ & $+5.69$ \\
All participants ($HD_P$) & $-4.53$ (0.84) & $-4.83$ (0.96) & $-29.38$ (0.88) \\
Staff ($HD_S$) & $-1.80$ (0.70) & $-2.58$ (0.86) & $-5.44$ (0.83) \\
Voting members ($HD_V$)$^{a}$ & $-3.24$ (0.77) & $-3.86$ (0.92) & $-29.48$ (0.88) \\
Non-voting participants ($HD_{NV}$) & $-4.54$ (0.89) & $-5.02$ (0.98) & $-25.07$ (0.90) \\
\midrule
\multicolumn{4}{l}{\emph{Paired loss differentials (two-sided $p$)}} \\
Participants $-$ staff & 0.014 (0.050) & 0.014 (0.007) & -- \\
Voters $-$ non-voters & $-0.007$ (0.33) & $-0.007$ (0.24) & 0.014 (0.50) \\
Minutes index $-$ voters' transcript scores & $-0.037$ (0.20) & $-0.064$ (0.010) & -- \\
\bottomrule
\end{tabular}}
\par\vspace{4pt}\begin{flushleft}\footnotesize
$^{a}$ $HD_V$ reproduces the transcript index published by the authors ($r=1.000$).
Percentage reduction in mean pinball loss relative to the benchmark (positive favors the signal); in parentheses, one-sided HLN $p$-values with HAC (Bartlett, lag 3). Moving-block bootstrap intervals ($L=4$, $B=10{,}000$, seed 20260903) include zero for every focal signal over the whole distribution. Holm-adjusted $p$-values of the two primary conjunctive tests are 1.00. Paired differentials are focal-minus-comparison loss, so a positive value favors the second signal. Design dated 18 October 2026, forecasts sealed before scoring. Long-sample complement, 1976Q2--2018Q4 (OLS of mean growth over $t+1,\dots,t+4$ on the expansively standardized signal, current growth and the NFCI, HAC lag 3): $HD_P$ $-0.04$ ($p=0.86$), $HD_S$ $-0.03$ ($p=0.90$), $N=170$.
\end{flushleft}
\end{minipage}
\end{table}

None of the four transcript-based series lowers the loss of the public benchmark; all raise it, in every block. The Minutes index lowers it over the same targets. The paired comparison of the Minutes index with the voters' transcript scores favors the Minutes in the center of the distribution ($p=0.010$) and is not distinguishable from zero over the whole distribution. The staff's disagreement, by the paired contrast, raises the loss less than the participants' does; the pre-specified reading of that family is ``not distinguished'', since neither signal improves on the benchmark. The limits are those of the window (41 targets) and of the signals: the transcript scores are noisy in the lower tail, where the loss differentials are large and imprecise, and the Minutes and transcript indices come from the same text model applied to two documents, so the comparison is between documents under one reading of dissent, not between two measures of it.

\subsection{Voting and recorded disagreement within the person}\label{app:speakers-rotation}

Table~\ref{tab:speakers-rotation} asks whether the same person expresses more or less disagreement when he or she votes. The sample is the 5,822 observations of the 110 governors and Reserve Bank presidents on the roster, 1976--2018. The coefficient of a voting indicator in a regression of the score on person and meeting fixed effects is 0.001 on the 0--1 scale of the score (standard error clustered by person 0.006; wild-cluster $p=0.86$); it is indistinguishable from zero within every chair's tenure. The gap between the mean scores of voters and non-voters, negative since 1990 (non-voters score higher), is accounted for by composition: the difference in the mean person fixed effects of the two groups explains 88 percent of the absolute gap over the sample and between 82 and 101 percent in each decade since 1980.

\begin{table}[!htbp]
\centering
\caption{Voting and recorded disagreement within the person, 1976--2018}
\label{tab:speakers-rotation}
\small
\begin{minipage}{\textwidth}\centering
\begin{tabular}{lrrr}
\toprule
 & $\beta$(voting) & s.e. & wild-cluster $p$ \\
\midrule
All meetings (person and meeting fixed effects) & 0.0010 & 0.0057 & 0.86 \\
\quad Burns--Miller & 0.011 & & 0.38 \\
\quad Volcker & $-0.003$ & & 0.82 \\
\quad Greenspan & $-0.001$ & & 0.90 \\
\quad Bernanke & 0.013 & & 0.38 \\
\quad Yellen & $-0.002$ & & 0.94 \\
\midrule
\multicolumn{4}{l}{\emph{Share of the voters--non-voters gap accounted for by composition}} \\
1970s & 31\% & 1980s & 82\% \\
1990s & 83\% & 2000s & 101\% \\
2010s & 97\% & Whole sample & 88\% \\
\bottomrule
\end{tabular}
\par\vspace{4pt}\begin{flushleft}\footnotesize
$N=5{,}822$ speaker--meeting observations, 110 persons, 357 scheduled meetings; staff excluded. Wild-cluster bootstrap by person, $B=9{,}999$. The decomposition splits each year's mean gap into the difference in mean person fixed effects between the two groups (composition) and a residual; the share is the ratio of absolute magnitudes and can exceed 100 percent when the two components offset. In the 1970s the mean gap is close to zero and the share is not informative. Only coefficients and aggregates are stored in the repository.
\end{flushleft}
\end{minipage}
\end{table}

\subsection{First stages: rotation and Board vacancies}\label{app:speakers-firststage}

Two features of the Committee change who votes without changing the economy. If recorded disagreement depended on who votes, either would be an exogenous shifter of the index. Table~\ref{tab:speakers-firststage} reports the first stages. The composition instrument is the mean person fixed effect, from the regression of Table~\ref{tab:speakers-rotation}, of the voting members of each meeting; its rotating component uses only the four Reserve Bank presidents in the annual rotation. The vacancy count is the number of empty seats on the Board at the date of each meeting, from the official membership list. The dependent variable is the Minutes index; controls are current growth, the NFCI and chair dummies; standard errors are clustered by year.

\begin{table}[!htbp]
\centering
\caption{First stages for the Minutes index}
\label{tab:speakers-firststage}
\small
\begin{minipage}{\textwidth}\centering
\footnotesize\setlength{\tabcolsep}{4pt}\begin{tabular}{llrr}
\toprule
Shifter & Sample & $\beta$ (s.e.) & Effective $F$ \\
\midrule
Composition (mean voter FE), with controls & 1994--2018, quarterly & 0.17 (0.33) & 0.27 \\
\quad without controls & 1994--2018, quarterly & 0.36 (0.30) & 1.48 \\
Rotating component only, with controls & 1994--2018, quarterly & 0.08 (0.11) & 0.53 \\
\quad without controls & 1994--2018, quarterly & 0.05 (0.11) & 0.17 \\
Board vacancies (per empty seat) & 1994--2025 & $-0.004$ (0.004) & 1.3 \\
Board vacancies & 1994--2007 & $+0.002$ (0.010) & 0.0 \\
Board vacancies & 2008--2025 & $-0.006$ (0.003) & 5.1 \\
\bottomrule
\end{tabular}
\par\vspace{4pt}\begin{flushleft}\footnotesize
Controls: current growth, NFCI, chair dummies; standard errors clustered by year (25 clusters, 1994--2018; 32 clusters, 1994--2025). The effective $F$ of a single instrument is the cluster-robust Wald statistic $\beta^2/\mathrm{s.e.}^2$. The composition instrument averages the person fixed effects of the members recorded as voting in each year; the meeting-level specifications give $F$ of 0.32 with controls and 1.59 without. The rotating component is reconstructed from the voting record (the four Reserve Bank districts with the most voting meetings in the year, each district averaged over its participants), since the roster carries no explicit rotation flag; it is the specification most sensitive to that reconstruction ($F$ between 0.16 and 0.53 across levels and controls). One standard deviation of the index is about 0.05, so one vacancy corresponds to roughly a tenth of a standard deviation. The person fixed effects, the instruments and the first stages are computed in the replication package from the speaker-level scores, which are not redistributed; no second stage is estimated, and the paper claims no exogenous shifter of the index.
\end{flushleft}
\end{minipage}
\end{table}

\subsection{Two kinds of disagreement}\label{app:speakers-quadrants}

\begin{figure}[!htbp]
\centering
\includegraphics[width=\textwidth]{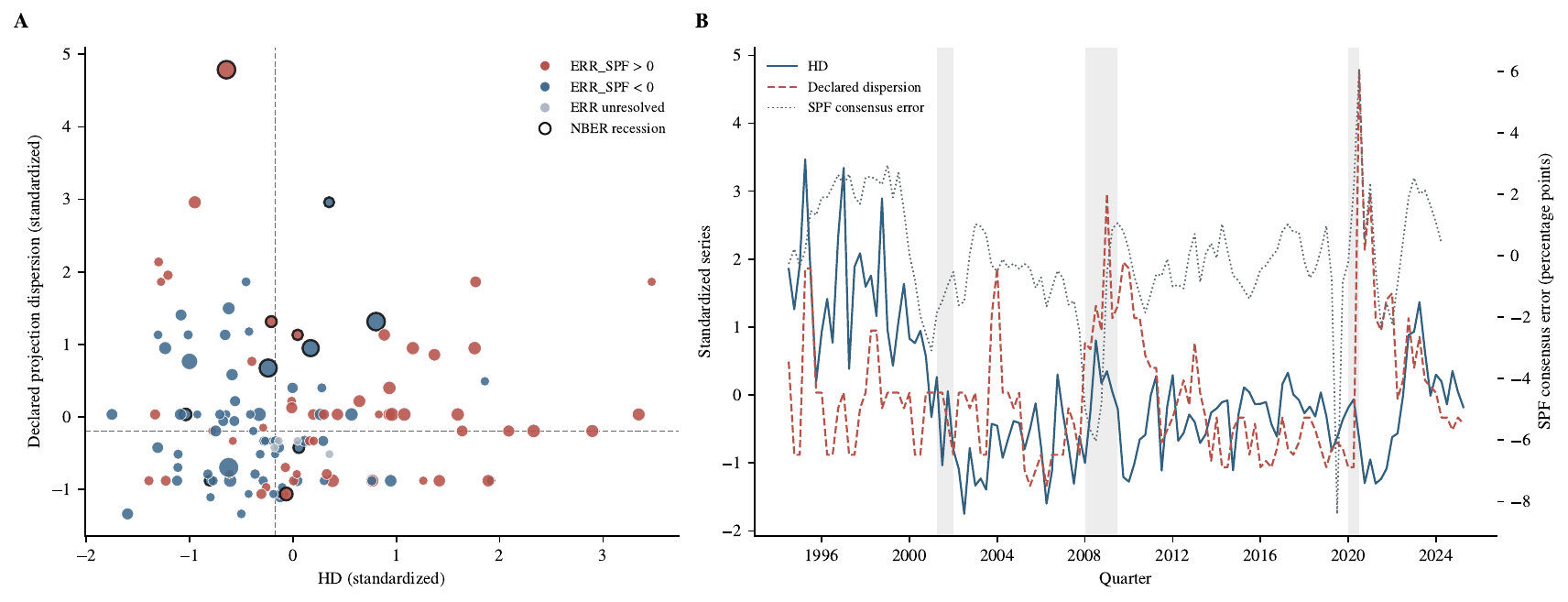}
\caption{Two kinds of disagreement. Left: the Minutes index against the declared dispersion of growth projections, both standardized, one point per quarter, colored by the sign of the consensus error (realized four-quarter growth minus the SPF median) and sized by its magnitude; the median lines define the cells of Table~\ref{tab:speakers-quadrants}; recession quarters outlined. Right: the two series over time, with the consensus error on the secondary axis and NBER recessions shaded.}
\label{fig:two-kinds}
\end{figure}

The last exercise is descriptive and was fixed, with its expected reading, before any quadrant was computed. Over 1994Q2--2025Q1 the quarters are sorted by the median of the Minutes index and by the median of the declared dispersion of growth projections (the series of Table~\ref*{tab:declared}; the SPF dispersion is used as a check). For each of the four cells, Table~\ref{tab:speakers-quadrants} reports the error the consensus went on to make, $y_{t+4}-\mathrm{SPF}_t$ with $y$ the mean of growth over $t+1,\dots,t+4$, the public measures of uncertainty and the share of quarters in or within four quarters before a recession. Figure~\ref{fig:two-kinds} plots the two series. The design assigned the figure to the body of the paper if the cell ``high index, low dispersion'' concentrated the quarters in which the consensus turned out too low and the index's gain was largest; it does not (the gain is spread, and the cells with the index above its median have a positive consensus error whatever the dispersion), so by that rule the exercise is reported here and summarized in one sentence in Section~\ref*{sec:interpretation}.

\begin{table}[!htbp]
\centering
\caption{Quarters sorted by the Minutes index and by declared dispersion, 1994Q2--2025Q1}
\label{tab:speakers-quadrants}
\small
\begin{minipage}{\textwidth}\centering
\resizebox{\textwidth}{!}{%
\begin{tabular}{llrrrrrrrr}
\toprule
Dispersion & Cell & $n$ & \multicolumn{2}{c}{Consensus error (pp)} & Share $>0$ & EPU & VIX & NFCI & Recession \\
 & & & mean & median & & \multicolumn{3}{c}{(standardized)} & or pre-4 \\
\midrule
Projections & Index low, dispersion low   & 28 & $-0.71$ & $-0.44$ & 0.21 & $-0.11$ & $-0.43$ & $-0.33$ & 0.25 \\
            & Index low, dispersion high  & 34 & $-0.68$ & $-0.58$ & 0.29 & 0.41 & 0.30 & 0.06 & 0.21 \\
            & Index high, dispersion low  & 30 & $+0.18$ & $+0.17$ & 0.47 & $-0.28$ & $-0.36$ & $-0.24$ & 0.10 \\
            & Index high, dispersion high & 32 & $+0.79$ & $+1.55$ & 0.75 & $-0.09$ & 0.40 & 0.45 & 0.19 \\
\midrule
SPF         & Index low, dispersion low   & 31 & $-0.99$ & $-0.68$ & 0.19 & $-0.14$ & $-0.37$ & $-0.29$ & 0.26 \\
            & Index low, dispersion high  & 31 & $-0.40$ & $-0.44$ & 0.32 & 0.49 & 0.31 & 0.06 & 0.19 \\
            & Index high, dispersion low  & 31 & $+0.55$ & $+0.80$ & 0.58 & $-0.42$ & $-0.28$ & $-0.26$ & 0.06 \\
            & Index high, dispersion high & 31 & $+0.48$ & $+0.78$ & 0.65 & 0.06 & 0.34 & 0.49 & 0.23 \\
\bottomrule
\end{tabular}}
\par\vspace{4pt}\begin{flushleft}\footnotesize
Cells defined by the full-sample medians of the Minutes index (0.219) and of the dispersion of growth projections (0.44 pp; SPF interquartile range 0.78 pp). Consensus error is realized mean growth over the next four quarters, final vintage, minus the SPF median, in percentage points, for the quarters with a resolved target. EPU, VIX and NFCI are quarterly means in standard deviations over the sample. The last column is the share of quarters in an NBER recession or in the four quarters preceding one. Thresholds, windows and series were fixed before the cells were computed; no $p$-values are reported, by design.
\end{flushleft}
\end{minipage}
\end{table}

Sorting by the index orders the sign of the consensus error: below the median the consensus turned out too high by about 0.7 percentage points on average, above it too low by about 0.5. Sorting by declared dispersion orders the public measures of uncertainty, higher in both high-dispersion cells, but not the sign of the error, which has the same sign on either side of the dispersion median within each half of the index. The SPF dispersion gives the same picture.
\section{Inflation as the target}\label{app:inflation}

This appendix reports the exercise summarized in the paragraph ``Not about prices'' of Section~\ref*{sec:consensus}. It asks whether the Minutes index carries information about inflation, with the design of the paper transferred to a second target and fixed, in two dated records, before any forecast was scored. Nothing in the growth exercise is touched.

\subsection{Design}

The target is the mean of the annualized quarterly inflation rate of the PCE price index over $t+1,\dots,t+4$, computed from the quarterly mean of the monthly index (three complete months) as $400\,\Delta\log$, for the same 70 targets, 2008Q4--2026Q1, and the same origins $t=q-4$ as the paper. Core PCE (\texttt{PCEPILFE}) is the primary target and headline PCE (\texttt{PCEPI}) the secondary one; each is estimated separately. The benchmark I-M2 is the public panel of 123 FRED-MD series with the paper's transformations and the exclusions of the sealed contract, with current inflation of the same index as the unpenalized conditioning variable in place of current growth; I-M3 adds the NFCI unpenalized, and I-SPF adds the SPF median forecast of the same inflation measure, averaged over the four quarters ahead (horizons 2 to 5 of the survey), as the growth consensus is built in Table~\ref*{tab:signals}. The index is added to each, standardized expansively, giving six models per target. The penalties are not inherited from the growth exercise: they are re-selected for I-M2 by the paper's rule (the largest penalty whose validation loss is within one HAC standard error of the minimum, on the 92 validation targets 1984Q4--2007Q3, grid $10^{-4}$ to $10^{2}$ in quarter-decades) and held fixed for the five other models of that target. The selection gave $\lambda=100$ for the lower and central blocks of both targets, at the upper edge of the grid, as for the central and upper blocks of the growth exercise, and $\lambda=1$ (core) and $0.32$ (headline) for the upper block; the grid was not extended, by rule. The data are the FRED-MD vintage of September 2026 and the final FRED series; the exercise is a backtest on one vintage, as the main exercise of the paper. The 228 forecast sets (6 models $\times$ 2 targets $\times$ 19 quantiles, 15,960 forecasts) were sealed and their hashes certified before any realized value was read; 875 fits (5.5 percent) used the mean-based numerical fallback of the protocol.

The primary tests, fixed before opening, are the three central-block comparisons of the index against its own benchmark for each target, I-M2+HD against I-M2, I-M3+HD against I-M3 and I-SPF+HD against I-SPF, with the paper's inference: one-sided HLN with HAC (Bartlett, lag 3), Holm within each target, moving-block bootstrap intervals ($L=4$, $B=10{,}000$, seed 20260903). The lower and upper blocks are secondary; the comparisons of I-M3 and I-SPF with I-M2 are descriptive; the partition of the targets by origin, 2021Q1--2023Q4 against the rest, was fixed in advance as a description without a test.

\subsection{Results}

\begin{table}[!htbp]
\centering
\caption{The Minutes index added to inflation benchmarks, 70 targets 2008Q4--2026Q1}
\label{tab:inflation}
\small
\begin{minipage}{\textwidth}\centering
\resizebox{\textwidth}{!}{%
\begin{tabular}{llrrrrr}
\toprule
Target & Comparison & Block & Gain (\%) & $p$ (HLN, one-sided) & Holm & MBB 95\% for $\Delta$ loss \\
\midrule
\multicolumn{7}{l}{\emph{Primary: center}} \\
Core PCE & I-M2+HD vs I-M2 & Center & $-1.28$ & 0.86 & 1.00 & $[-0.003,\ 0.009]$ \\
Core PCE & I-M3+HD vs I-M3 & Center & $-2.67$ & 0.94 & 1.00 & $[-0.002,\ 0.016]$ \\
Core PCE & I-SPF+HD vs I-SPF$^{a}$ & Center & $-15.61$ & 0.83 & 1.00 & $[-0.001,\ 1.016]$ \\
Headline PCE & I-M2+HD vs I-M2 & Center & $-0.65$ & 0.72 & 1.00 & $[-0.009,\ 0.014]$ \\
Headline PCE & I-M3+HD vs I-M3 & Center & $-0.60$ & 0.70 & 1.00 & $[-0.009,\ 0.013]$ \\
Headline PCE & I-SPF+HD vs I-SPF & Center & $-0.24$ & 0.57 & 1.00 & $[-0.018,\ 0.019]$ \\
\midrule
\multicolumn{7}{l}{\emph{Secondary: lower and upper blocks, index against I-M2}} \\
Core PCE & I-M2+HD vs I-M2 & Lower & $-1.29$ & 0.98 & -- & $[0.000,\ 0.004]$ \\
Core PCE & I-M2+HD vs I-M2 & Upper & $+2.63$ & 0.14 & -- & $[-0.025,\ 0.003]$ \\
Headline PCE & I-M2+HD vs I-M2 & Lower & $+0.93$ & 0.17 & -- & $[-0.011,\ 0.002]$ \\
Headline PCE & I-M2+HD vs I-M2 & Upper & $+4.31$ & 0.14 & -- & $[-0.070,\ 0.009]$ \\
\midrule
\multicolumn{7}{l}{\emph{Descriptive: what the NFCI and the consensus add to I-M2, center}} \\
Core PCE & I-M3 vs I-M2 & Center & $+0.35$ & 0.46 & -- & $[-0.021,\ 0.013]$ \\
Core PCE & I-SPF vs I-M2$^{a}$ & Center & $-1028$ & 0.84 & -- & $[-0.004,\ 5.948]$ \\
Headline PCE & I-M3 vs I-M2 & Center & $+1.09$ & 0.40 & -- & $[-0.060,\ 0.031]$ \\
Headline PCE & I-SPF vs I-M2 & Center & $-25.1$ & 0.93 & -- & $[-0.021,\ 0.352]$ \\
\midrule
\multicolumn{7}{l}{\emph{Added after opening: the SPF comparisons on the origins from 2010Q1 (61 targets)}} \\
Core PCE & I-SPF+HD vs I-SPF & Center & $-1.06$ & 0.78 & -- & $[-0.005,\ 0.009]$ \\
Headline PCE & I-SPF+HD vs I-SPF & Center & $+0.01$ & 0.49 & -- & $[-0.011,\ 0.008]$ \\
\bottomrule
\end{tabular}}
\par\vspace{4pt}\begin{flushleft}\footnotesize
Gain is the percentage reduction in mean pinball loss of the model with the index relative to the benchmark named (positive favors the index); mean loss of I-M2 in the center is 0.289 (core) and 0.574 (headline). $p$-values are one-sided HLN with HAC (Bartlett, lag 3); Holm over the three primary comparisons of each target; intervals from the moving-block bootstrap of the loss differential ($L=4$, $B=10{,}000$, seed 20260903), positive values favoring the benchmark. Two-sided normal $p$-values and HAC intervals are in the replication files.
$^{a}$ The SPF forecast of core PCE inflation starts in 2007Q1, so at the first origins its unpenalized coefficient is estimated on one or no training observations; one forecast (origin 2008Q1, quantile 0.95) is 1,738 points. By protocol nothing was trimmed, and the row is reported as it came. The last two rows, computed after opening on the same sealed forecasts and labelled as such, restrict both SPF comparisons to the origins from 2010Q1, where the SPF benchmark is well behaved (mean central loss 0.27 for core PCE); the cut at 2010Q4, the first origin with twelve SPF training observations, gives $-0.61$ and $+0.45$ percent.
\end{flushleft}
\end{minipage}
\end{table}

None of the six primary comparisons favors the index, and none of the secondary ones does either once the intervals are read. The 2021--2023 partition (Table~\ref{tab:inflation-partition}) is the only place where the sign turns: in the twelve targets whose origin falls in 2021Q1--2023Q4 the index lowers the central loss in all six comparisons, by 1.4 to 3.1 percent, and in the other 58 it raises it. We report it as a description. Empirical coverage of the central quantiles ranges from 0.31 to 0.79 across models and targets, a diagnostic of a one-vintage backtest over a window that contains the pandemic and the 2021--23 inflation, not a criterion.

\begin{table}[!htbp]
\centering
\caption{The 2021--2023 partition: central gain of the index by origin, fixed in advance, no test}
\label{tab:inflation-partition}
\small
\begin{minipage}{\textwidth}\centering
\resizebox{\textwidth}{!}{%
\begin{tabular}{llrr}
\toprule
Target & Comparison & Origins 2021Q1--2023Q4 ($n=12$) & Other origins ($n=58$) \\
\midrule
Core PCE & I-M2+HD vs I-M2 & $+2.86$ & $-2.42$ \\
Core PCE & I-M3+HD vs I-M3 & $+2.68$ & $-4.20$ \\
Core PCE & I-SPF+HD vs I-SPF & $+1.38$ & $-16.01$ \\
Headline PCE & I-M2+HD vs I-M2 & $+3.09$ & $-1.30$ \\
Headline PCE & I-M3+HD vs I-M3 & $+2.67$ & $-1.26$ \\
Headline PCE & I-SPF+HD vs I-SPF & $+1.68$ & $-1.05$ \\
\bottomrule
\end{tabular}}
\par\vspace{4pt}\begin{flushleft}\footnotesize
Percentage reduction in mean central pinball loss relative to the benchmark named, within each partition. The partition was fixed before opening; the subsamples are too short for a test and none is run.
\end{flushleft}
\end{minipage}
\end{table}
\section{Why the signal is left unpenalized}\label{app:regularization}

The main specification penalizes the public panel and leaves the signal, current growth and, where it defines the benchmark, the financial-conditions index unpenalized. The pre-specified robustness exercise R-2 penalizes the signal at the block penalty of the public panel---$\lambda_L=0.562$ in the lower block, $\lambda_C=100$ in the center and $\lambda_U=100$ in the upper block, on the signal and the panel alike, per observation---against the main forecast sets, which carry the same penalties, with everything else unchanged: the same 70 origins, 19 quantiles, targets, public history, nuisance predictors, transformation of the signal (standardized on its own available observations, zero before its first availability) and estimation code. Current growth and the financial-conditions index remain unpenalized. The exercise asks whether the result depends on exempting the signal from the shrinkage chosen for the 123-series block, and the answer, reported in Section~\ref*{sec:prespec}, is that it does: the grid gain against the public panel is 0.03 percent (one-sided $p=0.24$) and the lower-block gain against the panel with financial conditions 0.11 percent ($p=0.25$), with moving-block intervals containing zero at $L=4$, 6 and 8. The moving-block intervals at $L=4$, 6 and 8 contain zero in both comparisons (two-sided bootstrap $p$ between 0.38 and 0.64), and the signal wins 42 of 70 targets against the public panel and 41 of 70 in the lower block against the panel with financial conditions. The exercise comprises 2{,}660 forecasts; 91 used the estimator's pre-specified numerical fallback.

\paragraph{How much of the coefficient the penalty leaves.}
To read that result we reconstructed the signal's coefficient in both specifications at every origin and quantile of the lower and central blocks, accepting a reconstruction only if it reproduced the stored forecasts to $10^{-10}$. Because the signal enters both specifications on the same scale, the ratio of the penalized to the unpenalized coefficient measures the shrinkage directly. Table~\ref{tab:retention} reports it. In the lower block the penalized coefficient retains a median 7 percent of its unpenalized value (interquartile range 5 to 9 percent against the panel with financial conditions), less at the fifth percentile; in the center, where the block penalty is a hundred per observation, it retains 0.07 percent. No coefficient changes sign. The retention ratio is a diagnostic of the mechanism, computed after R-2 was known; it is not a third robustness exercise and it does not change R-2's verdict.

\begin{table}[!t]
\centering\small
\caption{Coefficient retention under R-2}\label{tab:retention}
\begin{tabular}{llcccc}
\toprule
Model & Block & $\lambda$ & Median $|\beta^{\mathrm{main}}|$ & Median $|\beta^{\mathrm{R\text{-}2}}|$ & Median retention \\
\midrule
\Mtwo{} + HD  & Lower  & 0.562 & 0.333 & 0.0129   & 6.78\% \\
\Mthree{} + HD & Lower  & 0.562 & 0.309 & 0.0175   & 7.05\% \\
\Mtwo{} + HD  & Center & 100   & 0.551 & 0.000353 & 0.07\% \\
\bottomrule
\end{tabular}
\begin{minipage}{0.9\textwidth}\footnotesize\vspace{4pt}
Notes: Retention is $|\beta^{\mathrm{R\text{-}2}}_{t,\tau}|/|\beta^{\mathrm{main}}_{t,\tau}|$, computed origin by origin and quantile by quantile (350 pairs per lower-block row, 630 in the center) and summarized by its median. Lower-block retention against \Mthree{} by quantile: 4.0, 11.3, 5.4, 6.6 and 9.0 percent at $\tau=0.05,\dots,0.25$. Same sign in 100 percent of pairs.
\end{minipage}
\end{table}

\paragraph{What the exercise can and cannot say.}
A penalty chosen to keep 123 correlated series from overfitting imposes, on a single regressor, a prior close to zero; any regressor treated that way loses most of its coefficient, whatever it contains. R-2 therefore shows that the result against the regularized benchmark depends on treating the signal as the variable of interest, which is the treatment the high-dimensional inference literature gives a parameter under test and the one the benchmark itself gives current growth and financial conditions. It cannot separate an absence of content from that attenuation. The canonical regressions of Table~\ref*{tab:hd-ladder}, in which no coefficient is penalized, bound the question from the other side: the signal's central and whole-distribution contributions are the same there as against the regularized panel, so what R-2 puts at issue is the lower-region result, which exists only against the panel with financial conditions. Two further facts bear on the objection that an unpenalized column is an advantage in itself. Every signal in the family and every external signal receives the same unpenalized column, and the placebo and most of the other measures worsen the forecast with it; and an unpenalized coefficient is estimated with more noise, not less, which works against an out-of-sample gain. No other regularization of the signal was specified in advance, and none is reported.
\section{Attributed heterogeneity and the second transparent measure}\label{app:ah}

\begin{table}[!t]
\centering
\caption{Attributed heterogeneity against the public panel and against the panel with financial conditions}
\label{tab:ah-anatomy}
\small
\resizebox{\textwidth}{!}{%
\begin{tabular}{l d{2.3} d{2.6} d{1.6} d{2.3} d{1.4} l c}
\toprule
Block & \multicolumn{1}{c}{Gain (\%)} & \multicolumn{1}{c}{$\bar d$} & \multicolumn{1}{c}{HAC SE} & \multicolumn{1}{c}{HLN $t$} & \multicolumn{1}{c}{$p$, one-sided} & \multicolumn{1}{c}{95\% HAC interval} & Wins \\
\midrule
\multicolumn{8}{l}{\emph{Panel A. Added to the public panel \Mtwo{}}} \\
All 19 quantiles   &  3.535 & -0.022017 & 0.009900 & -2.113 & 0.0191 & $[-0.041421,\ -0.002612]$ & 50/70 \\
Lower (Q05--Q25)   & -5.863 &  0.029799 & 0.022394 &  1.264 & 0.8948 & $[-0.014093,\ +0.073692]$ & \multicolumn{1}{c}{--} \\
Center (Q30--Q70)  &  4.858 & -0.035553 & 0.012017 & -2.811 & 0.0032 & $[-0.059107,\ -0.012000]$ & \multicolumn{1}{c}{--} \\
Upper (Q75--Q95)   &  9.136 & -0.049466 & 0.007142 & -6.580 & \multicolumn{1}{c}{$<$0.0001} & $[-0.063464,\ -0.035468]$ & \multicolumn{1}{c}{--} \\
\midrule
\multicolumn{8}{l}{\emph{Panel B. Added to the panel with financial conditions \Mthree{}}} \\
All 19 quantiles   &  2.643 & -0.016348 & 0.019101 & -0.813 & 0.2095 & $[-0.053787,\ +0.021091]$ & 46/70 \\
Lower (Q05--Q25)   & -11.533 &  0.049172 & 0.041481 &  1.126 & 0.8680 & $[-0.032131,\ +0.130474]$ & \multicolumn{1}{c}{--} \\
Center (Q30--Q70)  &  4.361 & -0.033145 & 0.039626 & -0.795 & 0.2148 & $[-0.110812,\ +0.044523]$ & \multicolumn{1}{c}{--} \\
Upper (Q75--Q95)   &  9.281 & -0.051634 & 0.006739 & -7.279 & \multicolumn{1}{c}{$<$0.0001} & $[-0.064842,\ -0.038426]$ & \multicolumn{1}{c}{--} \\
\bottomrule
\end{tabular}}
\begin{minipage}{0.97\textwidth}\footnotesize\vspace{4pt}
Notes: As Table~\ref*{tab:anatomy}. The pre-specified tests for attributed heterogeneity are Panel~A, all 19 quantiles, and Panel~B, lower block; Holm is applied across it and the companion measure of explicit directional disagreement: on the first the adjusted $p$ is 0.038 (raw 0.019), on the second neither measure improves the benchmark ($p$ of 0.868 and 0.932). Under the validation-minimizing penalties the Panel~A whole-distribution gain is 1.24 percent (one-sided $p=0.134$) and the Panel~B upper-block gain $-0.79$ percent (Supplementary Material~\ref{app:robust}).
\end{minipage}
\end{table}

\paragraph{The transparent reading of division.}
Table~\ref{tab:ah-anatomy} repeats the two comparisons for attributed heterogeneity, the member of the family added last (Section~\ref*{sec:setting}). Against the public panel the measure lowers the grid loss by 3.5 percent, and the reduction is established (one-sided $p=0.019$; moving-block intervals below zero at $L=4$, 6 and 8, Appendix~\ref{app:tables}). Its anatomy is not the index's. The center gain, 4.9 percent, and the upper-block gain, 9.1 percent, are both established, the latter with a $t$-statistic above six; the lower block is worsened by 5.9 percent, with an interval that contains zero. Against the panel with financial conditions the grid reduction, 2.6 percent, is no longer established ($p=0.21$), the lower block is worsened by 11.5 percent, and the center point estimate is unchanged at 4.4 percent but with a standard error three times larger, so that it is not established either; the upper-block gain is unchanged at 9.3 percent and remains established with a narrow interval. On the two tests the paper applies to every signal the verdict is therefore split: the grid against the public panel is met, and survives the Holm adjustment across the two transparent measures ($p=0.038$), while the lower block against the panel with financial conditions is not. The upper-block entry was not a pre-specified test and is reported as anatomy, not as a tested claim. It also does not survive re-tuning the benchmark. At the penalties that minimize validation loss, under which the public panel is at work in the upper block, the measure's upper-block gain against the panel with financial conditions falls from 9.3 to $-0.8$ percent (two-sided $p=0.84$), and its grid gain against the public panel---the test it met---falls from 3.5 to 1.2 percent (one-sided $p=0.13$); the same perturbation raises the index's grid gain from 4.1 to 5.5 percent (Section~\ref*{sec:rich}). What the transparent measure appeared to add in the upper region is, on this evidence, public information that the heavily penalized benchmark was not using. We therefore treat it as suggestive rather than established, and do not read the two measures as two dimensions of the deliberation that inform different parts of the distribution.

\begin{table}[!t]
\centering
\caption{Attributed heterogeneity under final-vintage and real-time public information}
\label{tab:ah-realtime}
\small
\resizebox{\textwidth}{!}{%
\begin{tabular}{l d{2.3} d{2.3} l l d{2.6}}
\toprule
 & \multicolumn{2}{c}{Gain (\%)} & \multicolumn{2}{c}{Real-time inference} & \multicolumn{1}{c}{Change in $\bar d$} \\
\cmidrule(lr){2-3}\cmidrule(lr){4-5}
Block & \multicolumn{1}{c}{Final} & \multicolumn{1}{c}{Real time} & \multicolumn{1}{c}{$p$, one-sided} & \multicolumn{1}{c}{95\% HAC interval} & \multicolumn{1}{c}{RT $-$ final} \\
\midrule
\multicolumn{6}{l}{\emph{Panel A. Against the public panel \Mtwo{}}} \\
All 19 quantiles  &  3.540 &  3.357 & 0.024 & $[-0.0402,\ -0.0016]$ &  0.001182 \\
Lower (Q05--Q25)  & -5.725 & -6.697 & 0.917 & $[-0.0112,\ +0.0793]$ &  0.004653 \\
Center (Q30--Q70) &  4.850 &  4.859 & 0.003 & $[-0.0592,\ -0.0120]$ & -0.000071 \\
Upper (Q75--Q95)  &  9.138 &  9.142 & $<0.001$ & $[-0.0635,\ -0.0355]$ & -0.000032 \\
\addlinespace
\multicolumn{6}{l}{\emph{Panel B. Against the public panel with financial conditions \Mthree{}}} \\
All 19 quantiles  &  2.840 &  2.703 & 0.205 & $[-0.0543,\ +0.0208]$ &  0.000828 \\
Lower (Q05--Q25)  & -10.384 & -11.088 & 0.873 & $[-0.0296,\ +0.1251]$ &  0.003296 \\
Center (Q30--Q70) &  4.363 &  4.366 & 0.214 & $[-0.1108,\ +0.0444]$ & -0.000032 \\
Upper (Q75--Q95)  &  9.281 &  9.295 & $<0.001$ & $[-0.0649,\ -0.0385]$ & -0.000094 \\
\bottomrule
\end{tabular}}
\begin{minipage}{0.97\textwidth}\footnotesize\vspace{4pt}
Notes: As Table~\ref*{tab:realtime}, with the same matched benchmarks, origins and inference. The final-vintage columns are matched quantities, which is why they differ slightly from Table~\ref{tab:ah-anatomy} (3.540 against 3.535 on the grid of Panel~A; $-10.384$ against $-11.533$ in the lower block of Panel~B, where the common-support restriction matters most).
\end{minipage}
\end{table}

\paragraph{Real-time public information.}
The transparent measure is put through the same two real-time arms as the index (Table~\ref{tab:ah-realtime}). Its profile survives the change of information set almost unaltered. Against the public panel the grid gain moves from 3.54 to 3.36 percent and remains established ($p=0.024$), the center from 4.85 to 4.86, and the upper block from 9.138 to 9.142. Against the panel with financial conditions the upper block moves from 9.281 to 9.295 percent, with a paired change in the loss differential of $-0.0001$, and remains established with a narrow interval, while the grid and the center remain unestablished and the lower block remains worsened, by 11.1 percent. Whatever the upper-region content of attributed heterogeneity is, it is not a product of revised public data, and it is as precise under vintage-specific information as under final data. The statement is descriptive: the upper block was not a pre-specified test for the measure, and the real-time arm does not change its status.

The lower-block deterioration has a precise source, which we report as a diagnostic formed after the results were known. It sits almost entirely in the five targets of the global financial crisis, 2008Q4 to 2009Q4: their loss differentials sum to more than the whole sample's, and without them the lower-block gain against the panel with financial conditions in real time is $+0.2$ percent rather than $-11.1$. The mechanism is visible in the forecasts. At the origins of 2008 the Minutes partitioned the room unusually often, and the model read that as a reason to raise the lower quantiles: at the origin of 2008Q2, the fifteenth-percentile forecast for growth over the following year rises from $-2.7$ to $+0.2$ percent when the measure is added, against a realized $-4.0$. The pandemic quarters work the other way, so the pattern is not one of crises in general. The same diagnostic applied to Hidden Dissent shows the opposite dependence: the crisis amplifies its lower-block gain but does not produce it, which is 3.6 percent without those five targets against 4.1 with them (Appendix~\ref{app:tables}).

\paragraph{The two readings together, and the second transparent measure.}
Two further results close the evaluation of the transparent measures. First, when both measures are in the model with financial conditions, neither absorbs the other where it acts. Hidden Dissent added to the benchmark that already contains attributed heterogeneity improves the lower block by 10.5 percent (HAC one-sided $p=0.065$; moving-block intervals below zero at $L=4$, 6 and 8), and attributed heterogeneity added to the benchmark that already contains Hidden Dissent improves the upper block by 10.8 percent, with a narrow interval. We read the first as a statement that the index's lower-region content is not reproduced by the transparent measure, not as a stronger contribution: part of the gain corrects the deterioration that attributed heterogeneity itself introduces in that block, and the model with three unpenalized focal terms estimates the tails with more variance, which the loss of HAC precision reflects. The second is anatomy under the same rule as before. The joint model answers a question recorded in the construct record before any forecasting outcome, whether the trained index adds to its transparent counterpart; its blocks and inference were fixed before it was opened. Second, the companion measure of explicit directional disagreement, the family's second hypothesis, carries no content: it worsens the grid against the public panel by 3.5 percent ($p=0.79$) and the lower block against the panel with financial conditions by 14.2 percent ($p=0.93$), and improves only the upper block, by 3.3 and 1.9 percent. Open conflict, as the Minutes record it, is rare and formalized, and it does not inform the distribution; what does is the latent tension the index reads and the plurality the Minutes acknowledge.
\section{Additional tables}\label{app:tables}

This appendix reports the per-quantile detail for Hidden Dissent and attributed heterogeneity, the moving-block inference for attributed heterogeneity, the lower block by episode, and the directional-disagreement measure. Three sets are not reported: the per-quantile detail for the other measures of Table~\ref*{tab:overview}, the full event-by-signal anatomy, and the horizon-two results.

\subsection*{Attributed heterogeneity: quantile anatomy and moving-block inference}

\begin{table}[H]
\centering
\caption{Gain by quantile from adding Hidden Dissent and attributed heterogeneity to the public panel \Mtwo{}}
\label{tab:ah-quantiles}
\small
\begin{tabular}{c d{2.3} d{2.3} c}
\toprule
$\tau$ & \multicolumn{1}{c}{Hidden Dissent} & \multicolumn{1}{c}{Attributed heterogeneity} & AH improves \\
\midrule
0.05 &  1.986 &  -4.898 & no \\
0.10 & -3.479 & -18.336 & no \\
0.15 & -1.999 &  -7.381 & no \\
0.20 & -0.761 &  -1.584 & no \\
0.25 &  3.677 &   0.365 & yes \\
0.30 &  2.010 &  -4.057 & no \\
0.35 &  0.073 &  -4.299 & no \\
0.40 &  3.561 &  -1.164 & no \\
0.45 &  8.504 &   3.504 & yes \\
0.50 &  6.214 &   3.404 & yes \\
0.55 &  5.422 &   3.801 & yes \\
0.60 &  6.967 &   9.910 & yes \\
0.65 &  6.331 &  12.130 & yes \\
0.70 &  9.688 &  17.501 & yes \\
0.75 &  7.369 &  18.643 & yes \\
0.80 &  8.094 &  10.537 & yes \\
0.85 &  1.726 &   7.972 & yes \\
0.90 &  1.123 &   5.961 & yes \\
0.95 & -1.697 &  -9.702 & no \\
\bottomrule
\end{tabular}
\begin{minipage}{0.8\textwidth}\footnotesize\vspace{4pt}
Notes: Percentage reduction in pinball loss at each quantile from adding the measure, unpenalized, to \Mtwo{}, on the 70 targets; the block entries of Tables~\ref*{tab:anatomy} and \ref{tab:ah-anatomy} average these. The two measures improve the forecast at different places: the index at 15 of 19 quantiles, spread over the body; the transparent measure at 11, concentrated between $\tau=0.45$ and $0.90$ with its largest gains at $\tau=0.60$ to $0.80$, and with losses at both ends of the grid that exceed any of the index's. Reported as description; no quantile or region was selected on the basis of this table.
\end{minipage}
\end{table}

\begin{table}[H]
\centering
\caption{Moving-block bootstrap for attributed heterogeneity added to \Mtwo{}}
\label{tab:ah-mbb}
\small
\begin{tabular}{l d{2.6} d{1.6} l d{1.4}}
\toprule
Block & \multicolumn{1}{c}{$\bar d$} & \multicolumn{1}{c}{Bootstrap SE} & \multicolumn{1}{c}{Percentile 95\% interval, $L=4$} & \multicolumn{1}{c}{$P(\bar d^{*}\le 0)$} \\
\midrule
All 19 quantiles  & -0.022017 & 0.009595 & $[-0.042210,\ -0.004820]$ & 0.9917 \\
Lower (Q05--Q25)  &  0.029799 & 0.020841 & $[-0.007286,\ +0.073253]$ & 0.0824 \\
Center (Q30--Q70) & -0.035553 & 0.012346 & $[-0.062488,\ -0.015296]$ & 1.0000 \\
Upper (Q75--Q95)  & -0.049466 & 0.007053 & $[-0.062043,\ -0.034349]$ & 1.0000 \\
\bottomrule
\end{tabular}
\begin{minipage}{0.9\textwidth}\footnotesize\vspace{4pt}
Notes: Non-circular moving-block bootstrap of the 70 target-level block differentials, 10{,}000 replications, block length $L=4$ primary. The bootstrap is not centered under the null, so the last column is a sampling-distribution diagnostic and not a $p$-value. At $L=6$ and $L=8$ the grid interval is $[-0.0431,\ -0.0088]$ and $[-0.0426,\ -0.0105]$; the center and upper intervals remain entirely below zero and the lower interval continues to contain it at every block length. Against \Mthree{} (Table~\ref{tab:ah-anatomy}, Panel~B) the $L=4$ percentile intervals, expressed as gains, are $[-26.7,\ +2.0]$ percent in the lower block, $[-7.9,\ +12.2]$ in the center and $[+6.6,\ +11.1]$ in the upper block; the last excludes zero at every block length.
\end{minipage}
\end{table}

\subsection*{Attributed heterogeneity in the lower block: the financial crisis}

\begin{table}[H]
\centering
\caption{Hidden Dissent and attributed heterogeneity added to the real-time panel with financial conditions: lower block with and without the financial crisis}
\label{tab:ah-gfc}
\small
\begin{tabular}{l d{2.2} d{2.2} d{2.2} d{2.2} d{2.2} d{3.2}}
\toprule
 & \multicolumn{1}{c}{$\tau=0.05$} & \multicolumn{1}{c}{$0.10$} & \multicolumn{1}{c}{$0.15$} & \multicolumn{1}{c}{$0.20$} & \multicolumn{1}{c}{$0.25$} & \multicolumn{1}{c}{Block} \\
\midrule
\multicolumn{7}{l}{\emph{Hidden Dissent}} \\
All 70 targets            & -1.59 &   1.51 &  -0.94 &   7.41 &  8.75 &   4.12 \\
Without 2008Q4--2009Q4 ($N=65$) & -1.51 &   1.28 &  -1.31 &   6.66 &  7.88 &   3.59 \\
\addlinespace
\multicolumn{7}{l}{\emph{Attributed heterogeneity}} \\
All 70 targets            & -6.78 & -18.64 & -21.63 & -11.92 & -0.50 & -11.09 \\
Without 2008Q4--2009Q4 ($N=65$) & -4.53 &   0.69 &   0.59 &   0.14 &  1.48 &   0.15 \\
\bottomrule
\end{tabular}
\begin{minipage}{0.92\textwidth}\footnotesize\vspace{4pt}
Notes: Percentage reduction in pinball loss; same forecasts, no re-estimation. The financial crisis amplifies the index's lower-block gain but does not account for it: without the five crisis targets the gain is 3.6 percent rather than 4.1, concentrated as before at $\tau=0.20$ and $0.25$, and the fifth and fifteenth percentiles remain slightly negative. For attributed heterogeneity the sum of the 70 target-level block differentials is $+3.345$; the five crisis targets contribute $+3.386$ and the remaining 65 contribute $-0.041$, so the crisis accounts for the whole deterioration and the rest of the sample slightly favors the measure. Removing the three pandemic targets instead moves the block from $-11.1$ to $-14.8$ percent. Diagnostic formed after the results were known; it is not a robustness check and no claim rests on it.
\end{minipage}
\end{table}

\subsection*{Explicit directional disagreement}

\begin{table}[H]
\centering
\caption{Explicit directional disagreement added to the two benchmarks}
\label{tab:divergence}
\small
\begin{tabular}{l d{2.3} d{2.3} d{2.3} d{2.3}}
\toprule
 & \multicolumn{1}{c}{All 19} & \multicolumn{1}{c}{Lower} & \multicolumn{1}{c}{Center} & \multicolumn{1}{c}{Upper} \\
\midrule
Added to \Mtwo{}   & -3.511 &  -3.831 & -6.203 & 3.338 \\
Added to \Mthree{} & -4.188 & -14.155 & -3.540 & 1.856 \\
\bottomrule
\end{tabular}
\begin{minipage}{0.8\textwidth}\footnotesize\vspace{4pt}
Notes: Percentage reduction in block pinball loss, 70 targets, same design as Table~\ref*{tab:anatomy}. Endpoints fixed before any forecast of the measure was generated: all 19 quantiles against \Mtwo{} (one-sided $p=0.789$) and lower block against \Mthree{} ($p=0.932$); moving-block intervals give the same reading. The other blocks are description.
\end{minipage}
\end{table}

\subsection*{Full notes to Tables \ref*{tab:signals}, \ref*{tab:overview}, \ref*{tab:realtime}, \ref*{tab:rt-lower} and \ref*{tab:events} of the paper}

The notes of these tables were shortened in the paper; the complete versions, with the estimation details of every row, follow.

\paragraph{Table \ref*{tab:signals}.} The seven Minutes-based measures form the deliberative family; the Minutes are public three weeks after the meeting. Deliberative measures are quarterly means over the meetings of the quarter, dated so that the value used at origin $t$ comes from meetings in quarter $t-1$. All cover the 70 test origins; the quantifier discount and the balance of risks use the preceding available quarter at 2 and 4 origins. The projection measures are dated by the publication date of each Monetary Policy Report or SEP; the regional dispersion by the Beige Book's, published two weeks before each meeting, with the twelve district sections coded on a five-point scale of activity (Section~\ref{app:properties}). Minutes-based signals that start after the public history are standardized on their own available observations and set to zero before them (Section~\ref*{sec:protocols}); the projection measures enter in levels and are set to zero before 1983Q1. Section~\ref{app:data} gives the construction of every series.

\paragraph{Table \ref*{tab:overview}.} Entries are $100\times[L(\Mtwo{})-L(\Mtwo{}{+}s)]/L(\Mtwo{})$ on the block pinball loss over the 70 targets (2008Q4--2026Q1); the ex-pandemic column drops 2020Q1--2020Q3 from evaluation only. Every measure enters unpenalized, with the pre-specified penalties on the public panel. Paired inference for Hidden Dissent is in Table~\ref*{tab:anatomy} and for the other rows in the Supplementary Material. Holm adjustment within each block, separately for the eight rows fixed in advance (Panel B without attributed heterogeneity, and the two reference rows) and for the five external signals: the Hidden Dissent entries for the whole distribution, center and upper block survive; among the external signals only the upper-block entry for the oil price does. Attributed heterogeneity was added after the other rows were known and is not in that family. $^{*}$Regional dispersion: dispersion across the twelve district sections of the Beige Book of the coded tone of activity, from 1983Q4 (Section~\ref*{sec:consensus}); its forecast sets were estimated on the common history from 1974Q4 with an upper-block penalty of 1, the value the design then stated, so no main-specification set differs from them only in that column, and its own content is reported as descriptive in Section~\ref{app:robust}, while Hidden Dissent's gain conditional on it, which compares two sets estimated alike, is in Table~\ref*{tab:robust}. Central tendency and dispersion: midpoint and width of the central tendency of the Committee's growth projections (Monetary Policy Reports 1983Q1--2007Q3, SEP from 2007Q4), in levels. Listed first because they precede the discussion. Each is measured against a public panel that differs from it only in that column, on the history from 1974Q4 (central tendency: pre-specified center penalty, center only; dispersion: the pre-specified lower-block penalty and 5.62 in the center, Section~\ref*{sec:rich}), so the rows are not strictly comparable with the others; the dispersion's whole-distribution and upper-block entries are not reported because the common-history benchmark's upper-block forecasts carry a different penalty from the dispersion's (Section~\ref{app:sep} gives them as descriptive). No entry is distinguishable from zero under HLN or under the moving-block bootstrap. Figure~\ref*{fig:sep} gives Hidden Dissent conditional on each measure.

\paragraph{Table \ref*{tab:realtime}.} 70 targets, pre-specified penalties in both arms of every comparison, Hidden Dissent unpenalized; in Panel B the NFCI is unpenalized in both arms and placed on the same real-time clock. Within each panel both benchmarks are restricted, at each origin, to the predictors constructible from the historical vintage under the release-lag rule, so the final-vintage columns are matched benchmarks and not the unrestricted specifications of Table~\ref*{tab:anatomy}. In Panel A they reproduce the unrestricted losses to within a thousandth in every block; in Panel B they reproduce the center and upper gains to within a hundredth of a percentage point but place the lower-block gain at 5.099 rather than 4.495, a difference due entirely to the common-support restriction (Section~\ref{app:realtime}). Real-time inference as in Table~\ref*{tab:anatomy}. The last column is the paired change in the mean loss differential, $\bar d^{\,\mathrm{RT}}-\bar d^{\,\mathrm{final}}$; positive values mean the signal's advantage weakens. It is a stability diagnostic, not an equivalence test. ``With GDP vintages'' replaces current growth, in benchmark and challenger alike, by its value in the vintage available at each origin, with the realized target unchanged; $^{a}$paired change against the real-time row above it (two-sided $p=0.389$).

\paragraph{Table \ref*{tab:rt-lower}.} Lower block (Q05--Q25), 70 targets, pre-specified penalties. The benchmark is the matched real-time public panel plus the NFCI on the same real-time clock, unpenalized; each signal enters unpenalized as the only difference between benchmark and challenger. Inference on the block loss differential as in Table~\ref*{tab:anatomy}. Moving-block percentile intervals lie below zero for Hidden Dissent and contain zero for every other signal. The paired columns test $d^{\HD}_t-d^{s}_t$ on the same benchmark; negative values favor Hidden Dissent. Holm adjustment within the two families of paired comparisons: among the Minutes measures only the comparison with verbosity survives ($p_{\text{Holm}}=0.0096$); among the five external signals, pre-specified as a set, none does (policy uncertainty $0.12$). Attributed heterogeneity was added to the family after the other rows were known. $^{*}$Beige Book and long-history projection measures (Table~\ref*{tab:overview}), added to the benchmark with financial conditions on final-vintage data and the history from 1974Q4, not to the real-time benchmark of the other rows; the paired columns compare each with Hidden Dissent on that benchmark. Hidden Dissent has the lower loss in 44 of 70 targets against the regional dispersion, 35 against the central tendency and 46 against the dispersion; the moving-block intervals of the first and last differences lie below zero.

\paragraph{Table \ref*{tab:events}.} Percentage reduction in block pinball loss from adding Hidden Dissent to \Mtwo{}, computed within each subset of targets on the forecasts of Table~\ref*{tab:anatomy}, without re-estimation. The seven windows and the non-event complement were fixed before any performance was examined; no window was redefined, shifted or dropped afterwards. The windows partition the 32 event targets. Targets won by the augmented model over the whole distribution: 23 of 32 inside the windows, 30 of 38 outside. The difference between the event and non-event gains over the whole distribution is $-1.6$ percentage points, with HAC/HLN two-sided $p=0.52$, HAC interval $[-6.5,\ 3.3]$ and moving-block interval $[-7.2,\ 3.3]$, computed from the same 70 target differentials; the third row is descriptive.
\section{Exploratory diagnostics on the model's response to the signal}\label{app:orient}

The diagnostics in this appendix were formed after the forecasting results were known. They use only the forecasts already generated---no realized outcomes, no losses, no re-estimation---and they are reported as illustrations of the reading offered in Section~\ref*{sec:interpretation}, not as evidence for it. All signals are reported, including the placebo.

\paragraph{The model's response to the signal.}
For each signal we classify the 70 origins by the value of the signal actually seen by the model at that origin, freezing the classification before examining any forecast, and compare the change in the predicted quantiles between the augmented and the benchmark model in high and low states. For the deliberative signal the response is positive at every block---$+0.53$ in the lower, $+0.74$ in the center and $+0.14$ in the upper, in percentage points of predicted growth---so a higher reading raises the predicted path and slightly compresses it. Economic policy uncertainty and implied volatility show the opposite signature, lowering the bottom of the distribution and raising the top ($-1.09$ and $+0.28$; $-1.63$ and $+0.81$), which is the widening one expects of an uncertainty measure. The placebo also produces a systematic response ($-0.49$, $-0.19$, $+0.29$), which is why this diagnostic cannot be read as a measure of predictive value: a fitted model responds to whatever it is given.

\paragraph{The state of the economy in high- and low-signal quarters.}
In the 70 evaluation origins, quarters in the highest fifth of measured disagreement begin with current growth of 1.84 percent against 3.67 percent in the lowest fifth, a gap of $-1.83$ percentage points; over the following four quarters the gap narrows to $-0.90$, so that growth relative to its starting point improves by 0.93 points more in high- than in low-disagreement quarters. The association between the signal and the contemporaneous state is not stable across samples: it is weakly positive over the full 1994--2026 record and negative over the evaluation origins. Measured disagreement is slightly lower in NBER recession quarters than in expansions over the full record.

\paragraph{What these diagnostics do not establish.}
The narrowing of the growth gap is consistent with ordinary mean reversion from a depressed starting point and cannot by itself distinguish that from the reading in Section~\ref*{sec:interpretation}. What distinguishes them is the design: the benchmark conditions on current growth as an unpenalized regressor, so reversion that is mechanically predictable from the observed state is already available to it, and the evidence for the reading is the incremental out-of-sample gain rather than anything in this appendix.

\end{document}